\documentclass[aps,pre,floats,twocolumn,superscriptaddress,reprint]{revtex4-1}

\usepackage{graphicx,epsfig}
\usepackage{graphics,dcolumn,bm,epic, eepic,fleqn,float}
\usepackage{amssymb,amsmath,amsfonts,multirow,rotate,color}

\newcommand{\up}[1]{^{(#1)}}
\newcommand{\avg}[1]{\langle #1 \rangle}

\newcommand{\mfptnorm}{\widetilde{\tau}}
\newcommand{\cctnorm}{\widetilde{\gamma}}

\usepackage[table]{xcolor}
\usepackage{flushend}

\DeclareMathOperator*{\argmin}{argmin}
\begin{document}

\title{Disproportionate incidence of COVID-19 in African Americans
  correlates with dynamic segregation}

\author{Aleix Bassolas}
\affiliation{School of Mathematical Sciences, Queen Mary University
  of London, London E1 4NS, United Kingdom}

\author{Sandro Sousa}
\affiliation{School of Mathematical Sciences, Queen Mary University
  of London, London E1 4NS, United Kingdom}

\author{Vincenzo Nicosia}
\affiliation{School of Mathematical Sciences, Queen Mary University
  of London, London E1 4NS, United Kingdom}

\date{\today}

\begin{abstract}
  Socio-economic disparities quite often have a central role in
  the unfolding of large-scale catastrophic events. One of the most
  concerning aspects of the ongoing COVID-19
  pandemics~\cite{Covid2020} is that it disproportionately affects
  people from Black and African American
  backgrounds~\cite{Millett2020,DiMaggio2020,Goldstein2020,Cyrus2020,UKCovidDisparity2020},
  creating an unexpected infection gap. Interestingly, the abnormal
  impact on these ethnic groups seem to be almost uncorrelated with
  other risk factors, including co-morbidity, poverty, level of
  education, access to healthcare, residential segregation, and
  response to
  cures~\cite{Raisi-Estabragh2020,Li2020,Prats-Uribe2020,Takagi2020,Khanijahani2020}. A
  proposed explanation for the observed incidence gap is that people
  from African American backgrounds are more often employed in
  low-income service jobs, and are thus more exposed to infection
  through face-to-face contacts~\cite{Bilal2020}, but the lack of
  direct data has not allowed to draw strong conclusions in this sense
  so far. Here we introduce the concept of dynamic segregation, that
  is the extent to which a given group of people is internally
  clustered or exposed to other groups, as a result of mobility and
  commuting habits. By analysing census and mobility data on more than
  120 major US cities, we found that the dynamic segregation of
  African American communities is significantly associated with the
  weekly excess COVID-19 incidence and mortality in those
  communities. The results confirm that knowing where people commute
  to, rather than where they live, is much more relevant for disease
  modelling.
\end{abstract}

\maketitle

The spread of a non-air-borne virus like COVID-19 is mostly mediated
by direct face-to-face contacts with other infected people. This is
why the first measures attempting at containing the spread of the
virus included the introduction of travel restrictions, social
distancing, curfews, and stay-at-home
orders~\cite{Chinazzi2020,Buckee2020,Kraemer2020,Aleta2020}.  However,
the distribution of the number of contacts per person is known to be
fat-tailed~\cite{Brown2013}, so that most of the infections are
actually caused by a relatively small set of individuals, called
\textit{super-spreaders}~\cite{Galvani2005,Stein2011}, who normally
have a disproportionately high number of face-to-face
contacts. Intuitively enough, super-spreaders are most commonly found
among service workers --cashiers, postmen, clerks, cooks, bus drivers,
waiters, etc.-- since their job involves being in direct contact with
a large number of people on a regular basis. This fact makes
super-spreaders more prone to catch diseases that propagate
preferentially through direct contacts, like COVID-19 does, and
--involuntarily-- more efficient at spreading them.

The fact that mainly African Americans seem to be affected by such a
markedly unusual COVID-19
incidence\cite{Pareek2020,Laurencin2020,Yancy2020}, rather than, say,
people with low-income, little access to healthcare, or with other
increased risk
factors~\cite{Raisi-Estabragh2020,Li2020,Prats-Uribe2020,Takagi2020,Khanijahani2020},
points to ethnic segregation, i.e., the tendency of people belonging
to the same ethnic group to live closer in space, as a possible
culprit~\cite{Dawkins2004,Dawkins2006,Brown2006,Cliff1981,Rey2013}. Indeed,
ethnic segregation is long-standing problem across the
US~\cite{Rajiv2004}, so the idea that the abnormal proportion of
COVID-19 infections among African Americans could be due to spatial
segregation does not sound unreasonable. However, the results
available so far confirm that, although there is a correlation between
ethnic segregation and overall incidence of COVID-19 in the
population, there seems to be little evidence of an association with
infection gap in African Americans~\cite{Hendryx2020}.

Our hypothesis is that the observed infection gap is most probably due
to a prevalence of \textit{super-spreading behaviours} in African
American communities, i.e., activities that contribute to increase the
typical number and variety of face-to-face contacts of individuals
---including for instance their job, habits, social life, commuting
and mobility patterns--- and that effectively make them more exposed
to the infection. In particular, we argue that these super-spreading
behaviours are connected to the presence of what we call
\textit{dynamic segregation}. By dynamic segregation we mean the
extent to which individuals of a certain class or group are either
preferentially exposed to other groups, or internally clustered, as a
result of their mobility patterns. In this sense, dynamic segregation
is somehow complementary to the classical notion of segregation based
on residential data, and is instead related with similar measures of
segregation based on the concept of activity space~\cite{Wong2011}. In
principle, the fact that a certain residential neighbourhood has an
overabundance of people belonging to a single ethnic group might have
\textit{per se} little or no role in increasing the probability that
those people catch COVID-19. Conversely, the fact that a group of
people works preferentially in specific sectors, or in specific areas
of a city, almost automatically increases the typical number of
face-to-face contacts they have during a day, e.g., by forcing them to
commute long distances in packed public transport services.

\section*{Results}

\subsection{Model}

\begin{figure*}[ht!]
\includegraphics[width=.95\textwidth]{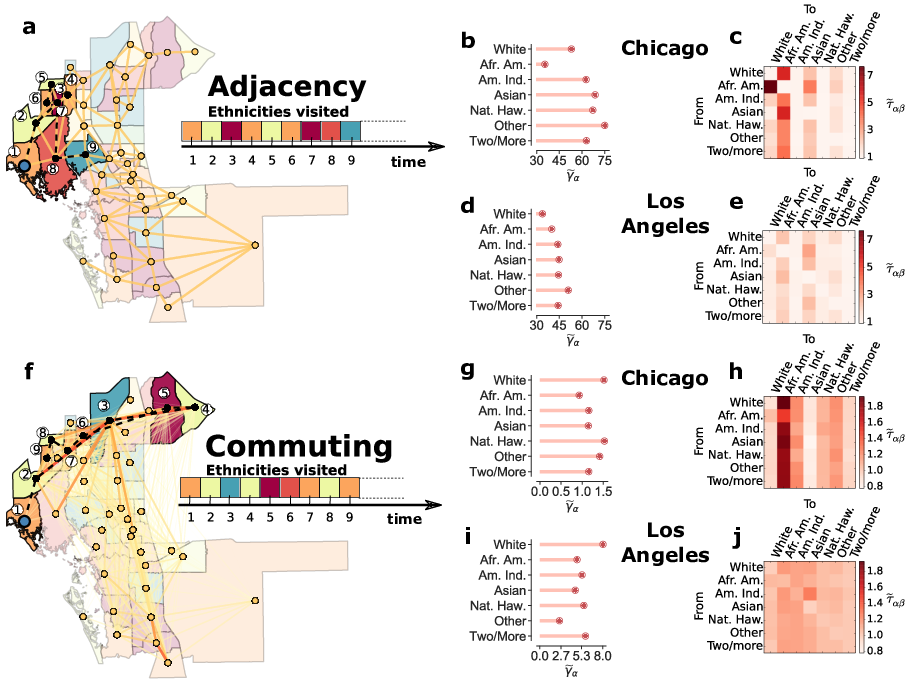}
\caption{\label{Figure1} \textbf{Using typical times of random walks
  to quantify urban dynamic segregation.}
  The sequence of ethnicities (here indicated
  by different colours) visited by a random walk over \textbf{a} the
  adjacency network or \textbf{f} the commuting network among census
  tracts of a city retains relevant information about the presence of
  spatial correlations in ethnicity distribution. Indeed, the
  normalised values of Class Coverage Time
  $\widetilde{\gamma}_{\alpha}$ (panels \textbf{b,d,g,i}) and Class
  Mean First Passage Time $\widetilde{\tau}_{\alpha\beta}$ (panels
  \textbf{c,e,h,j} of a random walk exhibit different patterns in
  different cities, and reveal different kinds of ethnic correlations
  in the adjacency and in the commuting network of the same city. We
  show here the values for Chicago or Los Angeles, since Illinois and
  California have, respectively, one of the highest and one of the
  lowest COVID-19 incidence gap. Indeed, the mean first passage time
  from African American to White neighbourhoods in the adjacency graph
  is much higher in Chicago than in Los Angeles, while the commuting
  graphs reveals that African Americans are much more exposed to all
  the other ethnicities in Chicago than in Los Angeles.}
\end{figure*}

We quantify the dynamic segregation of a certain group in a urban area
by means of the typical time needed by individuals of that group to
get in touch with individuals of other groups when they move around
the city. In our model, a city is represented by a graph $\mathcal{G}$
where nodes are census tracts and each edge indicates a relation
between two areas, namely either physical adjacency or the existence
of commuting flows between them. Each node is assigned to a class,
according to the ethnicity distribution in the corresponding area (see
Methods for details). Then, we consider a random walk on the graph
$\mathcal{G}$, and we look at the statistics of Class Mean First
Passage Times (CMFPT) and Class Coverage Times (CCT). The former is
the number of steps needed to a walker starting on a node of a certain
class $\alpha$ to end up for the first time on a node of class
$\beta$, while the latter is related to the time needed to a random
walk to visit all the classes in the system (see Methods for
details). The underlying idea is that a random walk through the graph
preserves most of the information about correlations and heterogeneity
of node classes~\cite{Nicosia2014}. Consequently, if a system is
dynamically segregated, the statistics of CMFPT and CCT will be
substantially different from those observed on a null-model graph
having exactly the same set of nodes and edges, but where a node is
assigned a class at random from the underlying ethnicity distribution.

\begin{figure*}[ht!]
\includegraphics[width=6in]{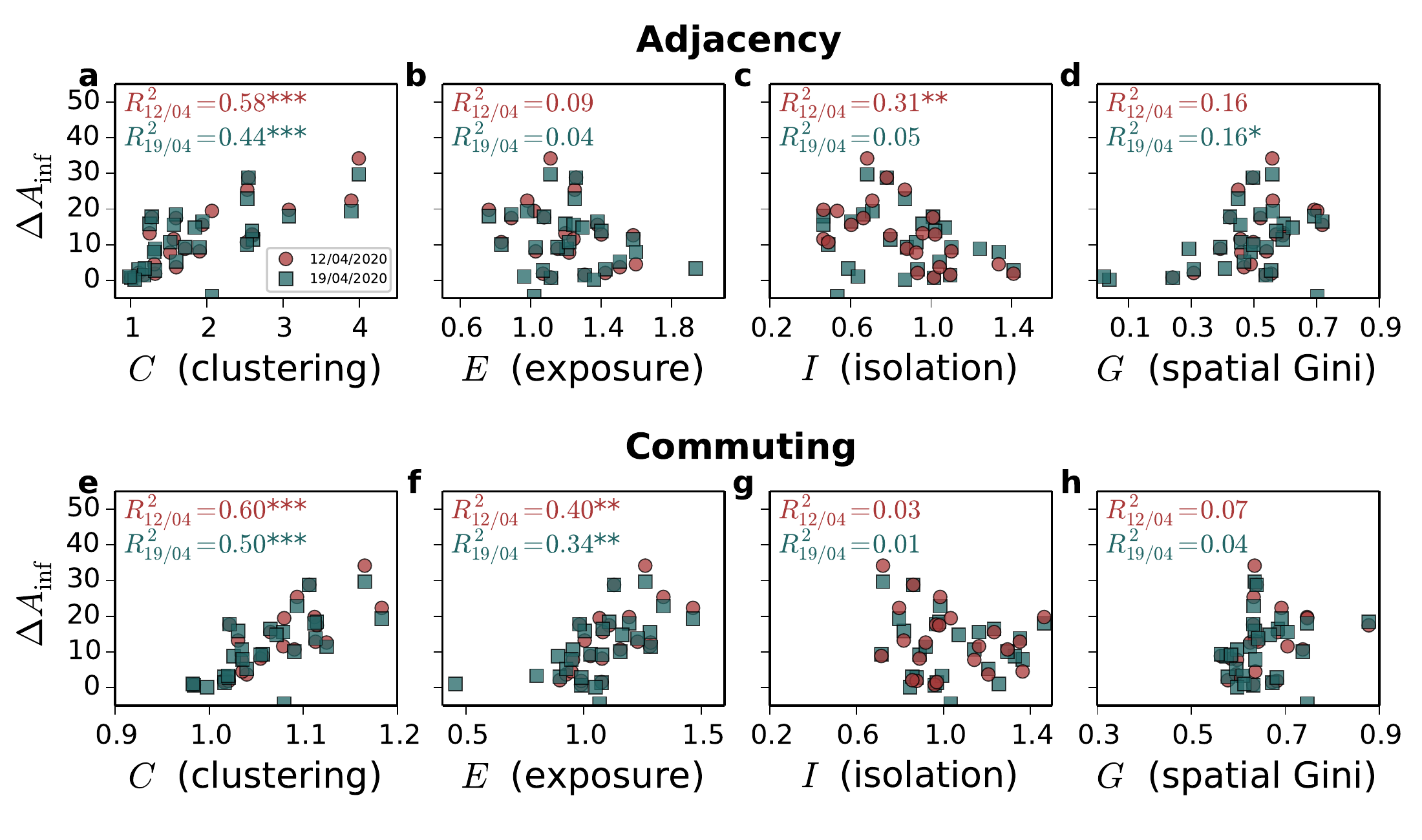}
\caption{\textbf{Correlation between incidence gap and dynamic
    segregation in the early stages of the epidemics.} The incidence
  gap $\Delta A_{\rm inf}$ across US states in the first two weeks
  after extensive lock-down measures were enforced exhibits somehow
  strong correlation with measure of segregation based on CMFPT and
  CCT on the adjacency (panels \textbf{a-d}) and on the commuting
  graphs (panels \textbf{e-h}). In particular, the dynamic clustering
  $C$ (\textbf{a,e}) is always positively correlated with $\Delta
  A_{\rm inf}$, the the dynamic exposure $E$ (\textbf{b,f}) is
  positively correlated with $\Delta A_{\rm inf}$ only in the
  commuting network, and the dynamic isolation $I$ (\textbf{c,g}) is
  negatively associated with incidence gap only in the adjacency
  network. Notice that classical measures of residential segregation,
  like the Spatial Gini coefficient (\textbf{d,h}), are instead poorly
  or not correlated at all with incidence gap. Each colour corresponds
  to a temporal snapshot of the data set, red for $12/04/2020$ and blue
  for $19/04/2020$. (*: $p<0.05$,**: $p<0.01$, ***: $p<0.001$)
\label{Figure2}}
\end{figure*}

In Fig.~\ref{Figure1} we provide a visual sketch of the model and we
show the distributions of CMFPT and CCT in Chicago and Los Angeles. We
chose these two specific cities since Illinois and California are two
states respectively characterised by a relatively high and a
relatively low incidence gap~\cite{Data1,Data2} (a detail of incidence
gap across US states is available in Supplementary Figures
18-19). Here each node is associated to one of the seven high-level
ethnic groups defined by the US Census Borough~\cite{Ethnicity}, with
a probability proportional to the abundance of that ethnicity in the
corresponding census tract.  The variables of interest are
$\mfptnorm_{\alpha\beta}$ and $\cctnorm_{\alpha}$. These are,
respectively, the ratio of the CMFPT from class $\alpha$ to class
$\beta$ in the real system and in the null-model, and the ratio of the
CCT when the walker starts from class $\alpha$ in the real system and
in the null model (see Eq.~\ref{eq:avg_CMFPT} and Eq.~\ref{eq:avg_CCT}
in Methods). In short, the farther away $\mfptnorm_{\alpha,\beta}$ is
from $1$, the higher the dynamic segregation from class $\alpha$ to
class $\beta$. Similarly, the higher the value of $\cctnorm_{\alpha}$
the more isolated ethnicity $\alpha$ is from all the other ones.

\begin{figure*}[!ht]
  \includegraphics[width=1\textwidth]{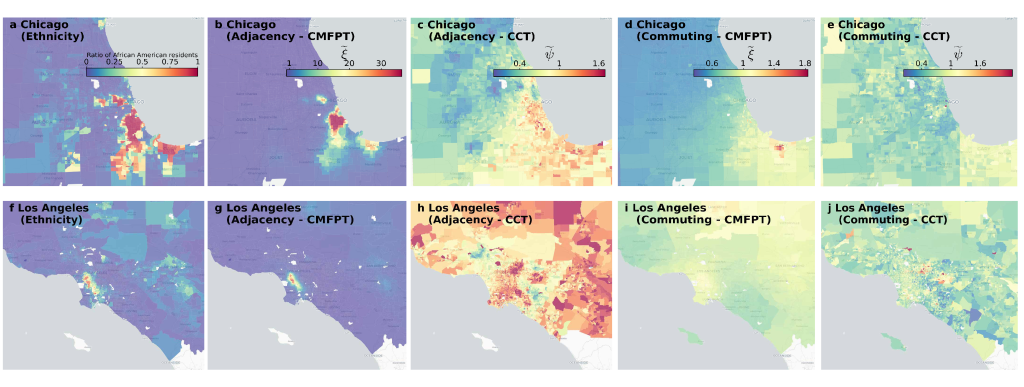}
  \caption{\textbf{Distribution of local dynamic segregation.} The
    distribution of the fraction of African American population living
    in each census tract (panels \textbf{a,f}) is mostly unrelated to
    the local clustering index $\widetilde{\xi}_i$ (panels
    \textbf{b,d,g,i}) and to the local isolation index
    $\widetilde{\psi}_i$ (panels \textbf{c,e,h,j}). The figure shows
    the result for Chicago (top panels) and for Los Angeles (bottom
    panels). Overall, there is little correlation between the density
    of African American residents and the dynamic segregation of
    African Americans in an area. This explains why dynamic
    segregation indices in a city correlate quite strongly with the
    COVID-19 infection gap, while no strong association with
    residential segregation has been found so far.}
    \label{Figure3}
\end{figure*}

The top panels of Fig.~\ref{Figure1} correspond to the unweighted
network $\mathcal{A}$ of physical adjacency between census tracts,
while the bottom panels are obtained on the weighted network
$\mathcal{C}$ of typical daily commute flows among the same set of
census tracts~\cite{Commuting} (see Methods for details). Notice that
the two graphs have quite different structures: the adjacency graph is
planar and each edge connects only nodes that are physically close,
while in the commuting graph long edges between physically separated
tracts are not only possible, but quite frequent. As a consequence,
the adjacency graph provides information about short trips, e.g., for
daily shopping and access to local services, while the commuting graph
represents long-range trips, e.g., related to commuting to and from
work. It is clear that each ethnicity has a peculiar pattern of
passage times to the other ethnicities, and this pattern varies across
cities. For instance, in Chicago the two largest values of
$\mfptnorm_{\alpha\beta}$ on the adjacency graph are observed between
African Americans and White, and between Asian and African
Americans. Conversely, in Los Angeles the two largest values of
$\mfptnorm_{\alpha\beta}$ are between African American and Asian and
between Other and Asian. As expected, the profile of
$\mfptnorm_{\alpha\beta}$ for a given class is quite different if we
consider the commuting network instead of the adjacency graph. In
Chicago, the largest value of $\mfptnorm_{\alpha\beta}$ is from White
to African American, while in Los Angeles there are a lot of pairs of
classes with pretty similar values of $\mfptnorm_{\alpha\beta}$,
indicating that in this city dynamic segregation for African Americans
is less prominent than in Chicago. The value of $\cctnorm_{\alpha}$
for African Americans is especially low in Chicago, but noticeably
different from that of the other ethnicities in Los Angeles. Results
for other cities are discussed in Appendix A
Supplementary Figures 1-4. As we shall see in a moment,
$\cctnorm_{\alpha}$ is related to the isolation of a class, so that
lower values correspond to increased exposure to all the other
classes.

\subsection{Dynamic segregation and infection gap}

Starting from the statistics of CMFPT and CCT at the level of each
city, we defined three indices of dynamic segregation, namely dynamic
clustering (C), dynamic exposure (E), and dynamic isolation (I), and
we associated to each state in the US the weighted average of each of
those indices across the largest metropolitan areas of the state (the
definitions of these measures are provided in Methods, while a ranking
of US states by each segregation index is reported in Appendix B and in Supplementary
Figure 5). We considered two temporal data sets of weekly percentage
of African Americans infected by and deceased due to COVID-19 for each
state in the US~\cite{Data1,Data2} (more details available in
Methods), and we calculated the incidence gap $\Delta A_{\rm inf}$ in
each state as the difference between the percentage of infected of
that state that are African Americans and the percentage of African
American population in the same state. Hence, Positive values of
$\Delta A_{\rm inf}$ correspond to a disproportionate incidence of
COVID-19 on African American communities.

In Fig.~\ref{Figure2} we show the scatter plots of the average dynamic
clustering, exposure, and isolation of African Americans at state
level, and of the corresponding COVID-19 infection gap in the first
two weeks after major lock-down measures were introduced across the
US. We chose these two temporal snapshots because the number of
confirmed infected individuals in a week actually depends on their
contacts up to two weeks before, due to the COVID-19 incubation
period~\cite{Linton2020}. The top panels report the results on the
adjacency networks of census tracts, while the bottom panels are for
the commuting graphs. Interestingly, there exists a quite strong
correlation between dynamic segregation and the disproportionate
number of infected in African American communities. In particular, the
dynamic clustering of African Americans in a state correlates
positively and quite strongly with the infection gap observed in that
state in the first two weeks of the data set, both on the adjacency
(respectively $R^2 = 0.58$ and $R^2=0.44$ in the first two weeks) and
in the commuting network (respectively $R^2=0.60$ and
$R^2=0.50$). This means that if African American citizens normally
require more time than citizens from other ethnic groups before ending
up in a non-African American neighbourhood, then the incidence gap
will be considerably higher.

\begin{figure*}[!htb]
\includegraphics[width=6in]{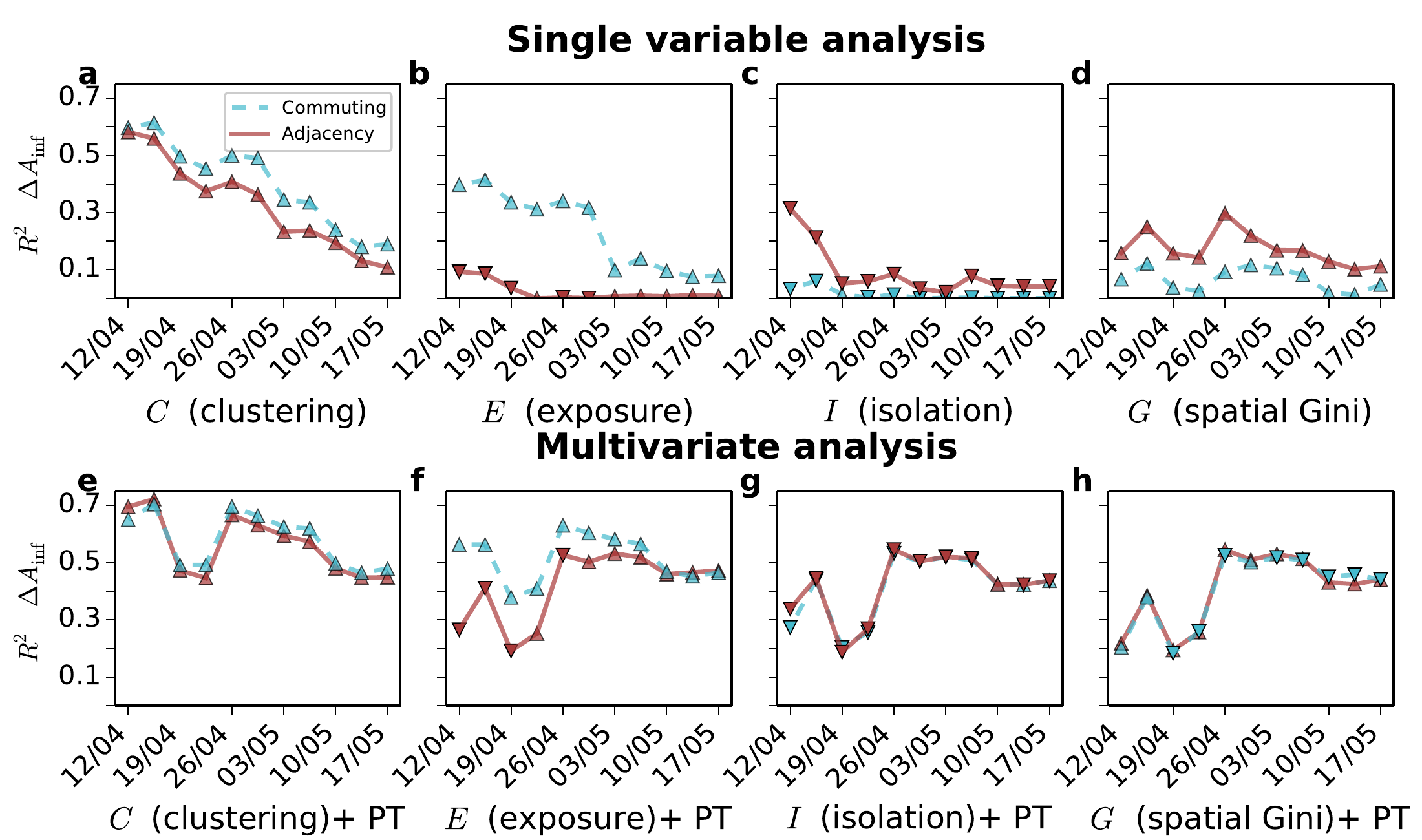}
\caption{\textbf{Temporal evolution of incidence gap correlations and
    multivariate analysis with public transport usage.} Evolution of
  the Pearson correlation ($R^2$) between African American incidence
  gap and \textbf{a} dynamic clustering, \textbf{b} dynamic exposure,
  \textbf{c} dynamic isolation, \textbf{d} Spatial Gini coefficient,
  respectively on the adjacency (solid red lines) and commuting graphs
  (dashed blue lines).  \textbf{e-h} Multivariate analysis of the same
  indices and usage of public transportation by African Americans for
  \textbf{e} dynamic clustering, \textbf{f} dynamic exposure,
  \textbf{g} dynamic isolation, \textbf{h} and spatial Gini
  coefficient. The type of marker indicates the sign of the
  correlation (triangles pointing up for positive correlations, and
  down for negative correlation).  Given the uneven temporal reporting
  of ethnicity data, each temporal snapshot has a slightly different
  number of US states (details provided in Supplementary Figure
  17). We have also tried alternative formulations for $C$ and $E$
  obtaining significant correlations, as shown in Supplementary Figure
  27.}
\label{Figure4}
\end{figure*}

The role of dynamic exposure is even more interesting. Indeed, the
dynamic exposure on the adjacency network is not correlated at all
with incidence gap, while it is a good predictor of incidence gap in
the commuting graph (respectively $R^2=0.40$ and $R^2=0.36$).
Conversely, the dynamic isolation of African Americans in the
adjacency graph is negatively correlated with incidence gap in the
early stages of the epidemics ($R^2=0.31$). Similar results are
obtained when we consider the correlation with the death gap $\Delta
A_{\rm dec}$, and the ratios of infection/deaths incidence instead of
the difference (see Supplementary Figures
8-10). In particular, dynamic isolation exhibits a somehow stronger
correlation with death gap ($R^2=0.27$). It is worth noting that the
isolation of other ethnicities has poor or no correlation with
incidence gap (see Supplementary Figures 11-14).

The fact that ethnic segregation does not correlate with infection gap
as much as dynamic segregation indices can bet better explained by
looking at how residential data and dynamic segregation are
distributed across a city. In Fig.~\ref{Figure3} we show the heat-maps
of abundance of African American residents in Chicago and Los Angeles
together with the local segregation indices $\widetilde{\xi}_i$ and
$\widetilde{\psi}_i$, respectively derived from passage times and
coverage times, on the adjacency and on the commuting graph of census
tracts (see the definitions provided in Methods, additional maps for
Detroit and Houston are reported in Appendix C and Supplementary Figure 15). It is
true that in Chicago $\widetilde{\xi}_i$ in the adjacency graph is
still somehow correlated with the fraction of African American
population (see Supplementary Figure 16). But the distribution of
$\widetilde{\xi}_i$ in the commuting graph is totally different. In
particular, the regions characterised by residential clusters of
African Americans exhibit lower values of $\widetilde{\xi}_i$, meaning
that the commuting patterns make those neighbourhoods overall less
isolated. Conversely, new hot-spots are identified in the
South-Eastern region of Gary, likely due to the fact that people in
this region do not commute much to the city centre anyway. Similarly,
the areas of Los Angeles with the largest local isolation are not the
neighbourhoods with a higher percentage of African Americans
residents, rather the suburbs characterised by high commuting.

\subsection{Combined effects of dynamic segregation and use of public transport}

Finally, in Fig.~\ref{Figure4}{\bf a-d} we show the correlation
between the infection gap and the different segregation measures as
the pandemic progresses. Unsurprisingly, the correlation with any
single measure decreases over time for all the indices, and both on
the adjacency and on the commuting graph. Similar results are found
for the correlation with death gap and with ratios of incidence and
deaths in African Americans (see Appendix D and Supplementary Figures 20-22) as well
as with a second dataset we had access to~\cite{Data2} (see
Supplementary Figures 23-26). The main reason for the observed
decreases is that once large-scale mobility restrictions are put in
place ---as it happened between the end of March and the beginning of
April across all the US states with stay-at-home orders and curfews---
the overall mobility structure of each city is massively disrupted. As
a result, super-spreading behaviours due to usual commuting patterns
are massively reduced, and the contagion progresses mainly through
face-to-face interactions happening close to the residential place of
each individual, and are not captured well by CMFPT and CCT on the
commuting graph.

In order to capture the focus on local transport after lock-downs are
enforced, in Fig.~\ref{Figure4}{\bf e-h} we show the results of the
multivariate analysis of the same set of segregation indices shown in
Fig.~\ref{Figure2} and of the fraction of African American population
using public transport in each city (see Methods for details). The
combination of dynamic segregation and use of public transport
correlates quite consistently with the incidence gap. These findings
are made more relevant by the fact that the incidence gap in African
Americans in the same period is quite poorly correlated with the
overall usage of public transport in the population, as well as with a
variety of other socio-economic indices, as shown in Supplementary
Figures 30-31. Since cities are complex interconnected systems, it is
plausible to hypothesise that segregation and public transport usage
are related in subtle and intricate ways, so that it is practically
impossible to establish whether the former has caused the latter, or
instead the two phenomena have co-evolved over time.

\section*{Discussion}
The vulnerability of African American communities and their higher
socio-economic disparities has been a standing issue
in the US long before the pandemic, the disproportional infection
rates simple highlighted and amplified the problem.  The presence of a
COVID-19 incidence gap in Black and African American population is
somehow unexpected, since no specific biological risk factor has been
strongly associated with an increased vulnerability to the virus of
any specific ethnic group. Hence, the most unbiased assumption to
explain such a disproportionate incidence, which in some areas is
three to five times higher than the fraction of African American
population, is that it should be related to behavioural and social
factors, rather than to biological ones. The most frequently whispered
theory is that African Americans are more exposed to COVID-19 because
they are more frequently employed in service works. This explanation
is indeed reasonable, since service workers normally have hundreds of
face-to-face interactions during a day. Indeed, some recent studies
have estimated that the switching to remote-working was mainly
available to people employed in non-essential services, and amounted
to 22\%-25\% of the work force before April~\cite{Remote2020}. As
expected, service workers are one of those categories to which the
option to switching to remote-working during the lock-down was not
available at all, especially in sectors deemed vital for the
functioning of a country during lock-down, including food production
and retailers, healthcare, transportation, and logistics. According to
the US Labor Force Statistics~\cite{LaborStats}, the occupations with
the highest concentration of African Americans are indeed jobs
characterised by face-to-face interaction, and most of them fall in
the area of \textit{essential} jobs: postal service sorters/processors
(42\%), nursing (37\%), postal service clerks (35\%), protective service
workers (34\%) and barbers (32\%). It would not
then come as a surprise to discover that one of the major early
COVID-19 outbreaks happened in South Dakota, in a meat-processing
plant, whose workers were mainly of African American
background~\cite{MeatPlant}.

The potential relation between ethnicity and mobility was somehow
hinted to in a recent study~\cite{Sy2020} which found that the
decrease in the usage of subway transport in New York during the
lock-down was uneven across ethnicities, with African Americans
experiencing the smallest relative drop. But unfortunately, the
publicly available data about COVID-19 incidence do not contain
detailed information about socio-economic characteristics of infected
individuals, so drawing an association between African Americans,
employment in essential service jobs, availability of remote-working
options, and increased COVID-19 exposure is very hard.

An interesting finding of the present work is that the combination of
dynamic segregation and use of public transport seems to explain the
persistence of infection gap throughout the early phases of the
pandemic. Indeed, before lock-downs are put in place, African Americans
are found to be more exposed to the virus, mainly due to the structure
of their daily commuting patterns.  After lock-downs are enforced,
instead, they are more likely to pass the virus over to other African
Americans, as a result of the high levels of clustering and isolation
of these communities measured in the adjacency graphs of census
tracts, which are a more reliable proxy for face-to-face interactions
when long-distance commuting is disrupted. In general, the states
where African Americans are more exposed with respect to long-distance
trips are also those where they are more clustered with respect to
short-range mobility (the rank correlation between the two measures is
$0.62$, as shown in Supplementary Figure 6-7 and Supplementary Table
I).

The importance of considering the interaction of different classes due
to mobility through the urbanscape has recently received some
attention~\cite{Farber2012,Ballester2014,Olteanu2019,Kawabata2007,Wong2011,Graif2017,Wang2018}. In
this sense, it is quite interesting that the simple diffusion model we
used here to quantify the presence of dynamic segregation, and the
corresponding indices of clustering, exposure, and isolation, are able
to unveil a relatively strong correlation between the structure of
mobility in a metropolitan area and the excess incidence of COVID-19
infections and deaths in African Americans. Although the model we
consider uses relatively small and coarse-grained information about a
city ---placement of census tracts, local ethnicity distribution, and
commuting trips among them--- the strong correlation between dynamic
segregation and incidence gap allows to conclude that when it comes to
predicting the exposure of a group to a non-airborne virus, knowing
the places where the members of that group commute for work is more
important and more relevant than knowing where they actually
live. This is also confirmed by the quite poor association of
incidence gap with other classical measures of racial segregation (see
Supplementary Figures 28-29).

The results presented in this work suggest that policy makers should
definitely take into account mobility patterns when modelling the
spread of a disease in a urban area, and in predicting the impact of
specific countermeasures. In particular, a strategy to mitigate
incidence gap should focus on reducing as much as possible
long-distance trips for people that are naturally more exposed to
face-to-face contacts, e.g., due to their occupation, and enforcing
stricter measures of social distancing on local activities.

\section*{Methods}

\subsection*{Geographic network data sets}

Ethnicity data was obtained from~\cite{Ethnicity} and includes the
data from the 2010 decennial census.  Commuting trips data comes from
the 2011 US census~\cite{Commuting}, focusing on the seven
highest-level ethnicity classes, namely: White, Black or African
American, American Indian and Alaska Native, Asian, Native Hawaiian
and Other Pacific Islander, Some Other Race, Two or More
Races. Population is updated to the latest American Community Survey
2014-2018 5-yeas Data Release~\cite{Population}.

For each metropolitan area we constructed two distinct spatial
networks. The first one is the \textit{adjacency network}, denoted by
$\mathcal{A}$ and obtained by associating each cell to a node and
connecting two nodes with a link if the corresponding cells border
each other. Notice that $\mathcal{A}$ is an undirected and unweighted
graph,. The second graph is the \textit{commuting network}, denoted as
$\mathcal{C}$. In this network each node is a tract and the directed
and weighted link $\omega_{ij}$ between node $i$ and $j$ indicates the
number of commuting trips from $i$ to $j$ as obtained from census
information. To reconstruct a mobility network that resembles the real
one (which amounts to something between $30\%$ and $40\%$ of the total
mobility in a city) we aggregated both the trips from home to work and
the corresponding return trip from work to home. 

Each node of the adjacency network $\mathcal{A}$ preserves information
about the ethnicity distribution on the corresponding census tract. We
use the $N\times \Gamma$ matrix $\mathcal{M}=\{m_{i,\alpha}\}$, where
$\Gamma$ is the number of ethnicities present in the city. The generic
element $m_{i,{\alpha}}$ of $\mathcal{M}$ indicates the number of
citizens of ethnicity $\alpha$ living on node $i$. We denote by
$M_i=\{m_{i,\alpha}\}$ the vector of population distribution at node
$i$, and by $M^{\alpha}=\sum_{i=1}^{N}m_{i,\alpha}$ the total number
of individuals of class $\alpha$ present in the system. In the
commuting network $\mathcal{C}$, instead, we attribute to each node
$i$ both the resident population at the corresponding tract and the
population commuting to node $i$, so that the abundance of individuals
of class $\alpha$ on node $i$ becomes:
\begin{equation}
  \widetilde{m}_{i,\alpha}=m_{i,{\alpha}}+\sum \omega_{ji}m_{j,{\alpha}}.
\label{eq:norm_popul}
\end{equation}
where $\omega_{ji}$ is the number of daily commuting trips from node
$j$ to node $i$.  By doing so we aim to capture the fact that a
commuter to cell $i$ will potentially have face-to-face interactions
with both residents in that area and other workers commuting to that
area every day. Moreover, since the commuting network $\mathcal{C}$
accounts for both work-home and home-work trips, the adjusted
population on the commuting network accounts for the potential
contacts that individuals had at the origin of a trip as well.

\subsection*{Class Mean First Passage Time (CMFPT)}

Let us consider a generic graph $\mathcal{G}(\mathcal{V},\mathcal{E})$
with $|\mathcal{V}|=K$ edges on $|\mathcal{V}|=N$ nodes, and a
colouring function $f:\mathcal{V}\rightarrow \chi$ that associates to
each node $i$ of $\mathcal{G}$ a discrete label $f(i)$ from the finite
set $\chi$ with cardinality $|\chi|=\Gamma$. Let us also consider a
random walk on $\mathcal{G}$, defined by the transition matrix $\Pi =
\{\pi_{ij}\}$ where $\pi_{ji}$ is the probability that the walk jumps
from node $i$ to node $j$ in one step. On the adjacency network
$\mathcal{A}$ we use a uniform random walk, i.e., $\pi_{ji} =
\frac{1}{k_i}$, while on the commuting graph $\mathcal{C}$ we have
$\pi_{ji}=\frac{\omega_{ij}}{s_i}$, where $s_i = \sum_j \omega_{ij}$
is the out-strength of node $i$.

Here we focus on the statistical properties of the trajectories $W_i =
\{f(i_0), f(i_1), \ldots\}$ of node labels visited by the random walk
$W$ at each time when starting from $i_0= i$ at time $t=0$. This
dynamics contains information about the existence of correlation and
heterogeneity in the distribution of colours. For instance, if the
graph $G$ is a regular lattice and the function $f$ associates colours
to nodes uniformly at random, we expect that, for long-enough time,
all the trajectories starting from each of the $N$ nodes will be
statistically indistinguishable.

We denote as $T_{i,\alpha}$ the Mean First Passage Time from a given
node $i$ to nodes of class $\alpha$, i.e., the expected number of
steps needed to a walk starting on $i$ to visit for the first time any
node $j$ such that $f(j) = \alpha$. We can write a self-consistent
forward equation for $T_{i,\alpha}$~\cite{Masuda2017}:
\begin{equation}
  T_{i,\alpha} = 1 + \sum_{j=1}^{N}
  \left(1-\delta_{f(j),\alpha}\right) \pi_{ji} T_{j,\alpha}
\end{equation}
The Mean First Passage Time $\tau_{\beta\alpha}$ from class $\alpha$
to class $\beta$ is defined as:
\begin{equation}
  \tau_{\alpha\beta} =
  \frac{1}{N_{\alpha}}\sum_{j=1}^{N}T_{j,\beta}\delta_{f(j), \alpha},
  \label{eq:MFPT}
\end{equation}
where $N_{\alpha}$ is the number of nodes in the graph associated to
class $\alpha$. Notice that in practice the value of
$\tau_{\alpha\beta}$ is obtained as an average over many realisations
of the random walk.

A notable issue of the MFPT defined in Eq.~(\ref{eq:MFPT}) is the fact
that its values might depend on the specific distribution of colours
(i.e., on their abundance) and on the size of the network under
consideration, which makes it difficult to compare Mean First Passage
Times computed on different systems. To obviate to this problem, we
define the normalised Class Mean First Passage Time between class
$\alpha$ and class $\beta$ as:
\begin{equation}
  \widetilde{\tau}_{\alpha\beta}=\frac{\tau_{\alpha\beta}}{\tau^{\rm
      null}_{\alpha\beta}}
  \label{eq:CMFPT}
\end{equation}
where $\tau^{\rm null}_{\alpha\beta}$ is the MFPT from class $\alpha$
to class $\beta$ obtained in a null-model graph. The null-model
considered here is the graph having the same topology of the original
one, and where node colours have been reassigned uniformly at random,
i.e., reshuffled by keeping their relative abundance.  Notice that
$\widetilde{\tau}_{\alpha\beta}$ is a pure number: if
$\widetilde{\tau}_{\alpha\beta}>1$ (resp.,
$\widetilde{\tau}_{\alpha\beta}<1$ it means that the expected time to
hit a node of class $\beta$ when starting from a node of class
$\alpha$ is higher (resp., lower) than in the corresponding
null-model. In general, a value different from $1$ indicates the
presence of correlations and heterogeneity.

\subsection*{Class Coverage Time (CCT)}

The coverage time is classically defined as the number of steps needed
to a random walk to visit a certain percentage of the nodes of a graph
when starting from a given node $i$~\cite{Masuda2017}. In the case of
a network with coloured nodes, a walk started at node $i$ will be
associated to the generic trajectory $W_i = \{f_{i_0}, f_{i_1},
f_{i_2}, \ldots\}$ of node labels visited by the walk at each
time. Since we are interested in quantifying the heterogeneity of
ethnicity distributions, we consider the time series
$\mathcal{W}_i=\{M_{i}, M_{i_1}, M_{i_2},\ldots\}$ where
$M_{i_{t}}=\{m_{i_t,\alpha}\}$ is the distribution of ethnicities at
node $i_{t}$ visited by the walk at time $t$. If we consider the
trajectory up to time $t$, the vector
$Q_{i_t}=\frac{1}{H_t}\sum_{\tau} M_{i_{\tau}}$ is the distribution of
ethnicities visited up to time $t$ by the walker started at $i$ (here
$H_t$ is a normalisation constant that guarantees
$\sum_{j}\{Q_{i_t}\}_j = 1$). We quantify the discrepancy between
$Q_{i_t}$ and the global ethnicity distribution across the city
$\mathcal{P} = \frac{1}{H'}\mathcal{M}^{\intercal} {\bf 1_{N}}$ by
means of the Jensen-Shannon divergence:
\begin{equation}
  J(\mathcal{P}\|Q_{i_t})= \frac{1}{2} \left[ D(\mathcal{P}\|\mu) +
  D(Q_{i_t}\|\mu) \right],
\end{equation}
where $\mu=\frac{1}{2}\left(\mathcal{P} + Q_{i_t}\right)$ and
$D(P\|Q)$ is the Kullback-Liebler divergence between $P$ and $Q$. We
define the Class Coverage Time from node $i$ at threshold
$\varepsilon$ as:
\begin{equation}
  \gamma^{i} = \argmin\limits_t \left\{ J(\mathcal{P}||Q_{i_t}) \leq
  \varepsilon \right\}
\end{equation}
and the associated normalised Class Coverage Time:
\begin{equation}
  \widetilde{\gamma}^{i} = \frac{\gamma^{i}}{\gamma^{i, {\rm null}}}
\end{equation}
where $\gamma^{i,{\rm null}}$ is the Class Coverage Time from node $i$
in a null-model where the colours associated to the nodes have been
reshuffled uniformly at random.

\subsection*{CMFPT and CCT in census networks}

In the case of ethnicity distributions in geographical networks, each
node is not uniquely associated to a colour, but it has instead a
local distribution of ethnicities. Nevertheless, the formalism for the
computation of Class Mean First Passage Times and Class Coverage Time
described above can still be used in this case as well. We consider a
stochastic colouring function
$\widetilde{f}:\mathcal{V}\rightarrow\mathcal{C}$ that associates to
each node $i$ of the adjacency graph one of the $\Gamma=7$ ethnicities
$\alpha$ with probability
$\frac{m_{i,\alpha}}{\sum_{\beta}m_{i,\beta}}$ (respectively, with
probability
$\frac{\widetilde{m}_{i,\alpha}}{\sum_{\beta}\widetilde{m}_{i,\beta}}$
in the commuting graph), i.e., proportionally to the abundance of
ethnicity $\alpha$ in node $i$.

To compute the CMFPT we consider $S$ independent realisations of the
stochastic colouring process for each network. On each realisation
$\ell$, we estimate the MFPT among all classes as in
Eq.~(\ref{eq:MFPT}), and the corresponding null-model MFPT. Then, we
compute the average Class Mean First Passage Time from class $\alpha$
to class $\beta$ as:
\begin{equation}
  \avg{\widetilde{\tau}_{\alpha\beta}} = \frac{\sum_{\ell=1}^{S}
    \tau_{\alpha\beta}\up{\ell}}{\sum_{\ell=1}^{S}
    \left(\tau_{\alpha\beta}^{\rm null}\right)\up{\ell}}
    \label{eq:avg_CMFPT}
\end{equation}
where $\tau_{\alpha\beta}\up{\ell}$ is the CMFPT computed on the
$\ell$-th realisation and $\left(\tau_{\alpha\beta}^{\rm
  null}\right)\up{\ell}$ is the corresponding value in the
null-model. For each system we computed $\tau^{\rm
  null}_{\alpha\beta}$ on 500 realisations of the null model, with 500
independent colour assignments per realisations, and 2000 walks per
node.

The computation of CCT works in a similar way. In order to take into
account the heterogeneous distribution of ethnicities across nodes,
before a walker starts from node $i$ we sample one of the ethnicities
present on $i$, according to their local abundance at $i$
$\{m_{i,\beta}\}$, and we attribute node $i$ to it. Then, we compute
the CCT from node $i$ of class $\alpha$ as the average CCT from node
$i$ across all the walks starting from $i$ where node $i$ was actually
assigned to class $\alpha$, and we call this quantity
$\gamma^{i}_{\alpha}$. Notice that in this case we consider the
trajectories $\mathcal{W}^{\alpha}_i=\{M^{\alpha}_{i},
M^{\alpha}_{i_1}, M^{\alpha}_{i_2},\ldots\}$ where
$M^{\alpha}_{i_{\ell}}$ is the distribution of ethnicities at the
$\ell$-th node visited by the walker, which does not include class
$\alpha$. The normalised Class Coverage Time from class $\alpha$ when
starting from node $i$ is defined as:

\begin{equation}
  \widetilde{\gamma}^{i}_{\alpha} =
  \frac{\gamma^{i}_{\alpha}}{\gamma_{\alpha}^{i, {\rm null}}}
  \label{eq:gamma_i_alpha}
\end{equation}
where $\gamma_{\alpha}^{i, {\rm null}}$ is the CCT from node $i$ of
class $\alpha$ in the null-model. Finally, the average CCT from class
$\alpha$ is simply obtained as:
\begin{equation}
  \widetilde{\gamma}_{\alpha} =\frac{1}{N}\sum_{i=1}^{N}
  \widetilde{\gamma}^{i}_{\alpha}
  \label{eq:avg_CCT}
\end{equation}
For all the computations of CCT shown in the paper we considered
averages over 5000 walks per node and we set $\varepsilon=
0.0018$. This value was obtained by using the Pinsker's inequality for
the Kullback-Leibler divergence and imposing a total variation
distance smaller than $6\%$.

\subsection*{Global indices of dynamic ethnic segregation}

We constructed three global indices of dynamic segregation based on
the values of CMFPT and CCT. In particular, we focused on the observed
discrepancies of CMFPT and CCT between African Americans and other
ethnicities. In the following the index $A$ will always indicate
African Americans, while the index $O$ will indicate all the other
ethnicities. We start by defining the following quantities:
\begin{align*}
  \overline{\tau}_{AA} & = \avg{\widetilde{\tau}_{AA}}\\
  \overline{\tau}_{AO} & = \frac{\sum_{\alpha \neq A} M^{\alpha}
    \avg{\widetilde{\tau}_{A\alpha}}}{\sum_{\alpha\neq A}M^{\alpha}} \\
  \overline{\tau}_{OA} & = \frac{\sum_{\alpha\neq A}
    M^{\alpha}\avg{\widetilde{\tau}_{\alpha A}}}{\sum_{\alpha \neq A}
    M^{\alpha}}\\
  \overline{\tau}_{OO} & = \frac{\sum_{\alpha,\beta\neq A}
    \avg{\widetilde{\tau}_{\alpha\beta}}M^{\alpha}
      M^{\beta}}{\sum_{\alpha,\beta \neq A}M^{\alpha}M^{\beta}}
\end{align*}
In practice: $\overline{\tau}_{AA}$ is the expected CMFPT from African
Americans to African Americans; $\overline{\tau}_{AO}$ is the expected
CMFPT from African Americans to all the other classes (weighted by
ethnicity distribution); $\overline{\tau}_{OA}$ is the expected CMFPT
from all the other classes to African Americans (again, weighted by
ethnicity distribution); and $\overline{\tau}_{OO}$ is the expected
CMFPT among all the other ethnicities. Notice that all these
quantities are pure numbers, since they are based on the corresponding
quantities defined in Eq.~(\ref{eq:avg_CMFPT}) which are correctly
normalised with respect to the null-model.

The clustering of African Americans is quantified as:
\begin{equation}
  C=\frac{\overline{\tau}_{AO}}{\overline{\tau}_{OO}}
  \label{eq:clustering}
\end{equation}
so that values of $C$ larger than $1$ indicate that for an African
American finding any other ethnicity is harder (i.e., requires more
time) than for all other ethnicities. Similarly, we define the
exposure of African Americans to other ethnicities as:
\begin{equation}
  E=\frac{\overline{\tau}_{OA}}{\overline{\tau}_{AO}}
  \label{eq:exposure}
\end{equation}
where values of $E$ larger than $1$ indicate that it is easier for
African American to be found in touch with any other ethnicity than
for people from all the other ethnicities to be found in touch with
African Americans.

We define similar quantities for the $CCT$ of African Americans and
other ethnicities, namely $\widetilde{\gamma}^{i}_{A}$ as in
Eq.~(\ref{eq:gamma_i_alpha}), and:
\begin{equation}
  \widetilde{\gamma}_O^{i} = \frac{1}{\Gamma-1}
    \sum_{\alpha\neq A} \widetilde{\gamma}_\alpha^{i}
  \label{eq:gamma_O}
\end{equation}
Finally, we define the isolation of African Americans for the whole system
by the average ratio:
\begin{equation}
  I = \frac{1}{N} \sum_{i=1}^{N}\frac{\widetilde{\gamma}_A^{i}}
  {\widetilde{\gamma}_O^{i}}
  \label{eq:isolation}
\end{equation}
over all the nodes. Notice that values of $I$ larger than $1$ indicate
that the normalised CCT from nodes of class $A$ (African American) is
higher than the CCT from nodes of all the other classes. The
State-level value of each index is obtained as an average of the
corresponding index on the cities of the state, weighted by the
population of each city.

\subsection*{Local dynamic ethnic segregation}

We define two local segregation indices for African Americans in a
census tract $i$. The first index is based on CMFPT:
\begin{equation}
\widetilde{\xi}_i=\frac{\sum_{\alpha \neq A}
  \widetilde{T}_{i,\alpha}}{ (\Gamma-1) \widetilde{T}_{i,A}}
\label{eq:xi}
\end{equation}
where $\widetilde{T}_{i,\alpha}$ corresponds to the normalised Mean
First Passage Time to a generic class $\alpha$ when a random walker
starts from node $i$, while $\widetilde{T}_{i,A}$ is the CMFPT to
African Americans tracts. Values of $\widetilde{\xi}_i$ larger than 1
indicate that the time to reach any other ethnicity is higher than
the time needed to reach African Americans, hence indicating a local
clustering of African Americans around node $i$. 

The local index of isolation is derived from CCT:
\begin{equation}
  \widetilde{\psi}_i = \frac{\widetilde{\gamma}_A^{i}}{\widetilde{\gamma}_O^{i}}
\end{equation}
where $\widetilde{\gamma}_A^{i}$ is the CCT from node $i$ for African
Americans and $\widetilde{\gamma}_O^{i}$ is the average CCT from node
$i$ for all the other ethnicities, as defined in
Eq.~(\ref{eq:gamma_O}). In general, if $\widetilde{\psi}_i$ is larger
than $1$ then African Americans living at node $i$ are isolated, since
they will require more time to visit all the other classes than
required by individuals from other ethnicities.

\subsection*{Data on COVID-19 incidence among the African American Population}
The data related to the percentage of infected African Americans was
obtained from two different sources \cite{Data1} and \cite{Data2}.
The first data set reports the number of infected and deceased of each
ethnicity along with those unknown.  To calculate the percentage of
African Americans we have removed first the unknown from the total,
otherwise our analysis would also capture the fraction of unknown.
For the other data set we just extract the data they provide in tables.

\subsection*{Public Transportation data set}
The public transportation data set was obtained from the 2018 American
 Community Survey from U.S. Census Bureau~\cite{Ethnicity}. It includes
information about the percentage of public transportation usage per
ethnicity and State.

\subsection*{Multivariate Analysis}
The multivariate analysis was performed using R and the ANOVA model in
the \texttt{car} package.

\section*{Acknowledgements}
A.B. and V.N. acknowledge support from the EPSRC New Investigator
Award Grant No. EP/S027920/1. This work made use of the MidPLUS
cluster, EPSRC Grant No. EP/K000128/1. This research utilised Queen
Mary's Apocrita HPC facility, supported by QMUL Research-IT.
\href{http://doi.org/10.5281/zenodo.438045}{doi.org/10.5281/zenodo.438045}.

\section*{Author Contributions}
All the authors devised the study. A.B. and S. S. performed the
simulations and computations. All the authors provided methods and
analysed the results. A. B. and S. S. prepared the figures and all the
visual material. All the authors wrote the paper and approved the
final submitted version.

\renewcommand\theequation{{S-\arabic{equation}}}
\renewcommand\thetable{{S-\Roman{table}}}
\renewcommand\thefigure{{S-\arabic{figure}}}
\renewcommand\thesection{{S-\arabic{section}}}

\setcounter{section}{0}
\setcounter{table}{0}
\setcounter{figure}{0}
\setcounter{equation}{0}

\onecolumngrid

\appendix                                   

\section{Quantifying ethnic segregation through CMFPT and CCT}
\subsection*{Ethnic segregation and COVID-19 incidence through CMFPT}
Ethnic segregation is quantified here through random walks, and more precisely, class mean first passage times (CMFPT). Given a set of classes present in a city -- ethnicities in this case-- we are interested in the number of steps you need to reach one as a function of the ethnicity at the origin.
Random walks start from each of the city cells --or tracts-- and move until they have visited each of the distinct
classes or ethnicities present in a city. If we average the passage times across all the city cells and then divide
by the same quantity from the null model we
obtain the normalised CMFPT between class $\alpha$ and $\beta$ $\widetilde{\tau}_{\alpha \beta}$.
It is important to note that this matrix is not necessarily symmetric and depends on the spatial distribution of classes.
We show in Supplementary Figure \ref{MFPT_adj} the CMFPT on the adjacency graph for four cities: Detroit, Chicago, Houston and Los Angeles.
On a first look, strong differences can be detected between those cities on the left and those on the right.
The normalised CMFPT are substantially higher in Detroit and Chicago when compared to Houston and Los Angeles.
Despite the difference in the maximum values, the shape of the matrix and curves is not so different across cities, with African Americans much more isolated than the rest of ethnicities. Reaching African Americans is much harder for any other class, while it is considerably low for other African Americans.  As can be seen, there are not only strong differences between both cities but also between the type of network used (See Supplementary Figure \ref{MFPT_com}). One significant change that appears in some cities when $\widetilde{\tau}_{\alpha \beta}$ is computed over the adjacency graph  is that for African Americans, Whites are more easy to reach than themselves, which means that mobility plays a crucial role on approaching African Americans to the rest of the population and exposing them. Additionally, the normalised CMFPT seems to be less dependent on the ethnicity of the origin and more on the ethnicity of the destination. Likely as a consequence of the higher mixing produced by the long-range links present in the mobility network. Not only that, but the differences between each ethnicity are also reduced. Overall, to properly quantify segregation we need to take into account not only the residences but also how the ethnicities move in cities.

\begin{figure}[!htbp]
\begin{center}
\includegraphics[width=4.3cm]{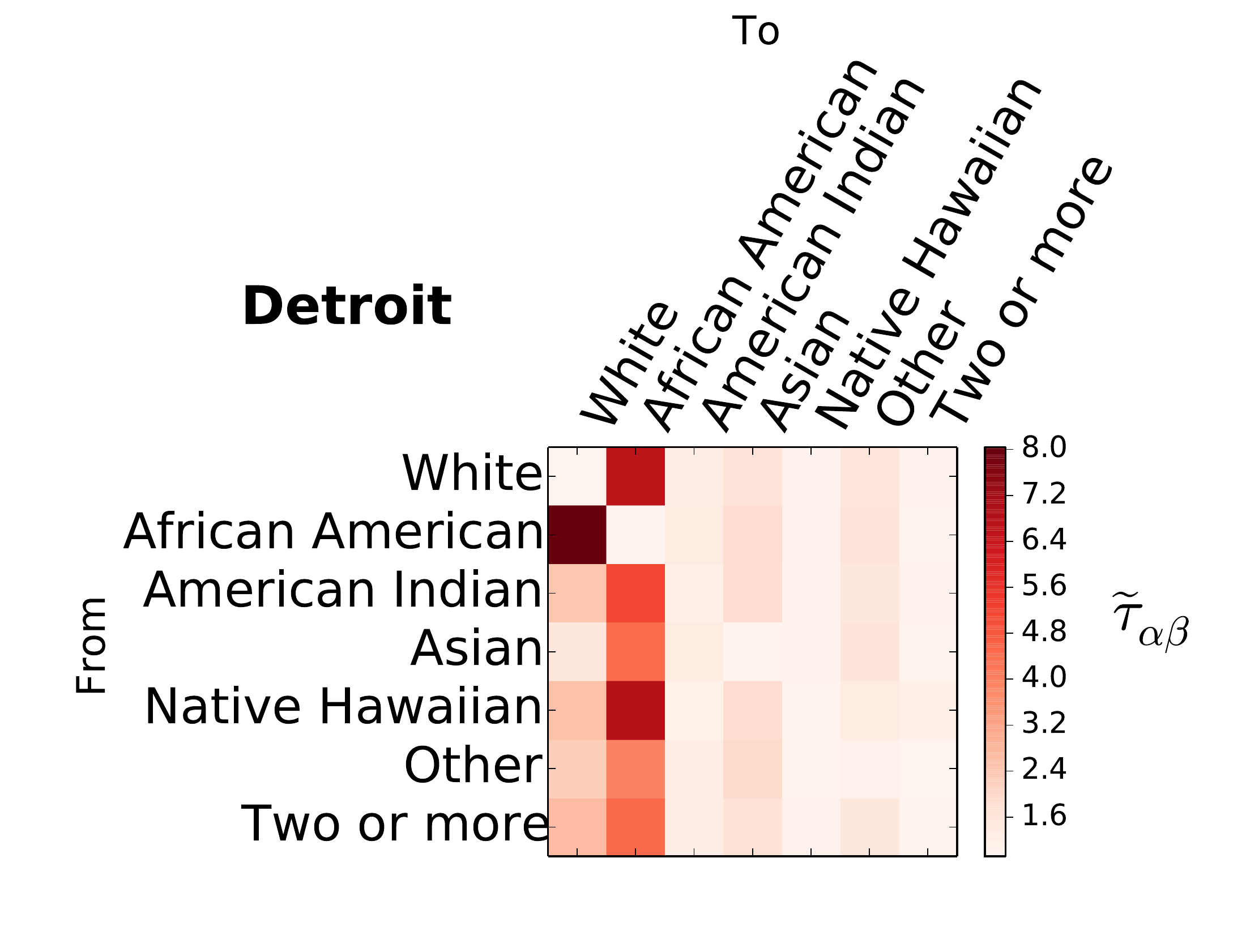}
\includegraphics[width=4.3cm]{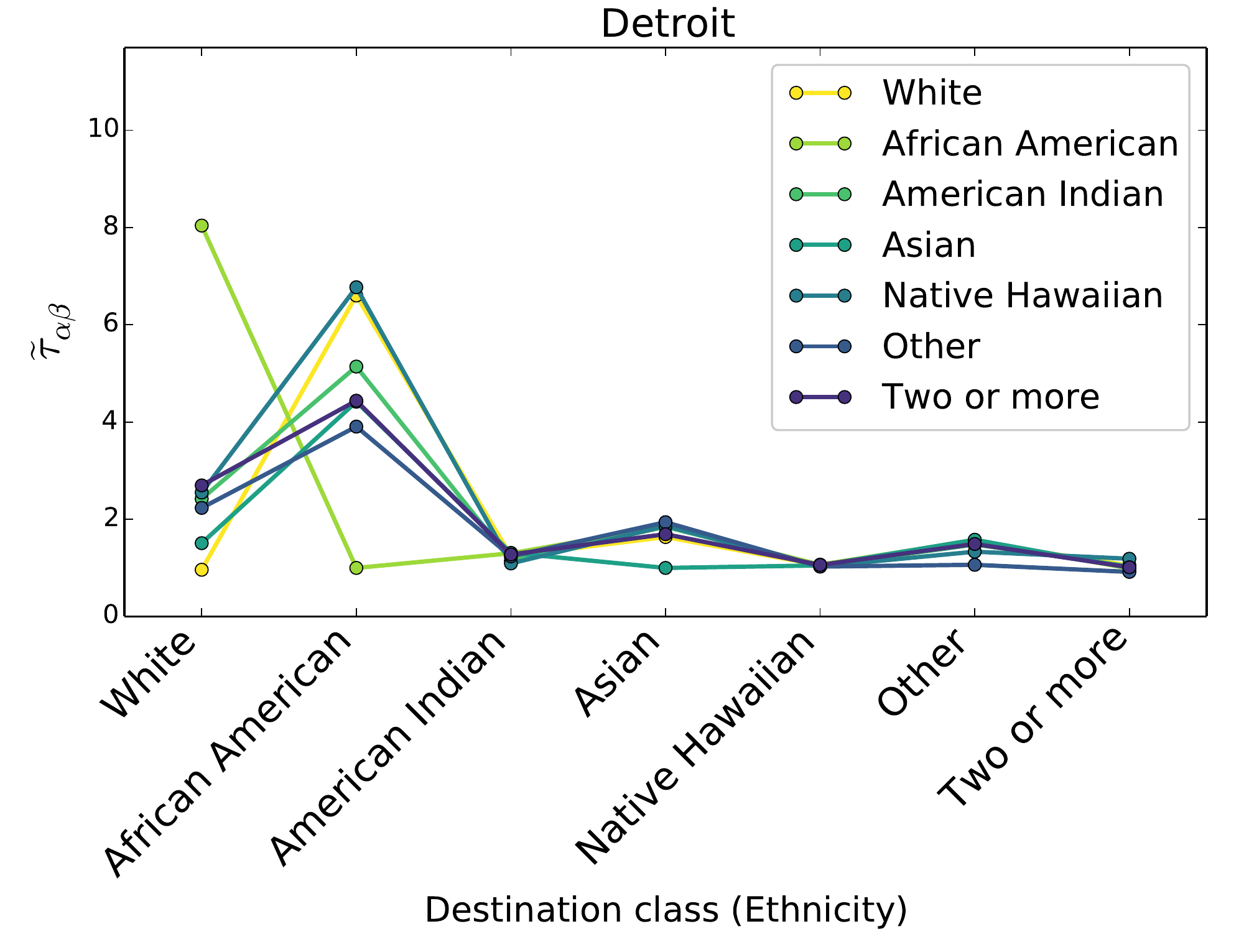}
\includegraphics[width=4.3cm]{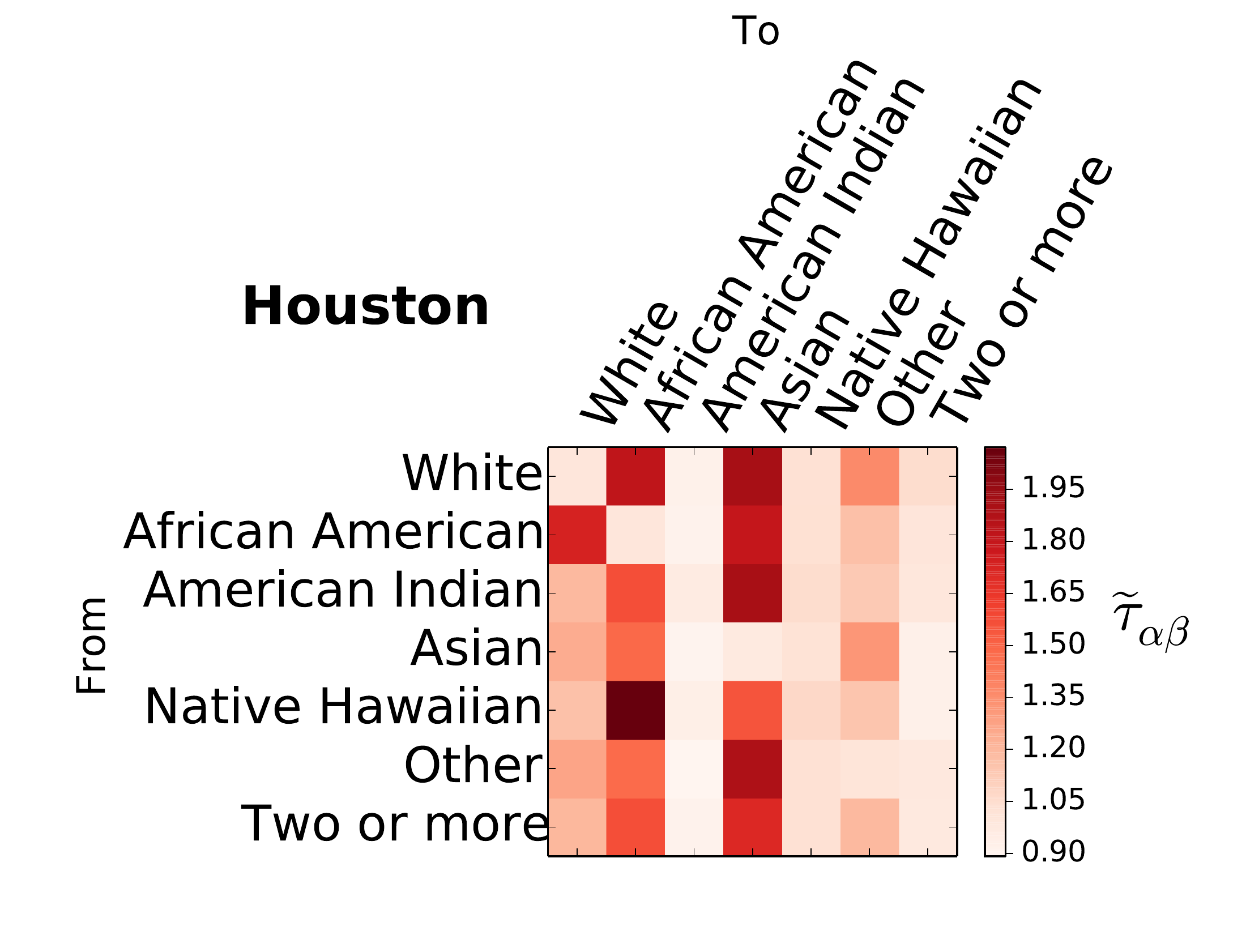}
\includegraphics[width=4.3cm]{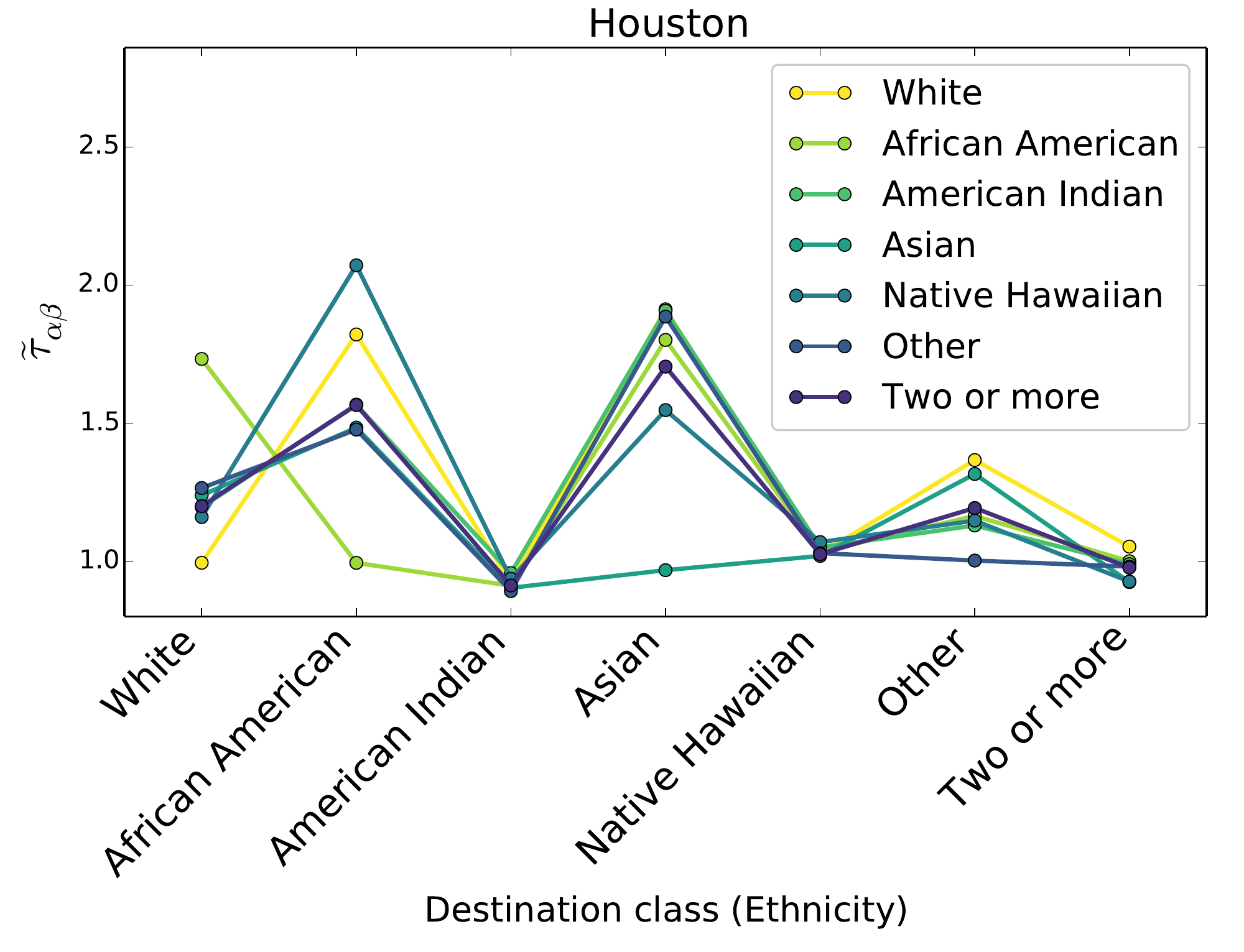}
\includegraphics[width=4.3cm]{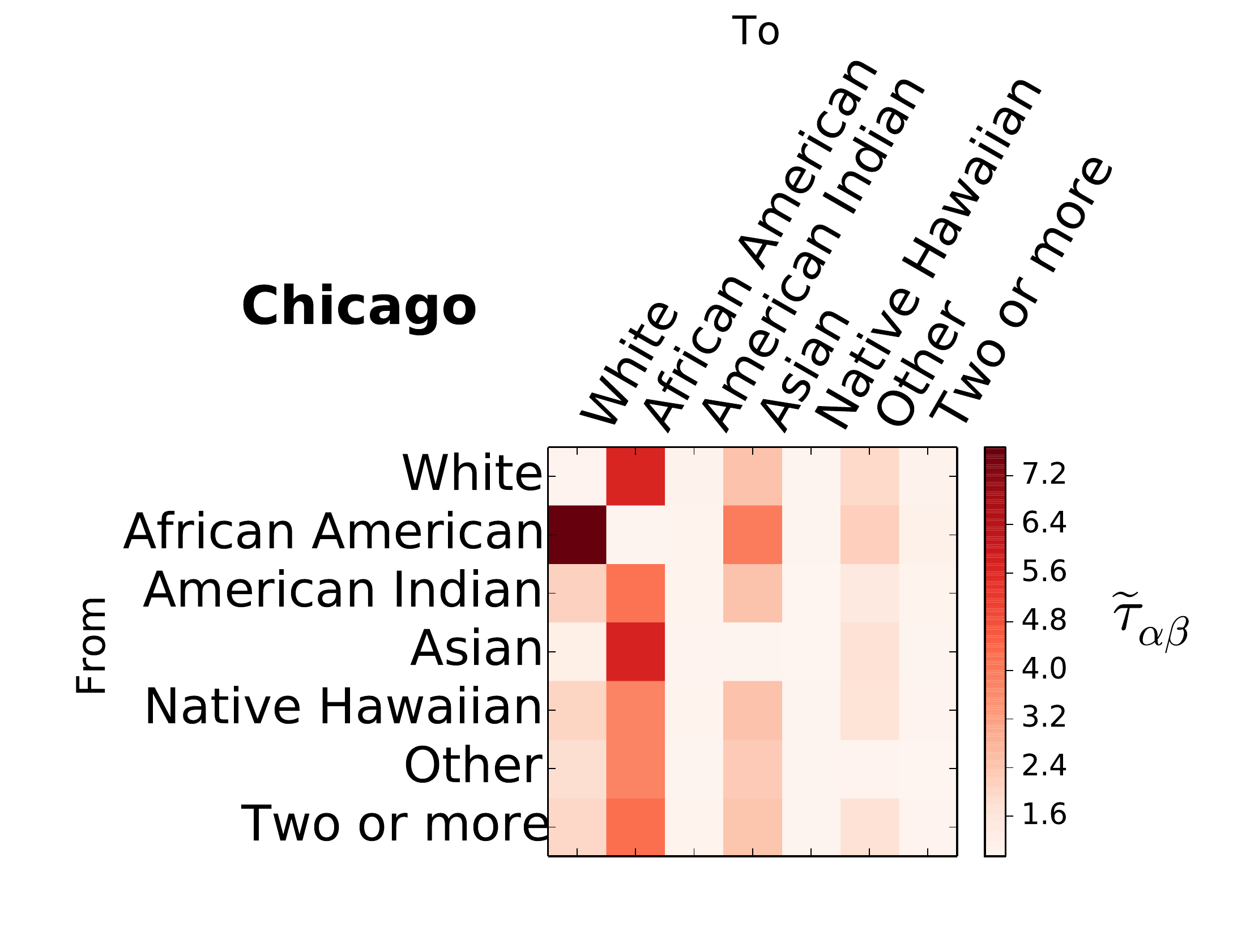}
\includegraphics[width=4.3cm]{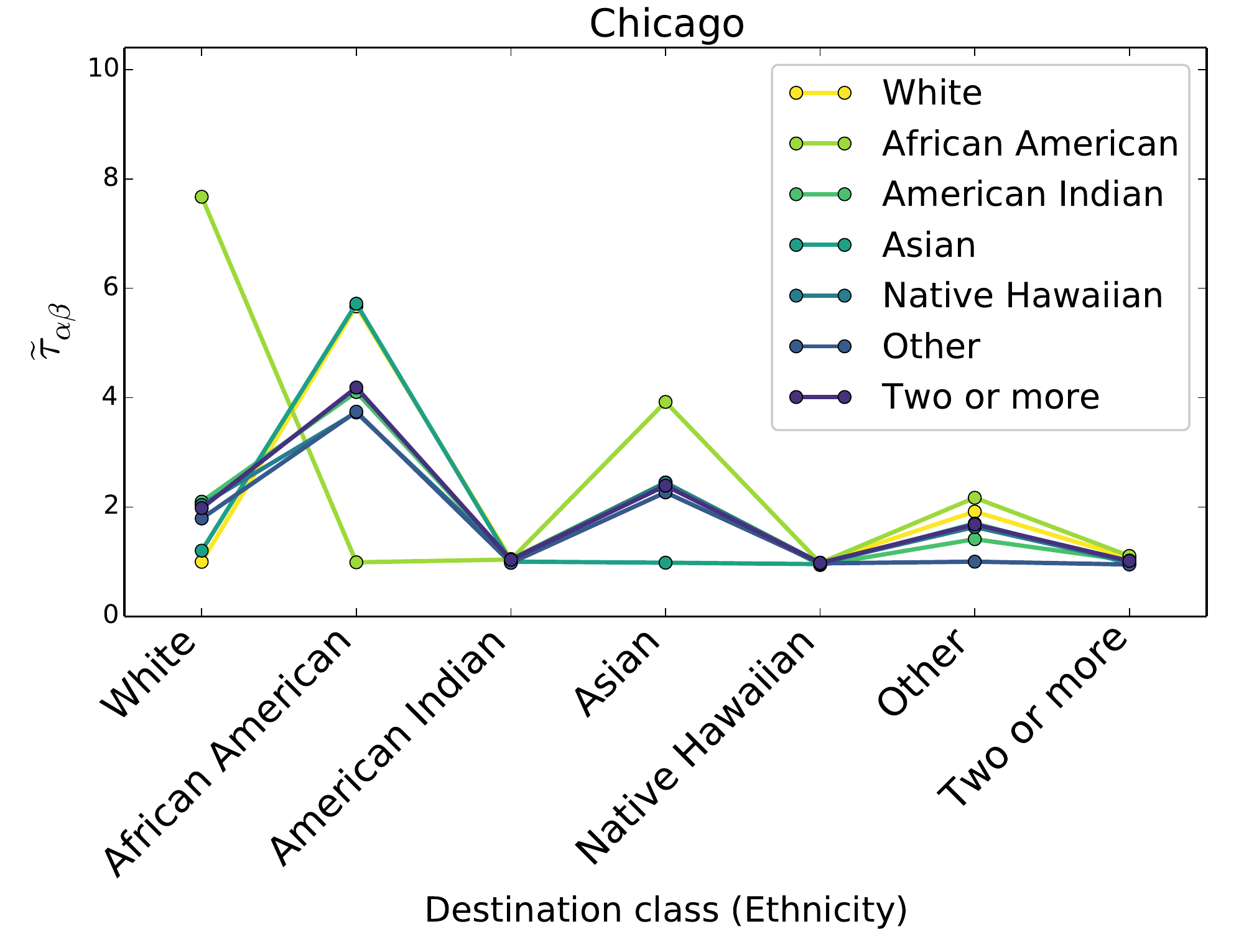}
\includegraphics[width=4.3cm]{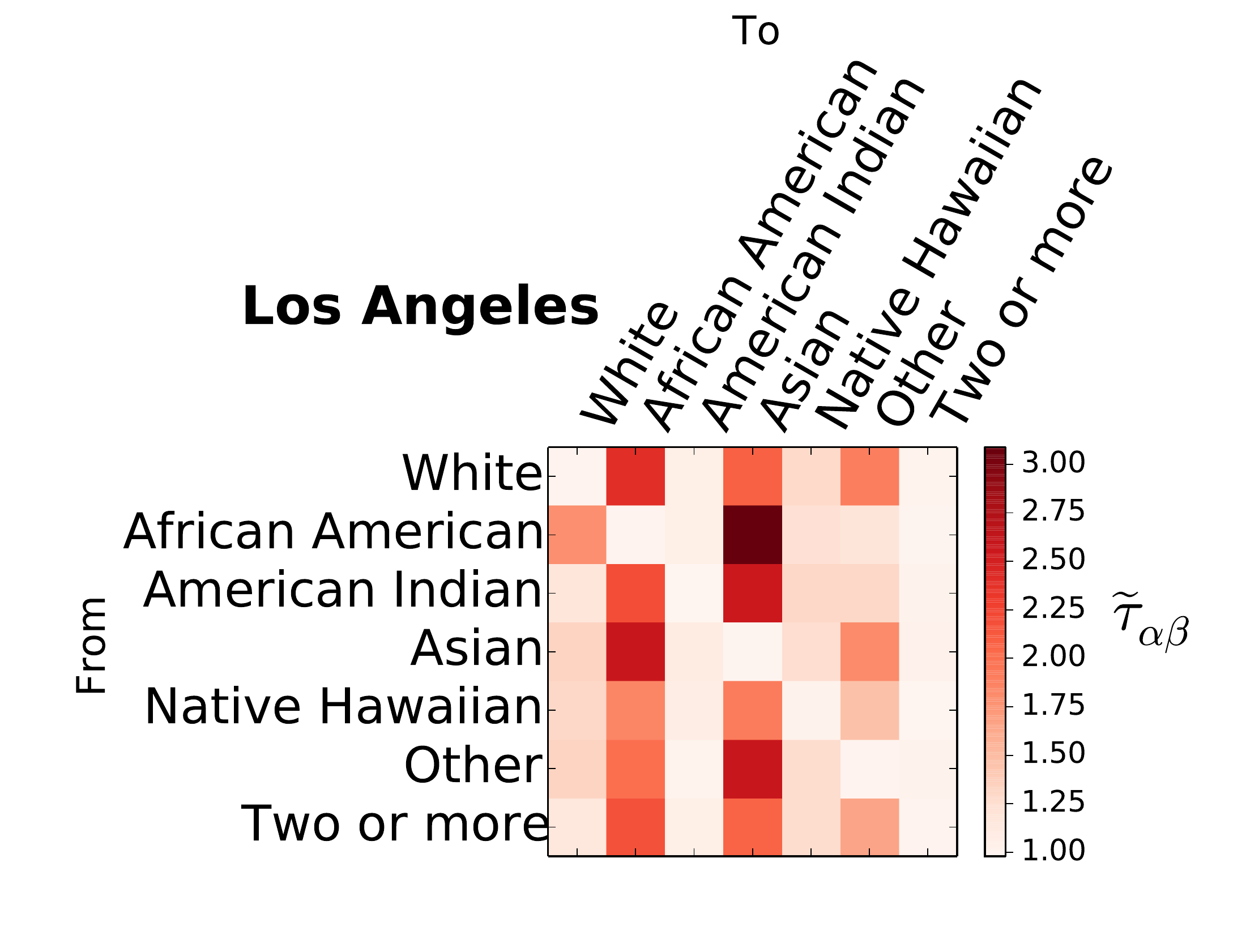}
\includegraphics[width=4.3cm]{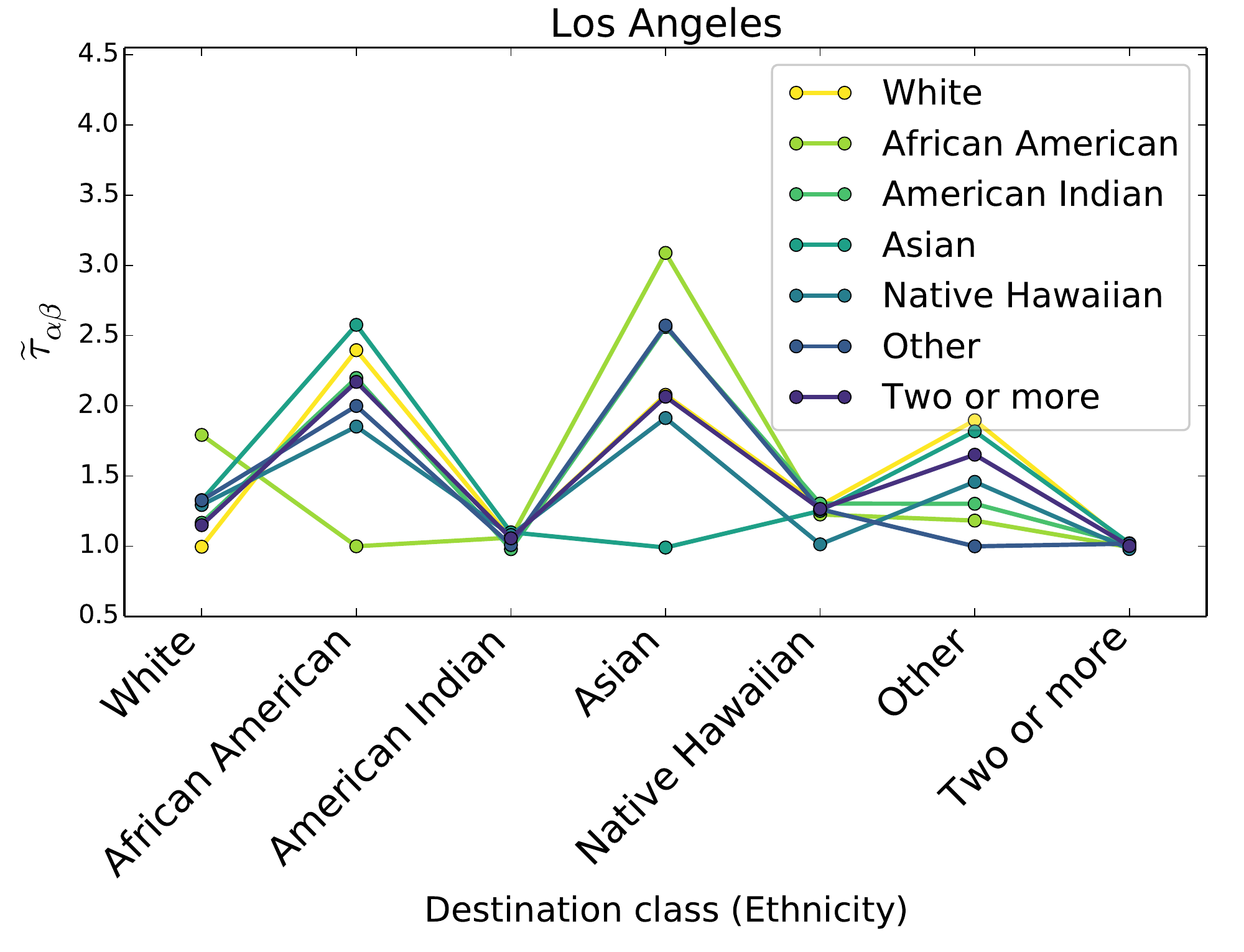}
\caption[Normalised inter-class mean first passage times among the different ethnicities contained in our data set when walkers move on the adjacency network
in two different visualisation styles for the following cities: Chicago, Detroit,
Houston and Los Angeles.]{Normalised inter-class mean first passage times among
the different ethnicities contained in our data set when walkers move on the
adjacency network in two different visualisation styles for the following
cities: Chicago, Detroit, Houston and Los Angeles.
\label{MFPT_adj}}
\end{center}
\end{figure}

\begin{figure}[!htbp]
\begin{center}
\includegraphics[width=4.3cm]{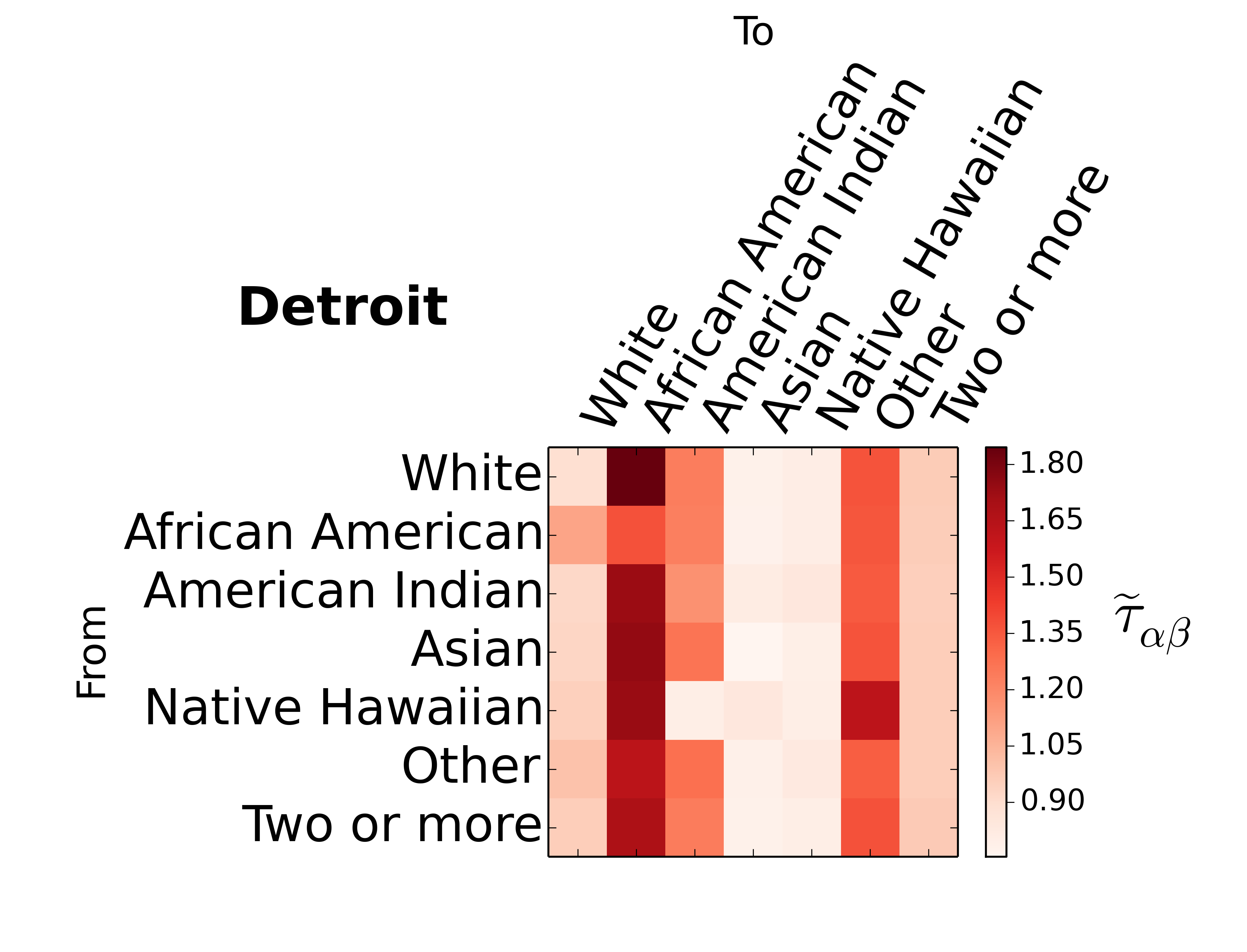}
\includegraphics[width=4.3cm]{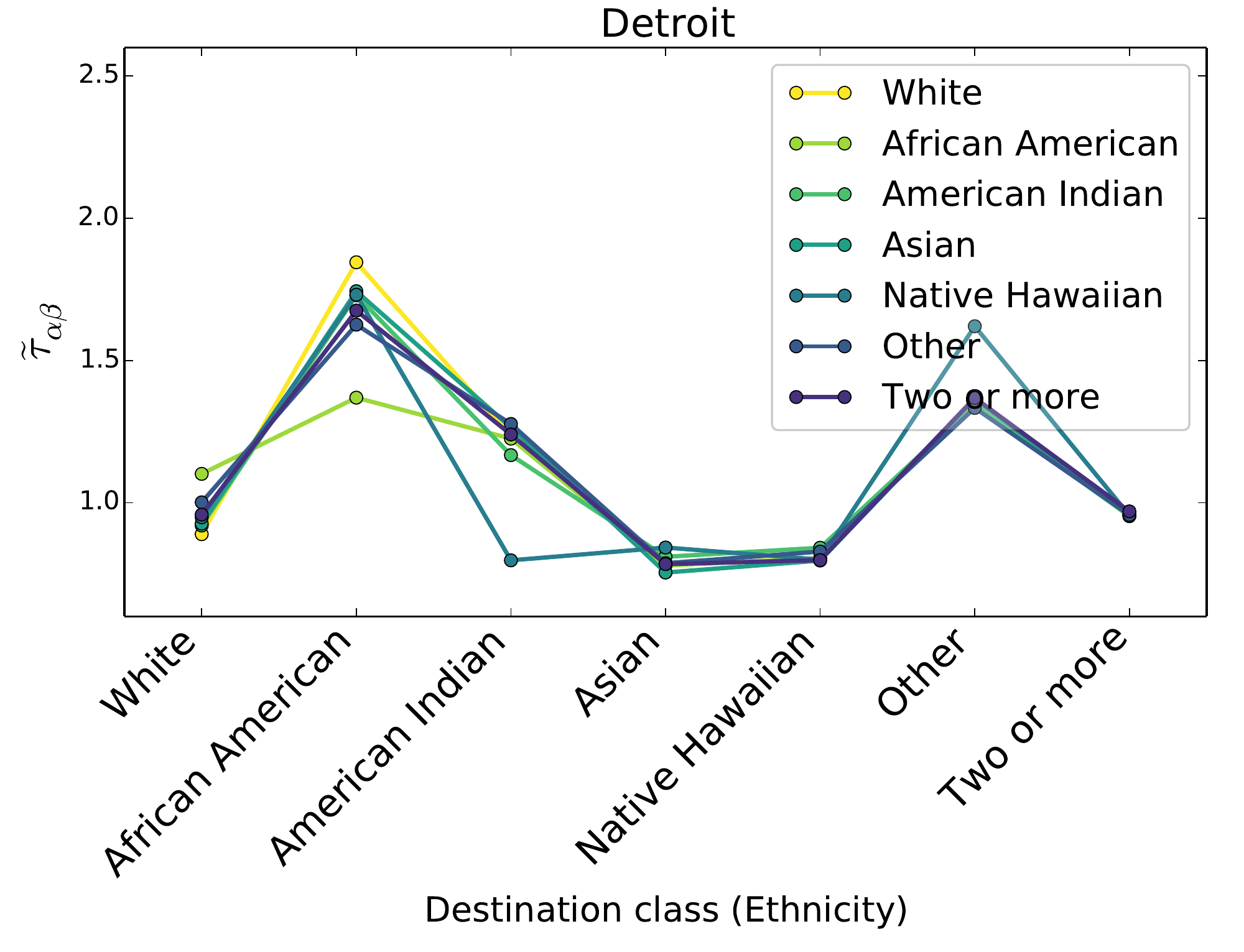}
\includegraphics[width=4.3cm]{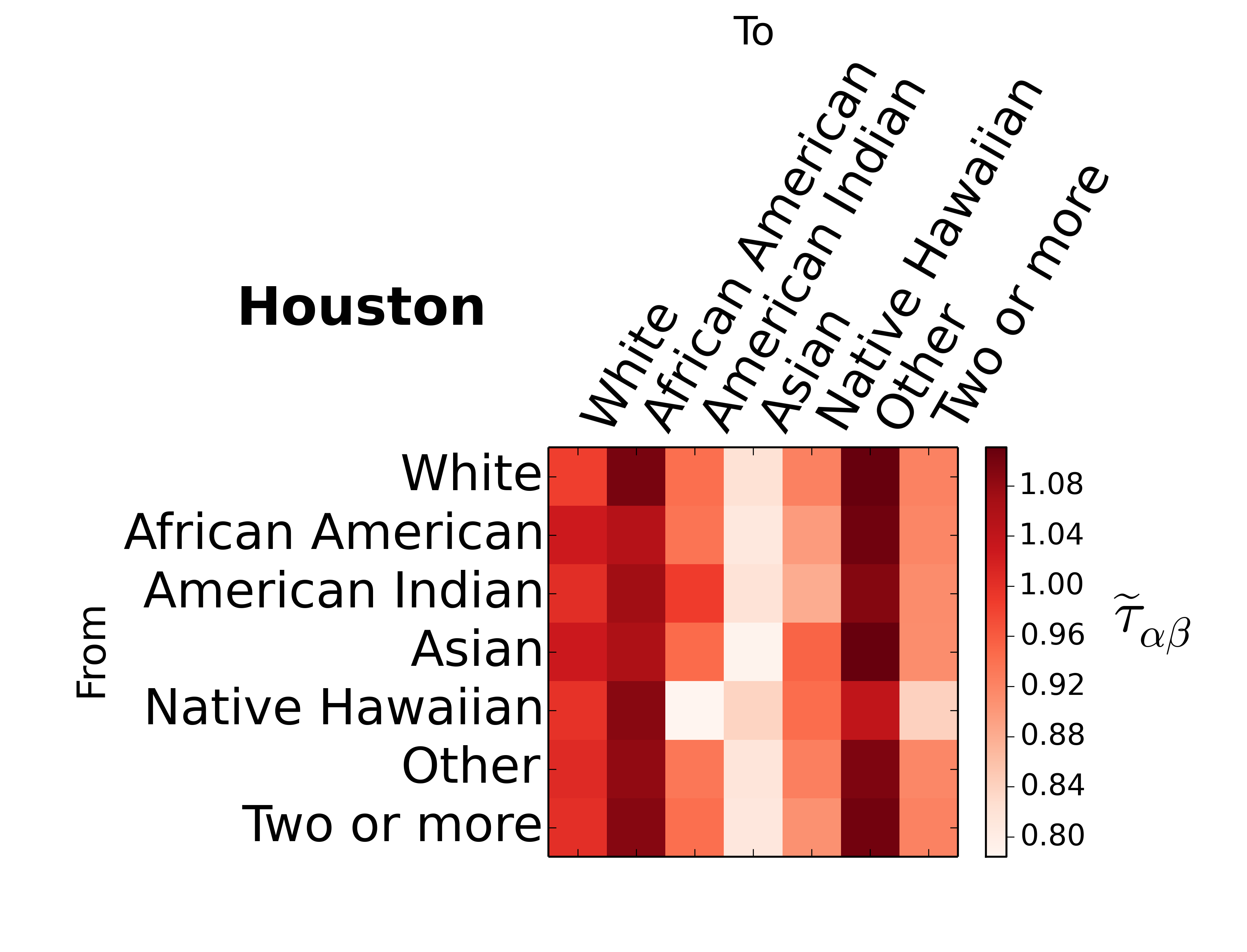}
\includegraphics[width=4.3cm]{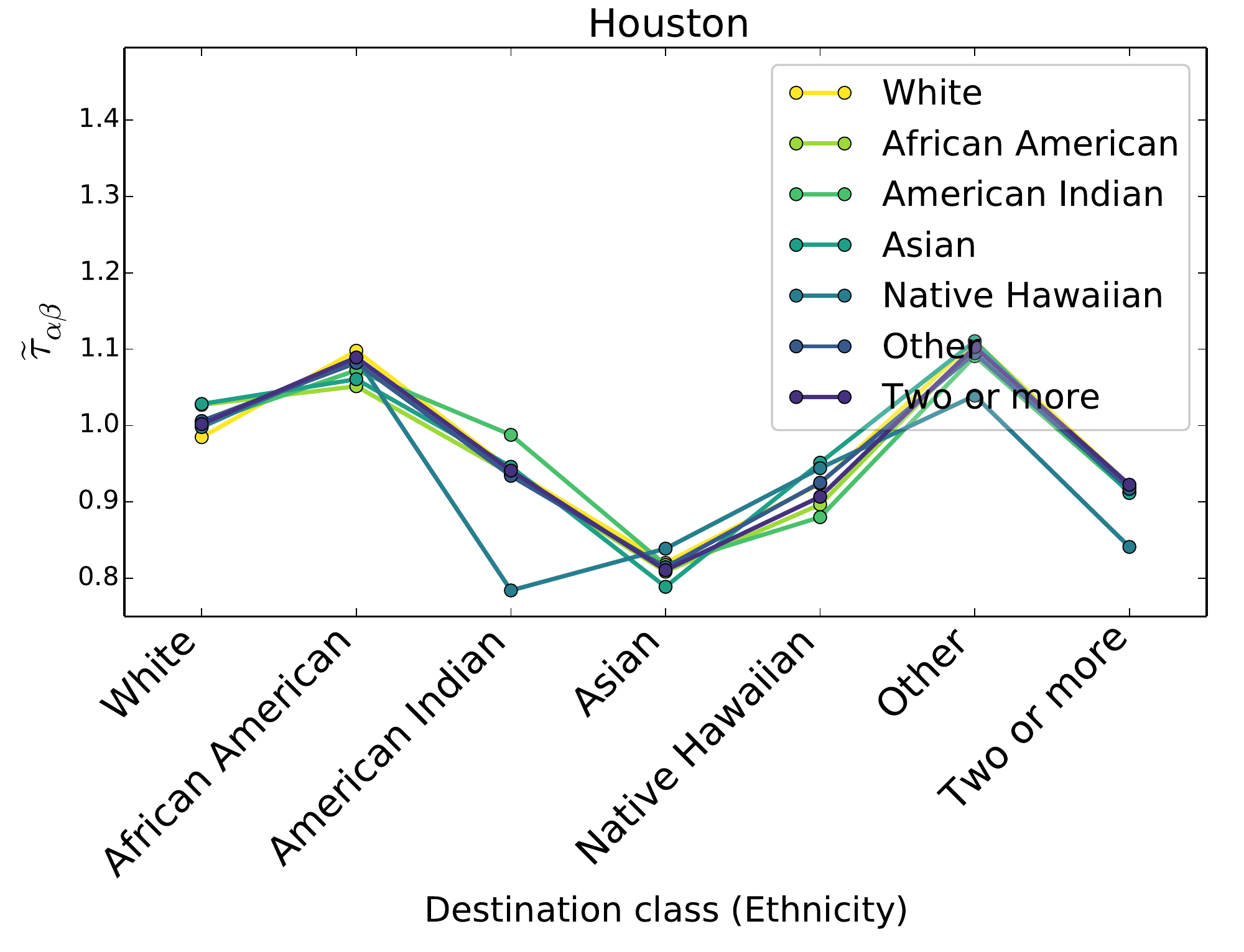}
\includegraphics[width=4.3cm]{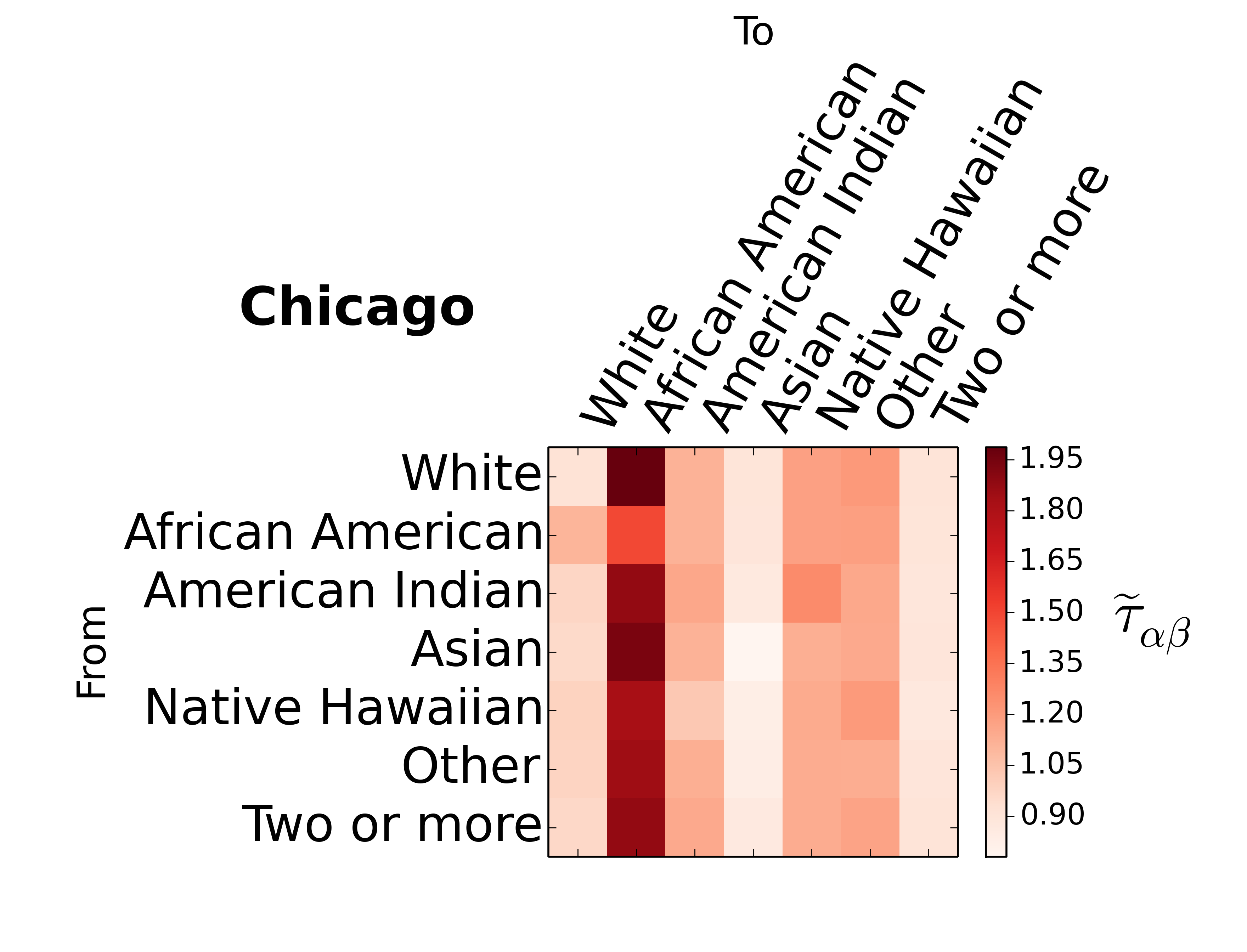}
\includegraphics[width=4.3cm]{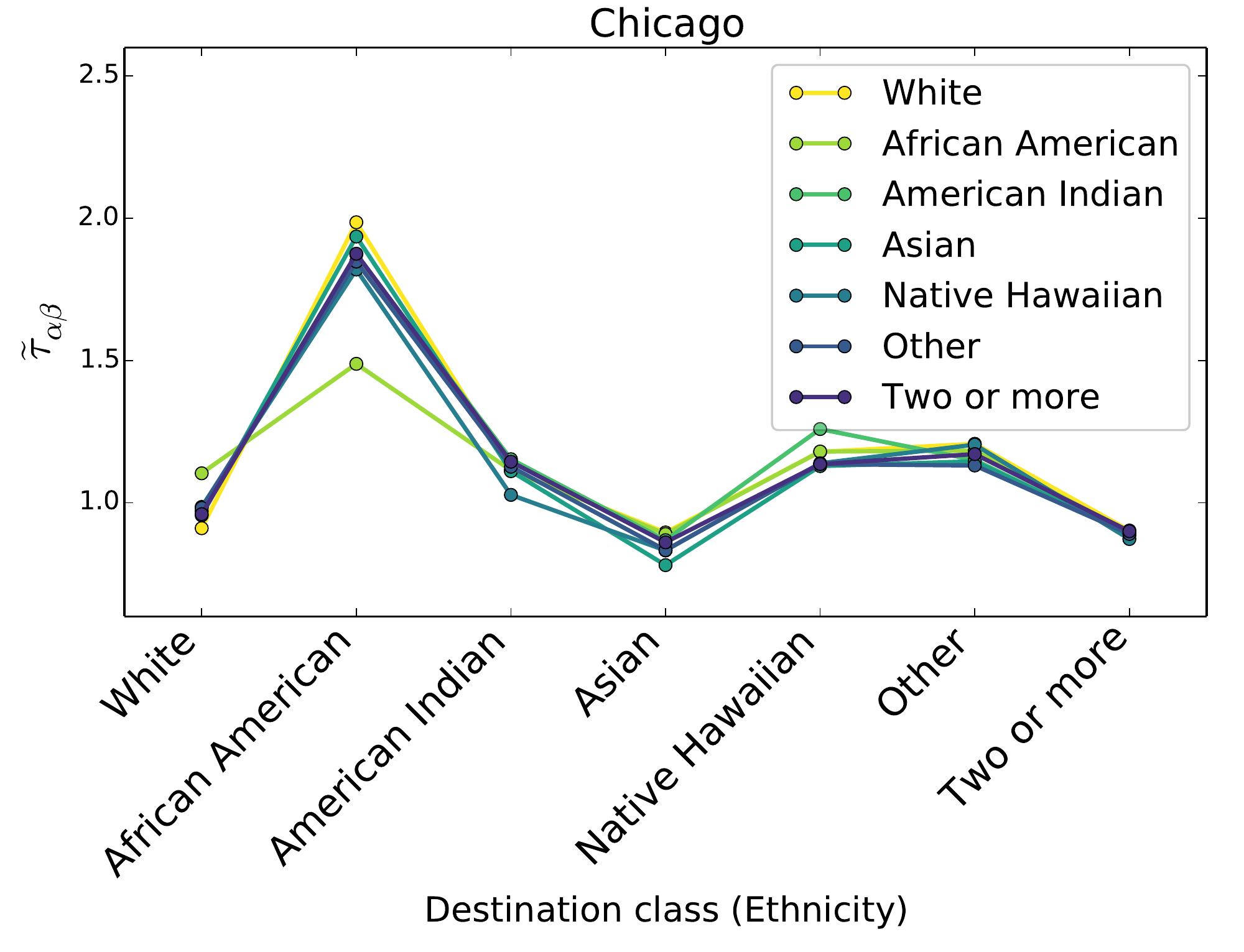}
\includegraphics[width=4.3cm]{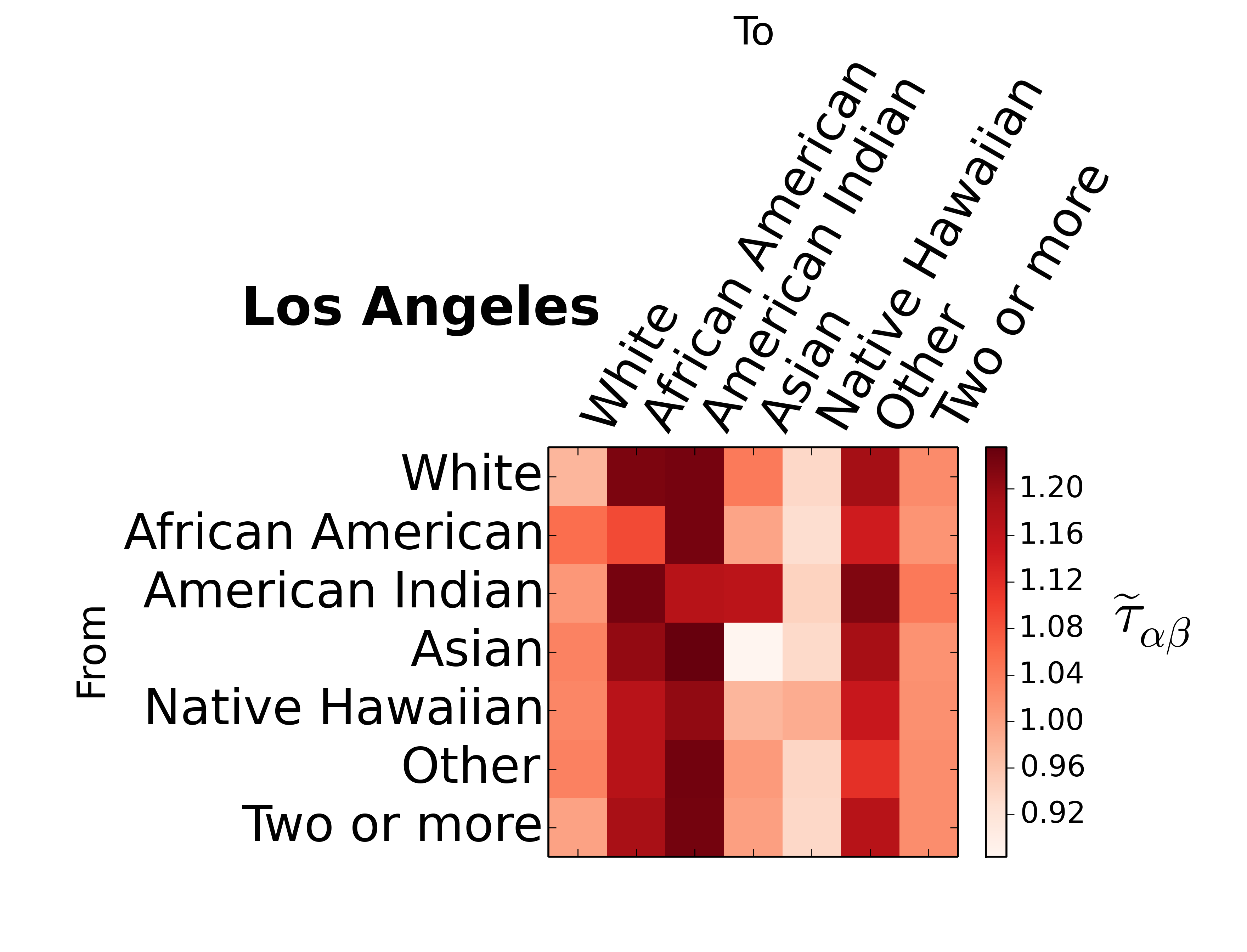}
\includegraphics[width=4.3cm]{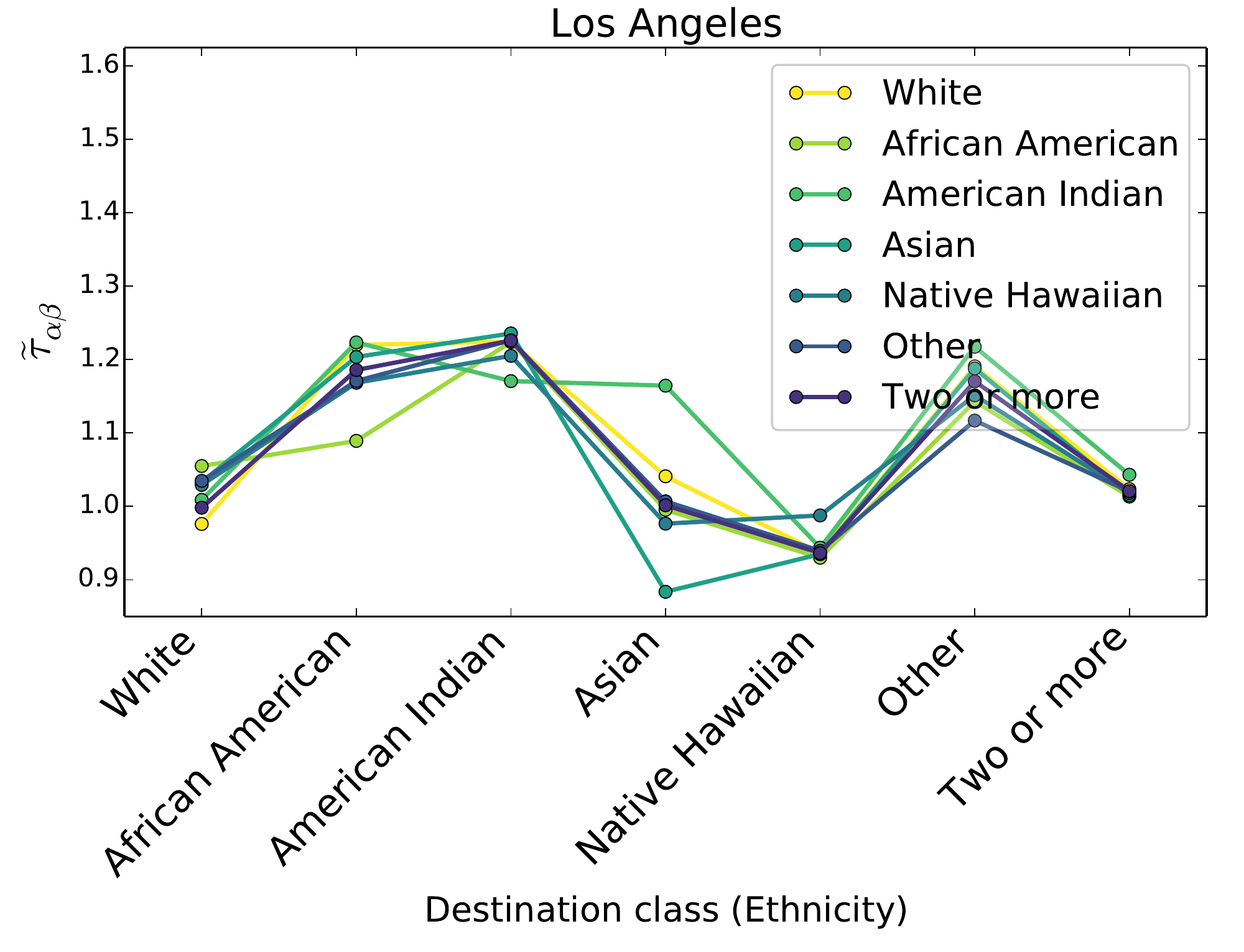}
\caption{Normalised inter-class mean first passage times among the different ethnicities contained in our data set when walkers move on the commuting
network in two different visualisation styles for the following cities:
Chicago, Detroit, Houston and Los Angeles.
\label{MFPT_com}}
\end{center}
\end{figure}

\subsection*{Ethnic segregation and COVID-19 incidence through CCT}

We consider a number of Consolidated Statistical Areas (CSA) in the
US, 128 networks are constructed based on adjacency while 171 networks
were constructed for commuting. These systems are represented as a
spatial graph $\mathcal{G}$ and we look at the statistical properties
of the trajectories of a random walk on $\mathcal{G}$. Each walk
starting at node $i$ is associated to an ethnicity sampled from
$Q_{i_t}$ and it stops at time $t$ when $J(\mathcal{P}||Q_{i_t}) \leq
\varepsilon$. When the ethnicity $\alpha$ is sampled, the
corresponding bin is removed from the computation
$J(\mathcal{P}||Q_{i_t})$ so that the effect that $\alpha$ has on the
coverage time at threshold $\epsilon$ can be quantified.

Trajectories from each node are averaged over 5000 repetitions for the
adjacency network and 2000 for commuting. The null model for $CCT$ is
obtained by randomly reassigning the vector $M_i$ to a new node
in $\mathcal{G}$ and the concentration of an ethnicity in a region - if
such spatial pattern is present - is dissolved across $\mathcal{G}$.
Here we consider 20 independent repetitions of
the null model for each CSA on both adjacency and commuting networks,
then, the Class Coverage Time ($CCT$) for the null model of a city is
the average behaviour over 20 independent realisations.

\begin{figure}[!htbp]
  \center
  \includegraphics[width=7.5cm]{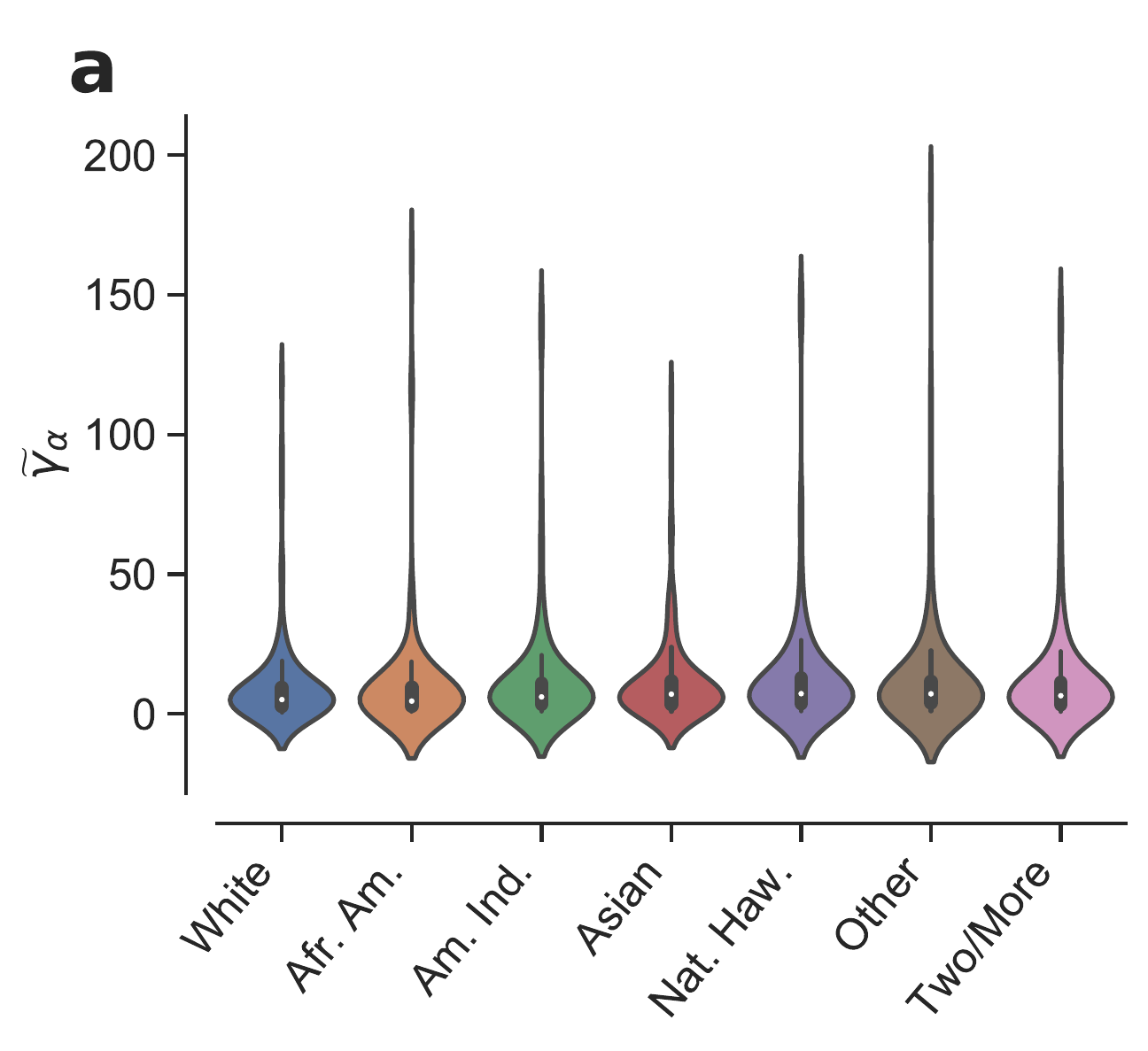}
  \includegraphics[width=7.5cm]{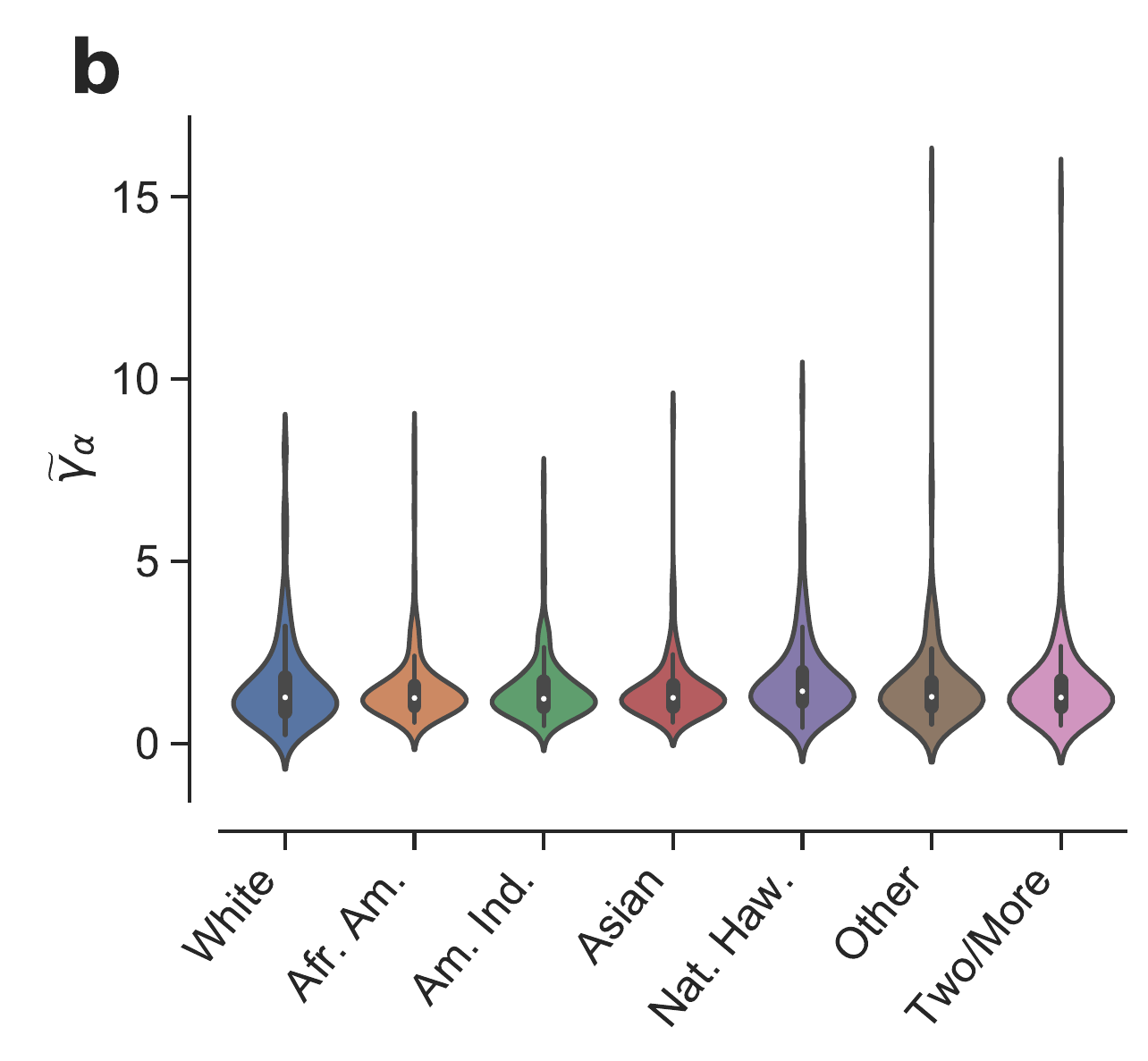}
  \includegraphics[width=7.5cm]{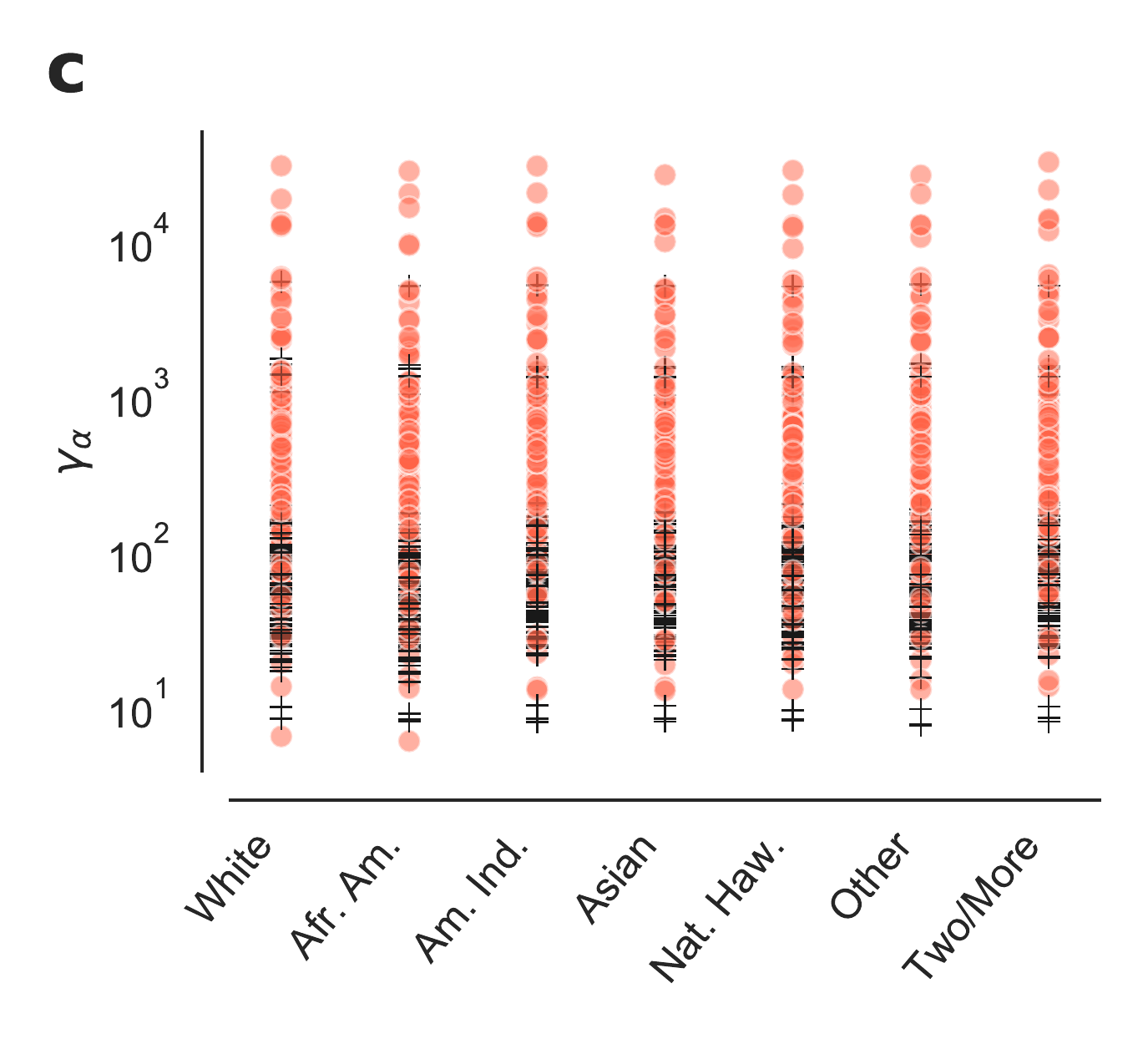}
  \includegraphics[width=7.5cm]{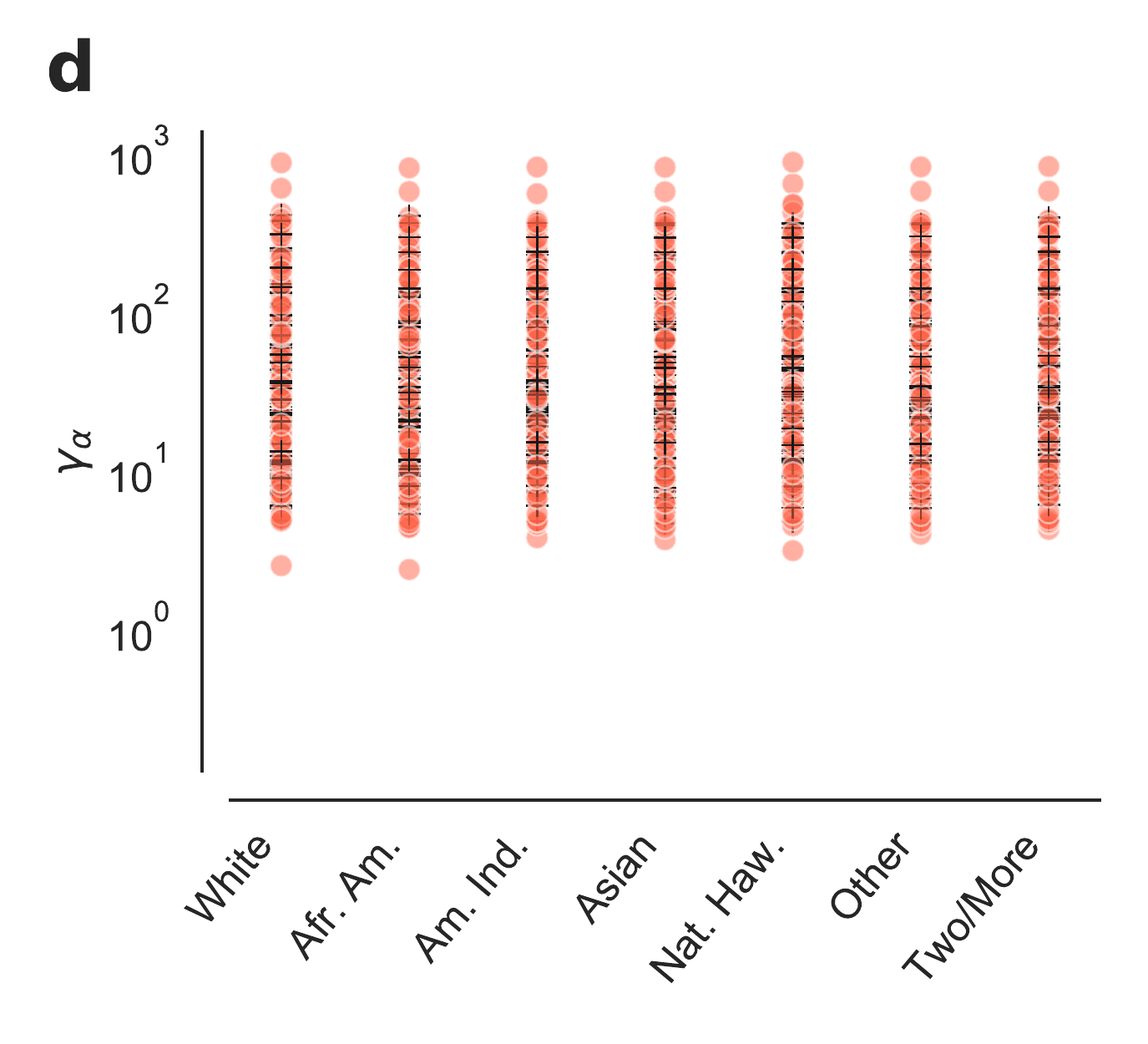}
  \caption[Class coverage times and the corresponding normalised values
  on the adjacency and commute networks of all CSA]{Class coverage times
  and the corresponding normalised values on the adjacency
  \textbf{a} and commute \textbf{b} networks for all CSA. Panels
  \textbf{c-d} report the coverage times (non-normalised) for adjacency
  and commute networks where the black crosses correspond to the values
  for the equivalent null model. Each data point contained in an ethnicity
  column is equivalent to $\widetilde{\gamma}_\alpha$ at a CSA.}
  \label{cct}
\end{figure}

The normalised $CCT$ is reported for the two spatial configurations in Fig.
\ref{cct} \textbf{a-b}. The coverage times obtained for the adjacency
networks \textbf{a} are considerably larger compared to commuting
\textbf{b}, where on the later, the majority of values spams in a small
range between 0 and 4. The distributions of $\widetilde{\gamma}_\alpha$
for each ethnicity have a comparable shape within the network type where
most values are contained in a common interval, yet, they are distinguishable
and differences between the ethnicities can be observed. The corresponding
non-normalised quantities can be read on panels \textbf{c-d} where the $CCT$
of the real system and the equivalent null model are reported.

It is important to note that the difference on the $CCT$ of the real system
and the null model is significantly large on the adjacency networks, with
the former having cases where coverage times are two orders of magnitude
larger (See Fig. \ref{cct} \textbf{c}). Although the null model corresponds
to the non-segregated counterpart of the city and coverage times are
expected to be smaller, these large differences suggest caution and open
an interesting question for further investigation. In particular, to
understand what factors influence the large differences, for instance if
it is mainly driven by the population distribution, the threshold $\epsilon$,
the network topology or the combination of two or more factors.

\begin{figure}[!htbp]
  \center
  \includegraphics[width=14cm]{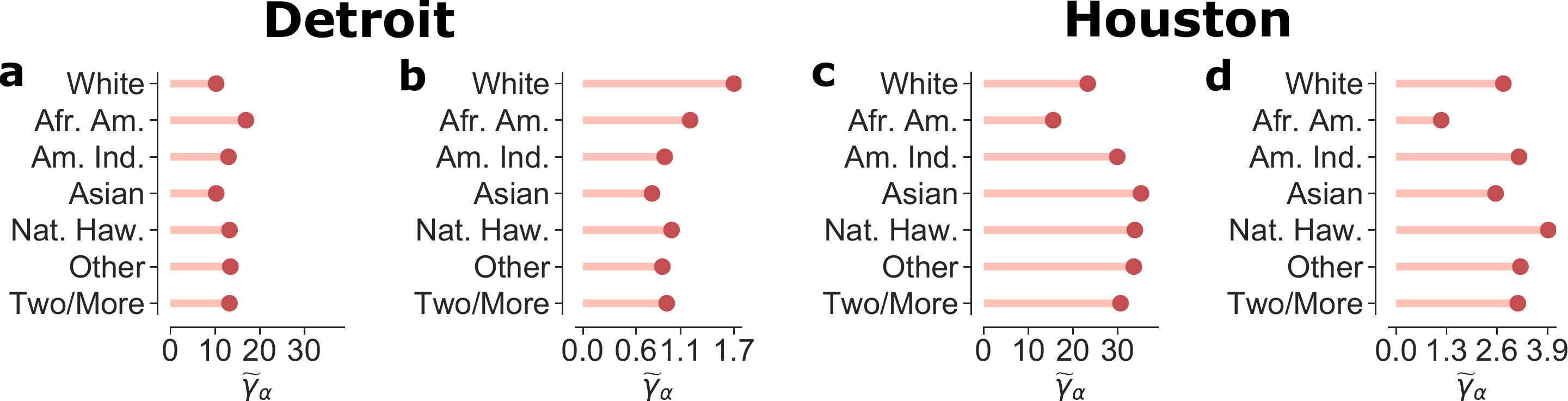}
  \caption[Normalised coverage time for the adjacency and commute network
  in Detroit and Houston]{Normalised coverage time on the adjacency and
  commute networks for Detroit \textbf{a-b} and Houston \textbf{c-d}.
  Values are larger on the adjacency network compared to commute for both
  cities while African Americans are significantly less isolated in
  Houston for both networks.}
  \label{gamma_extra}
\end{figure}

In addition to the cities discussed in the main manuscript, we report two
other systems in Fig. \ref{gamma_extra} where individual values of
$\widetilde{\gamma}_\alpha$ for each ethnicity can be observed on the
adjacency and commute networks. African Americans are substantially less
isolated in Houston compared to the other ethnicities in both adjacency
and commute networks. In Detroit, The adjacency information gives the
opposite picture for African Americas while Whites are the most isolated
on the commute network.

\section{Correlations between the incidence of COVID-19 among the African American population cases and ethnic segregation)}

\subsection*{Rankings and comparisons}

We detail the cities studied in Supplementary Table \ref{tablecities}, it is important to note that those are the cities studied disregarding if those states provide ethnic information on the impact of COVID-19.

\begin{table}
\begin{center}
\resizebox{\textwidth}{!}{\begin{tabular}{ |c|c|c|c| }
 \hline
City name & City name & City name & City name \\
  \hline
Albany-Schenectady & Albuquerque-Santa Fe-Las Vegas & Appleton-Oshkosh-Neenah & Asheville-Brevard  \\
Atlanta--Athens-Clarke County--Sandy Springs & Bend-Redmond-Prineville & Birmingham-Hoover-Talladega & Bloomington-Bedford  \\
Bloomington-Pontiac & Bloomsburg-Berwick-Sunbury & Boston-Worcester-Providence & Bowling Green-Glasgow  \\
Brownsville-Harlingen-Raymondville & Buffalo-Cheektowaga & Cape Coral-Fort Myers-Naples & Cape Girardeau-Sikeston  \\
Charleston-Huntington-Ashland & Charlotte-Concord & Chattanooga-Cleveland-Dalton & Chicago-Naperville  \\
Cincinnati-Wilmington-Maysville & Cleveland-Akron-Canton & Clovis-Portales & Columbia-Moberly-Mexico  \\
Columbia-Orangeburg-Newberry & Columbus-Auburn-Opelika & Columbus-Marion-Zanesville & Columbus-West Point  \\
Corpus Christi-Kingsville-Alice & Dallas-Fort Worth & Davenport-Moline & Dayton-Springfield-Sidney  \\
Denver-Aurora & DeRidder-Fort Polk South & Des Moines-Ames-West Des Moines & Detroit-Warren-Ann Arbor  \\
Dixon-Sterling & Dothan-Enterprise-Ozark & Eau Claire-Menomonie & Edwards-Glenwood Springs  \\
Elmira-Corning & El Paso-Las Cruces & Erie-Meadville & Fargo-Wahpeton  \\
Fayetteville-Lumberton-Laurinburg & Findlay-Tiffin & Fort Wayne-Huntington-Auburn & Fresno-Madera  \\
Gainesville-Lake City & Grand Rapids-Wyoming-Muskegon & Green Bay-Shawano & Greensboro--Winston-Salem--High Point  \\
Greenville-Spartanburg-Anderson & Greenville-Washington & Harrisburg-York-Lebanon & Harrisonburg-Staunton-Waynesboro  \\
Hartford-West Hartford & Hickory-Lenoir & Hot Springs-Malvern & Houston-The Woodlands  \\
Huntsville-Decatur-Albertville & Idaho Falls-Rexburg-Blackfoot & Indianapolis-Carmel-Muncie & Ithaca-Cortland  \\
Jackson-Brownsville & Jackson-Vicksburg-Brookhaven & Jacksonville-St. Marys-Palatka & Johnson City-Kingsport-Bristol  \\
Johnstown-Somerset & Jonesboro-Paragould & Joplin-Miami & Kalamazoo-Battle Creek-Portage  \\
Kansas City-Overland Park-Kansas City & Knoxville-Morristown-Sevierville & Kokomo-Peru & Lafayette-Opelousas-Morgan City  \\
Lafayette-West Lafayette-Frankfort & Lake Charles-Jennings & Lansing-East Lansing-Owosso & Las Vegas-Henderson  \\
Lexington-Fayette--Richmond--Frankfort & Lima-Van Wert-Celina & Lincoln-Beatrice & Little Rock-North Little Rock  \\
Longview-Marshall & Los Angeles-Long Beach & Louisville/Jefferson County--Elizabethtown--Madison & Lubbock-Levelland  \\
Macon-Bibb County--Warner Robins & Madison-Janesville-Beloit & Manhattan-Junction City & Mankato-New Ulm-North Mankato  \\
Mansfield-Ashland-Bucyrus & Martin-Union City & McAllen-Edinburg & Medford-Grants Pass  \\
Memphis-Forrest City & Miami-Fort Lauderdale-Port St. Lucie & Midland-Odessa & Milwaukee-Racine-Waukesha  \\
Minneapolis-St. Paul & Mobile-Daphne-Fairhope & Modesto-Merced & Monroe-Ruston-Bastrop  \\
Morgantown-Fairmont & Moses Lake-Othello & Mount Pleasant-Alma & Myrtle Beach-Conway  \\
Nashville-Davidson--Murfreesboro & New Bern-Morehead City & New Orleans-Metairie-Hammond & New York-Newark  \\
North Port-Sarasota & Oklahoma City-Shawnee & Omaha-Council Bluffs-Fremont & Orlando-Deltona-Daytona Beach  \\
Oskaloosa-Pella & Paducah-Mayfield & Parkersburg-Marietta-Vienna & Pensacola-Ferry Pass  \\
Peoria-Canton & Philadelphia-Reading-Camden & Pittsburgh-New Castle-Weirton & Portland-Lewiston-South Portland  \\
Portland-Vancouver-Salem & Pueblo-Canyon City & Pullman-Moscow & Quincy-Hannibal  \\
Raleigh-Durham-Chapel Hill & Rapid City-Spearfish & Redding-Red Bluff & Reno-Carson City-Fernley  \\
Richmond-Connersville & Rochester-Austin & Rochester-Batavia-Seneca Falls & Rockford-Freeport-Rochelle  \\
Rocky Mount-Wilson-Roanoke Rapids & Rome-Summerville & Sacramento-Roseville & Saginaw-Midland-Bay City  \\
St. Louis-St. Charles-Farmington & Salt Lake City-Provo-Orem & San Jose-San Francisco-Oakland & Savannah-Hinesville-Statesboro  \\
Seattle-Tacoma & Sioux City-Vermillion & South Bend-Elkhart-Mishawaka & Spokane-Spokane Valley-Coeur d'Alene  \\
Springfield-Branson & Springfield-Greenfield Town & Springfield-Jacksonville-Lincoln & State College-DuBois  \\
Steamboat Springs-Craig & Syracuse-Auburn & Tallahassee-Bainbridge & Toledo-Port Clinton  \\
Tucson-Nogales & Tulsa-Muskogee-Bartlesville & Tyler-Jacksonville & Victoria-Port Lavaca  \\
Virginia Beach-Norfolk & Visalia-Porterville-Hanford & Washington-Baltimore-Arlington & Wausau-Stevens Point-Wisconsin Rapids  \\
Wichita-Arkansas City-Winfield & Williamsport-Lock Haven & Youngstown-Warren  & \\
 \hline
\end{tabular}}
\caption{Table of cities studied}
\label{tablecities}
\end{center}
\end{table}

\begin{figure}[!htbp]
\begin{center}
a\includegraphics[width=8.5cm]{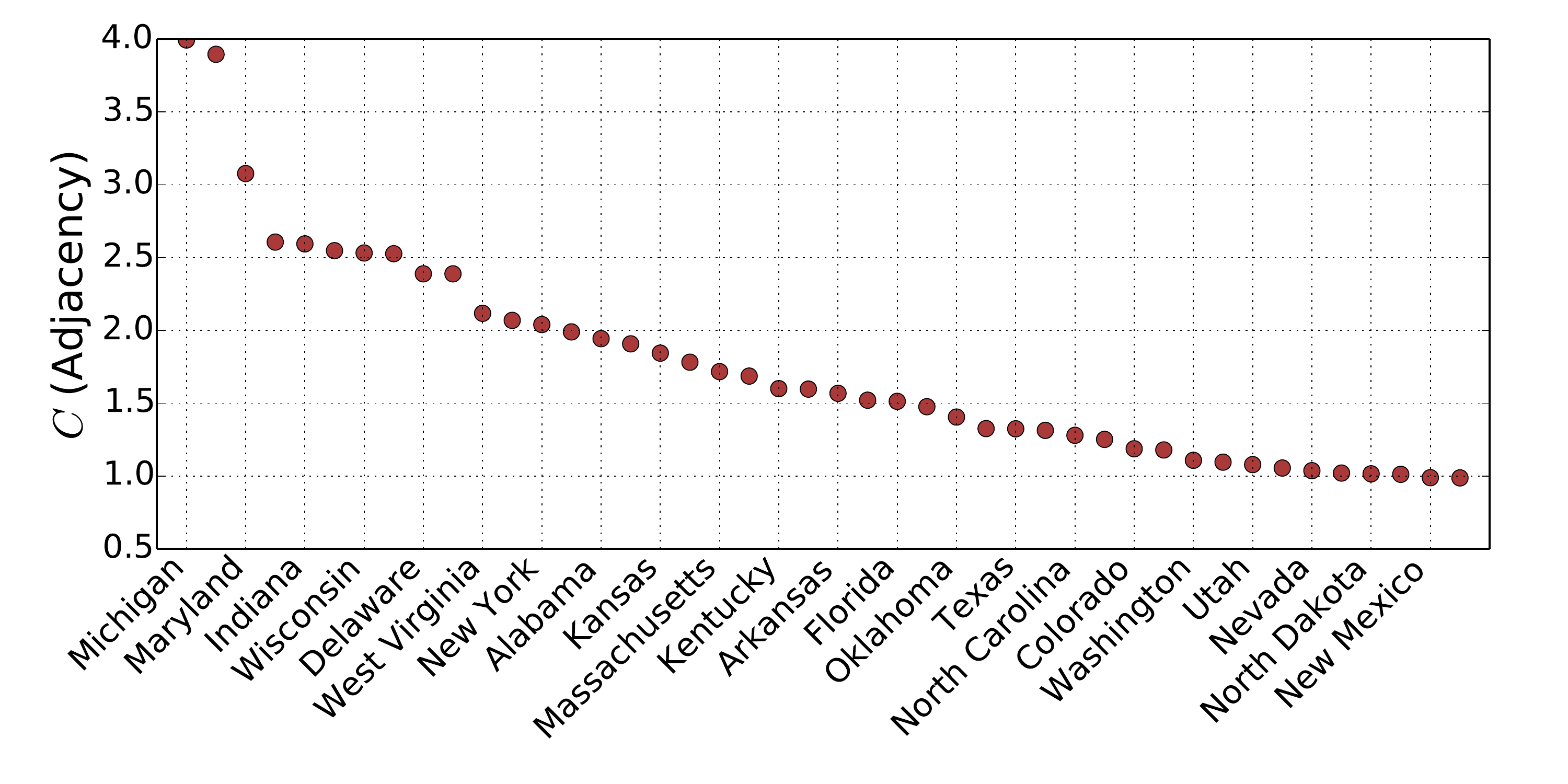}
b\includegraphics[width=8.5cm]{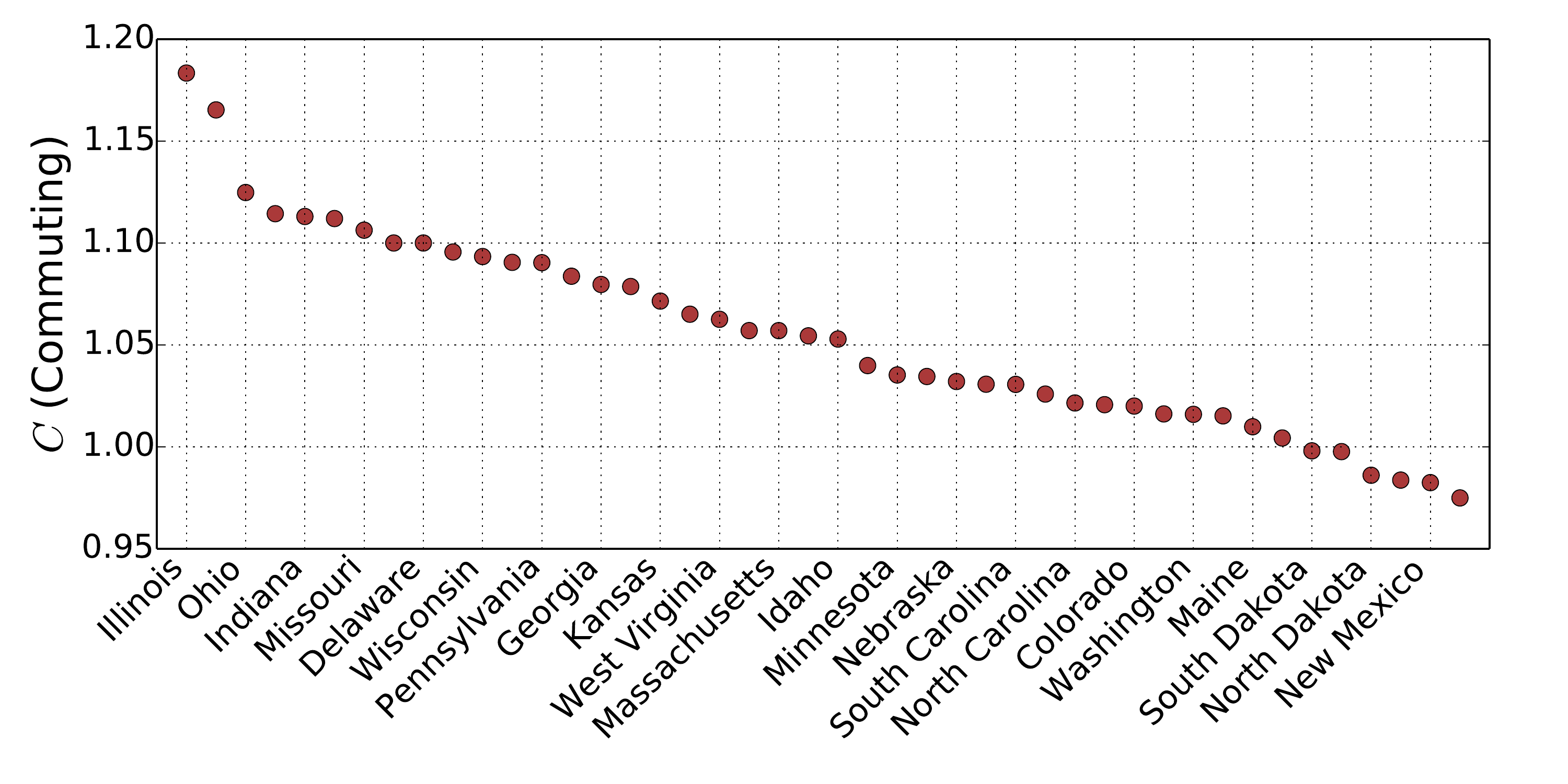}
c\includegraphics[width=8.5cm]{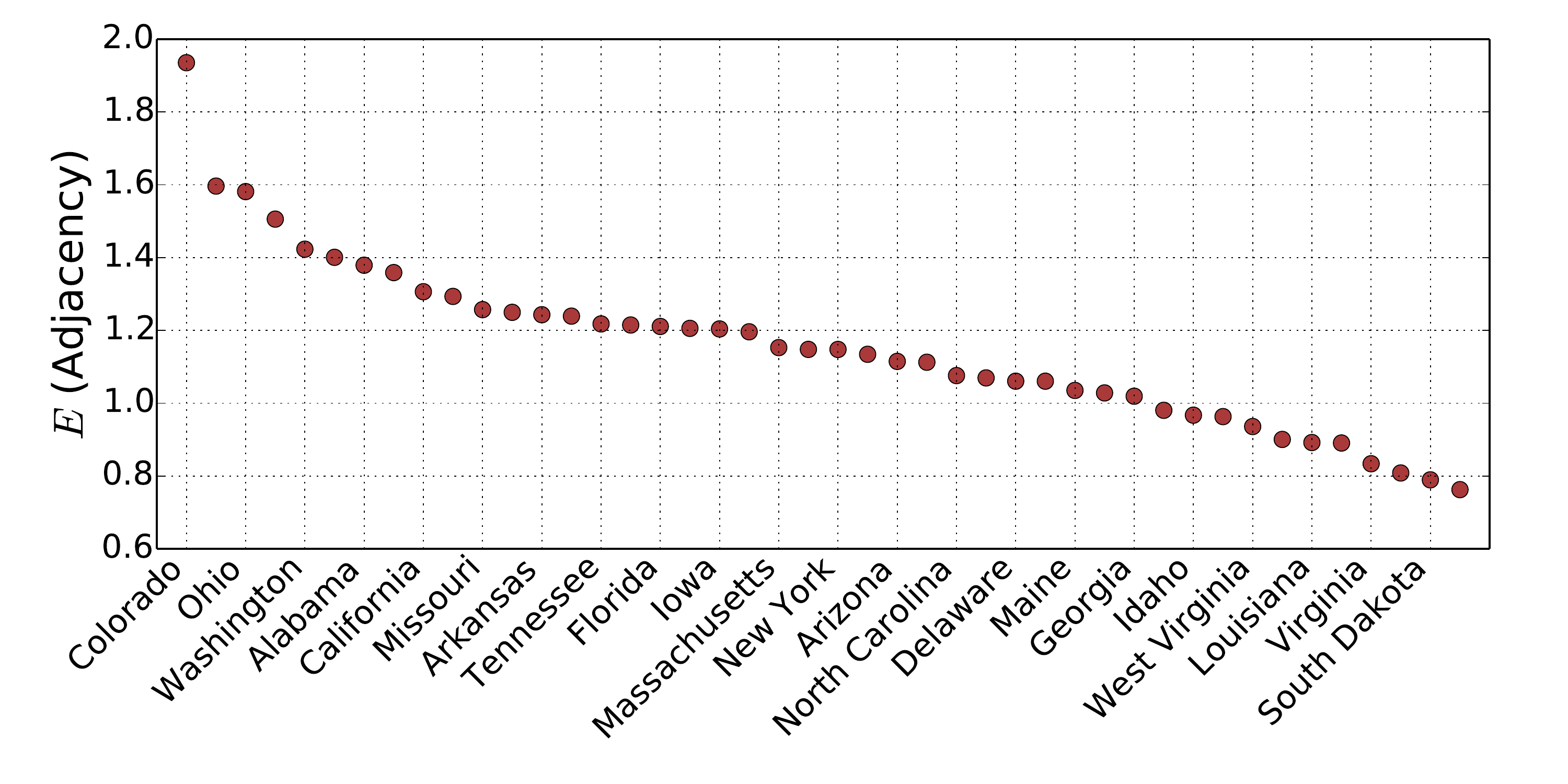}
d\includegraphics[width=8.5cm]{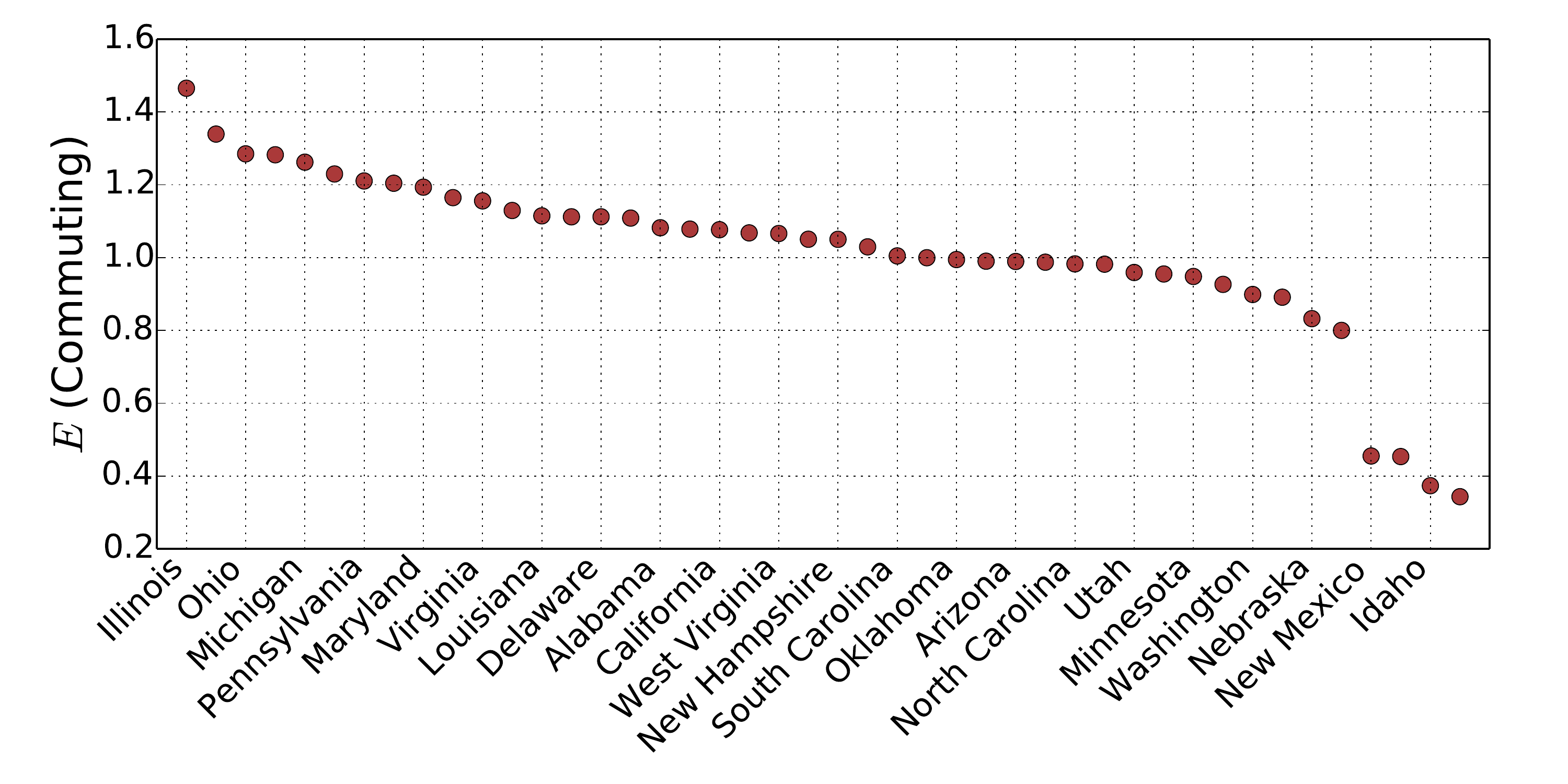}
e\includegraphics[width=8.5cm]{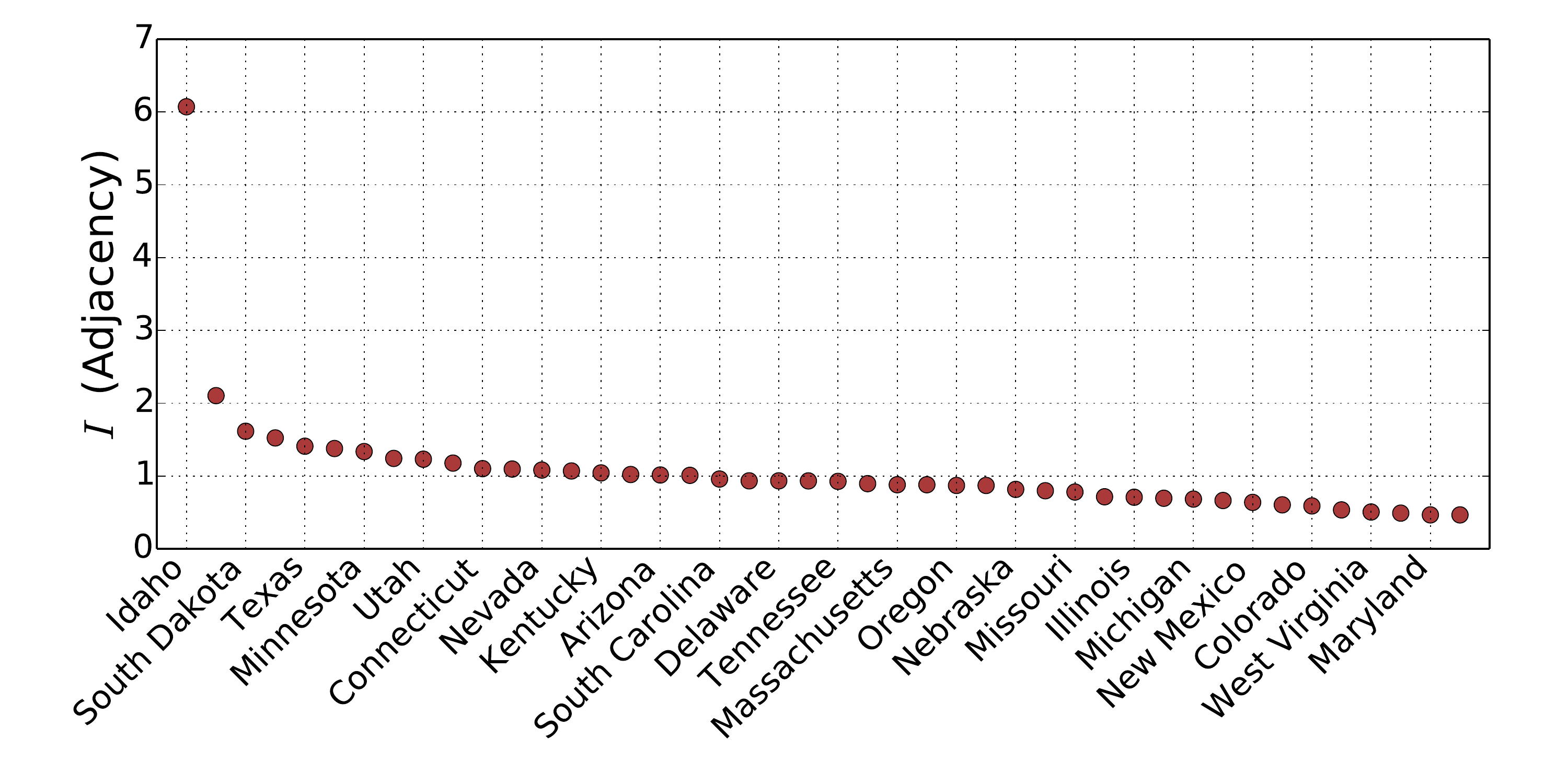}
f\includegraphics[width=8.5cm]{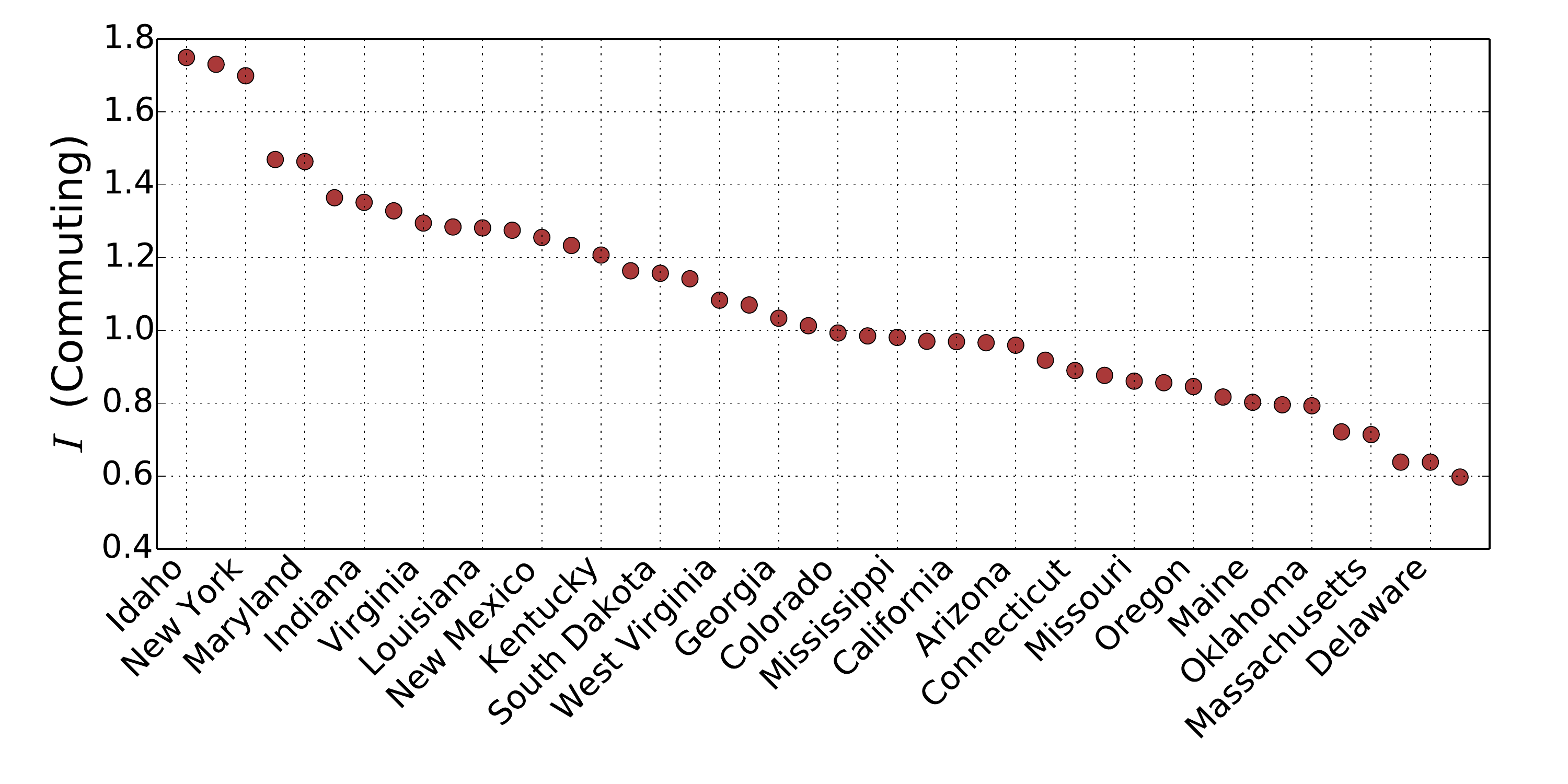}
g\includegraphics[width=8.5cm]{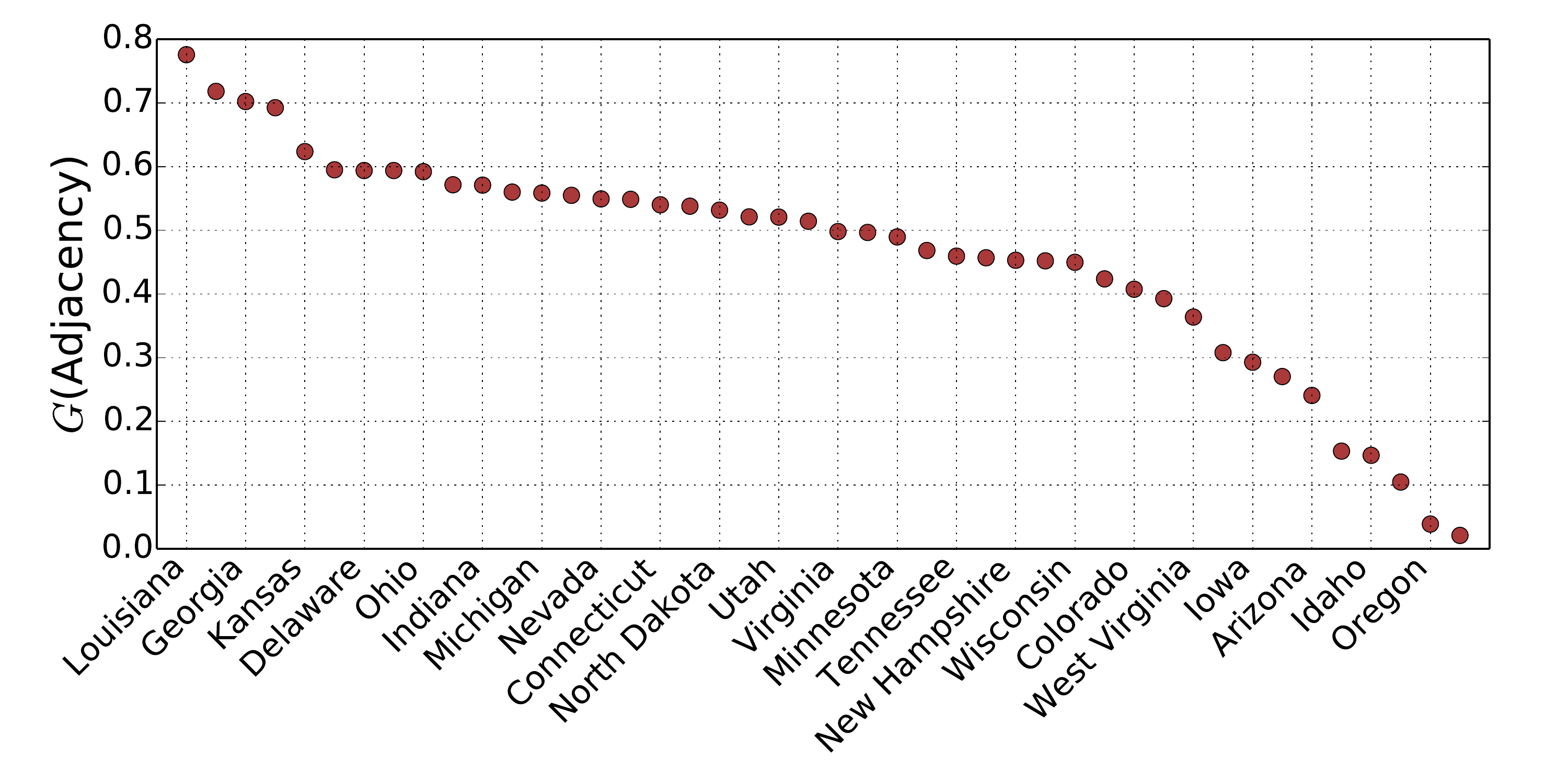}
h\includegraphics[width=8.5cm]{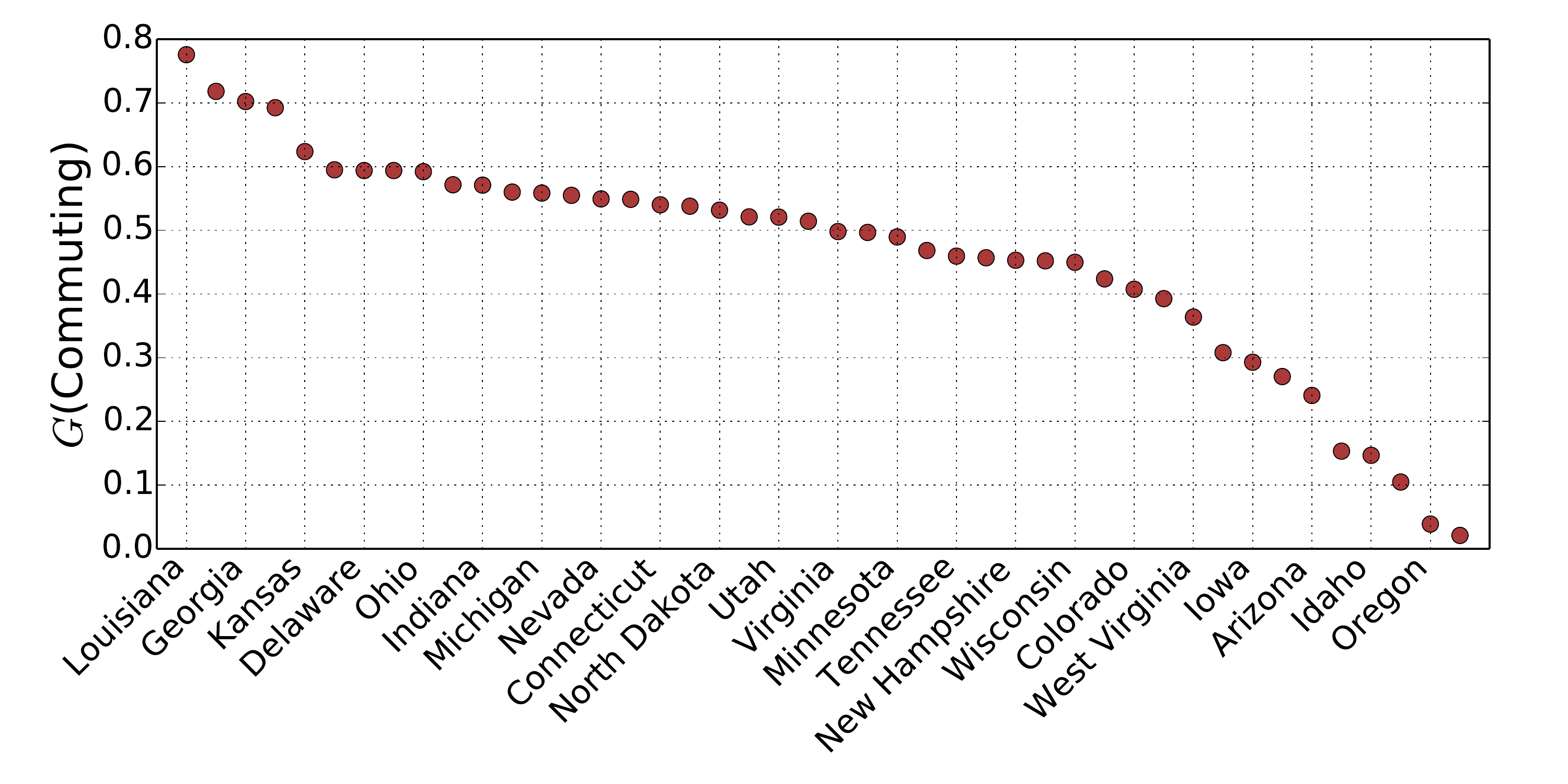}
\caption[Ranking for the four indices studied in the main manuscript: $C$, $E$, $I$ and$G$ computed in both the adjacency and commuting graph]{Ranking for the four indices studied in the main manuscript: $C$, $E$, $I$ and$G$ computed in both the adjacency and commuting graph.
\textbf{a} $C$ (Adjacency),
\textbf{b} $C$ (Commuting),
\textbf{c} $E$ (Adjacency),
\textbf{d} $E$ (Commuting),
\textbf{e} $I$ (Adjacency),
\textbf{f} $I$ (Commuting),
\textbf{g} $G$ (Adjacency),
\textbf{h} $G$ (Commuting).
\label{ranking}}
\end{center}
\end{figure}

Supplementary Figure \ref{ranking} displays the ranking of values for
each of the four metrics studied in the main manuscript computed over
the adjacency or commuting graphs. As can be seen, strong similarities
between rankings appear.

\begin{figure}[!htbp]
\begin{center}
\includegraphics[width=12cm]{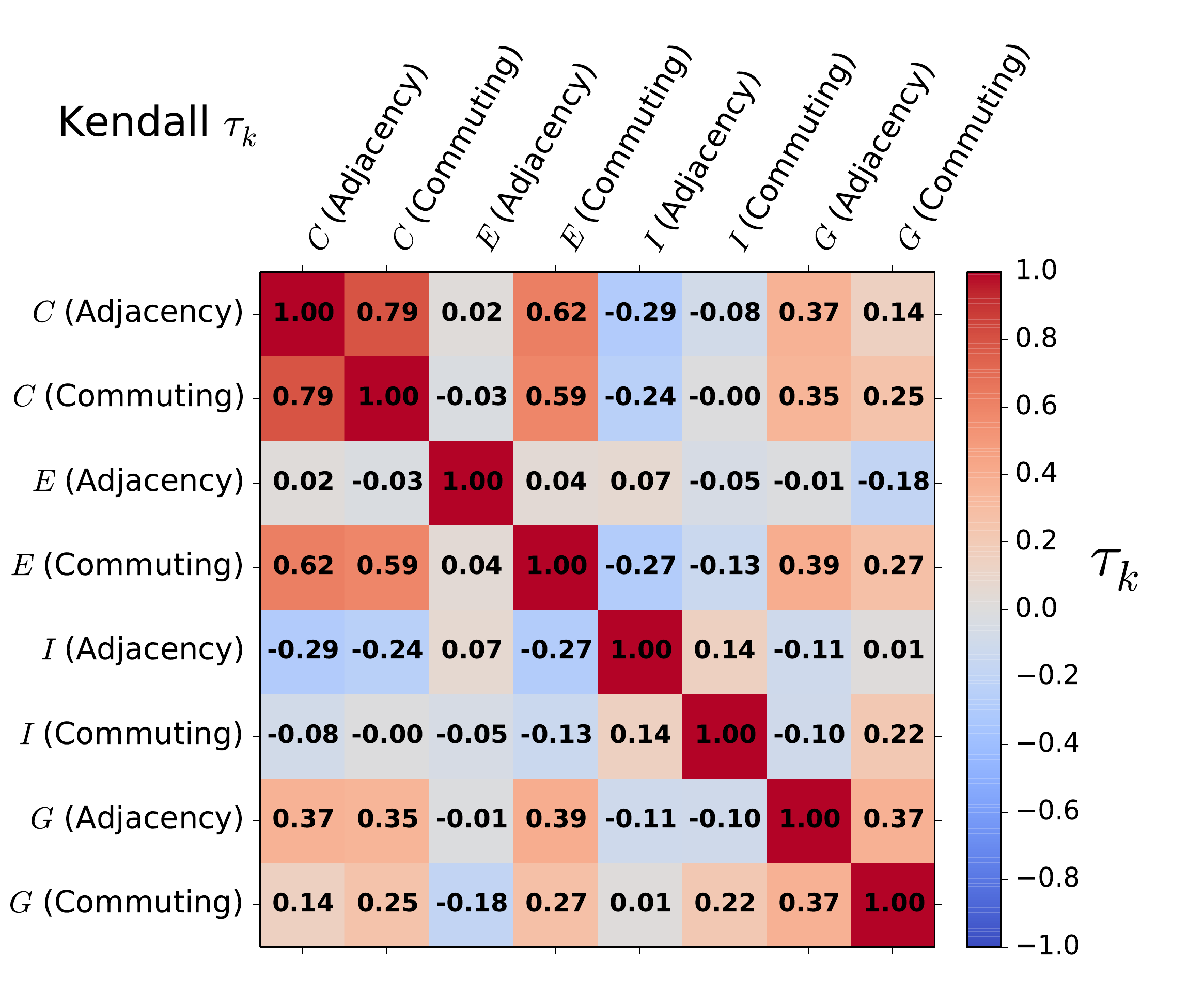}
\caption{Kendall tau $\tau_k$ correlation between each of the four indices studied in the main manuscript computed either over the adjacency or the commuting graphs.
\label{kendall}}
\end{center}
\end{figure}

To evaluate how similar are those rankings we have calculated the Kendall $\tau_k$ between each pair of rankings. Supplementary Figure \ref{kendall} displays the values of $\tau_k$ between each pair of the four metrics studied in the main manuscript computed in the adjacency and commuting graphs. For instance, there is a high correlation between the index $C$ computed in the adjacency and the commuting graphs while for the exposure index $E$ there is almost no correlation between both. Likely pointing out that exposure can only be effectively measured by including the commuting network. Another additional observation is the connection between $C$ and $E$ measured on the commuting graph, which seems to point out that in those states where African Americans are more segregated they are also more exposed.

\begin{figure}[!htbp]
\begin{center}
a\includegraphics[width=7cm]{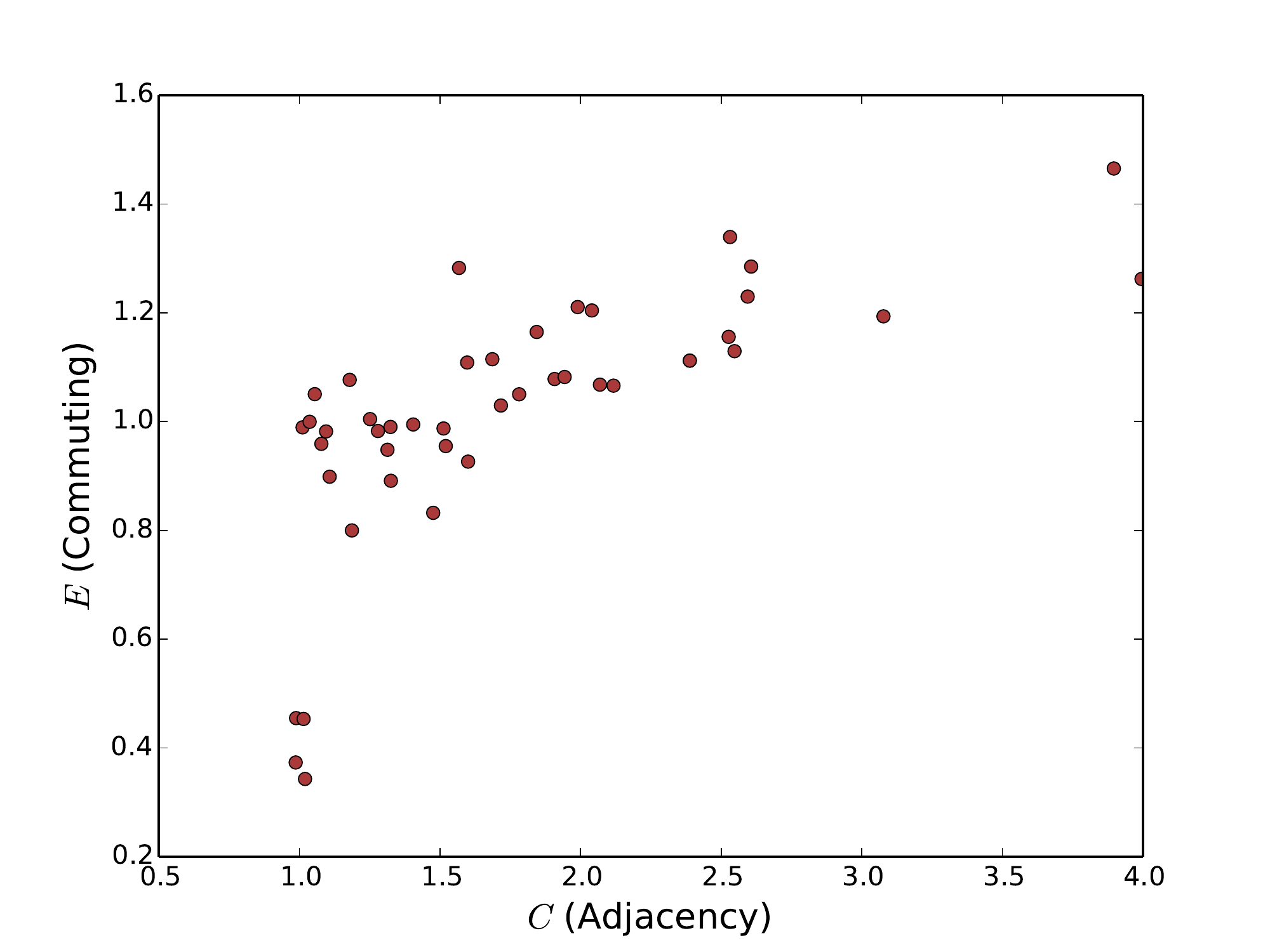}
b\includegraphics[width=7cm]{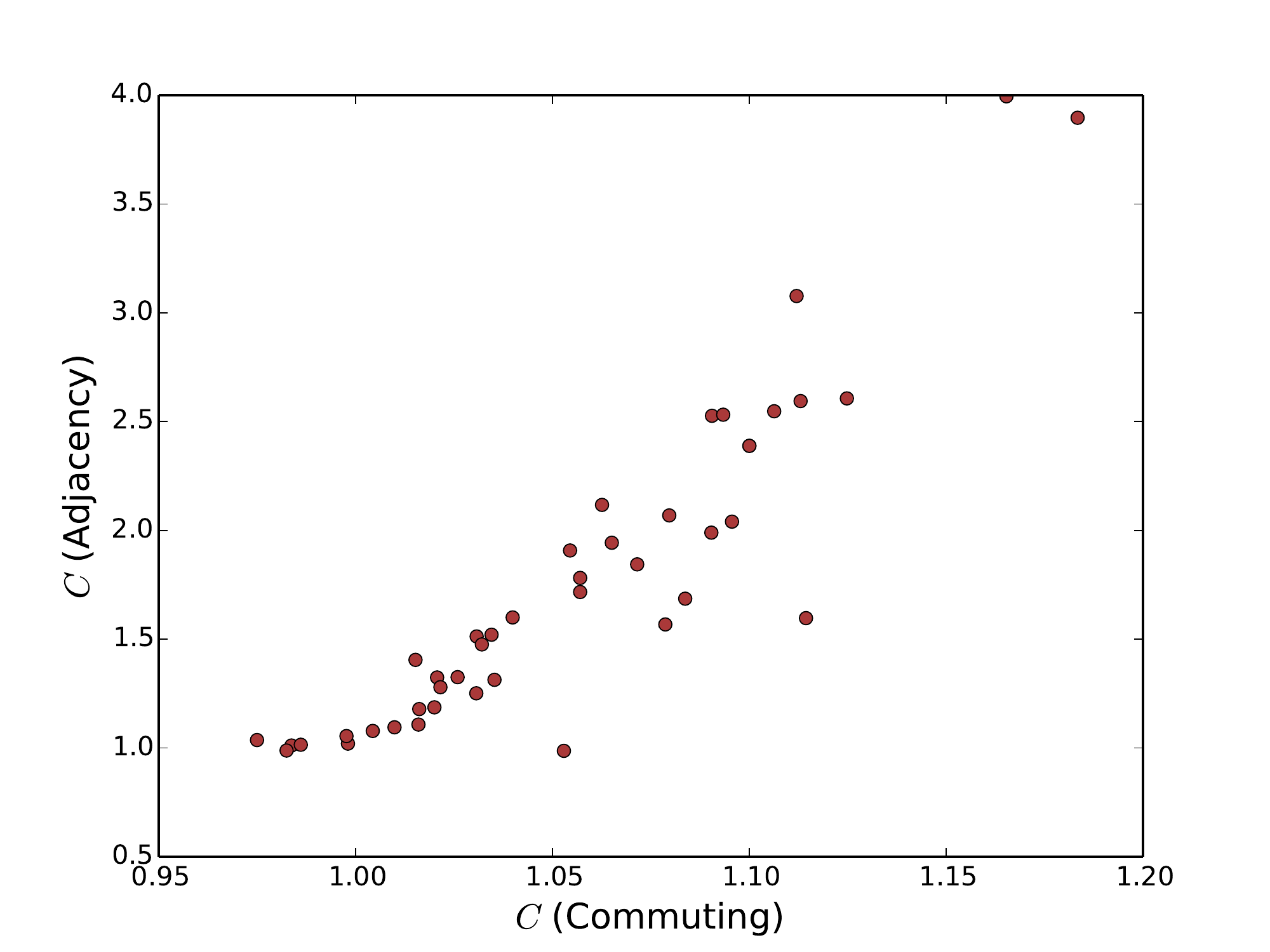}
c\includegraphics[width=7cm]{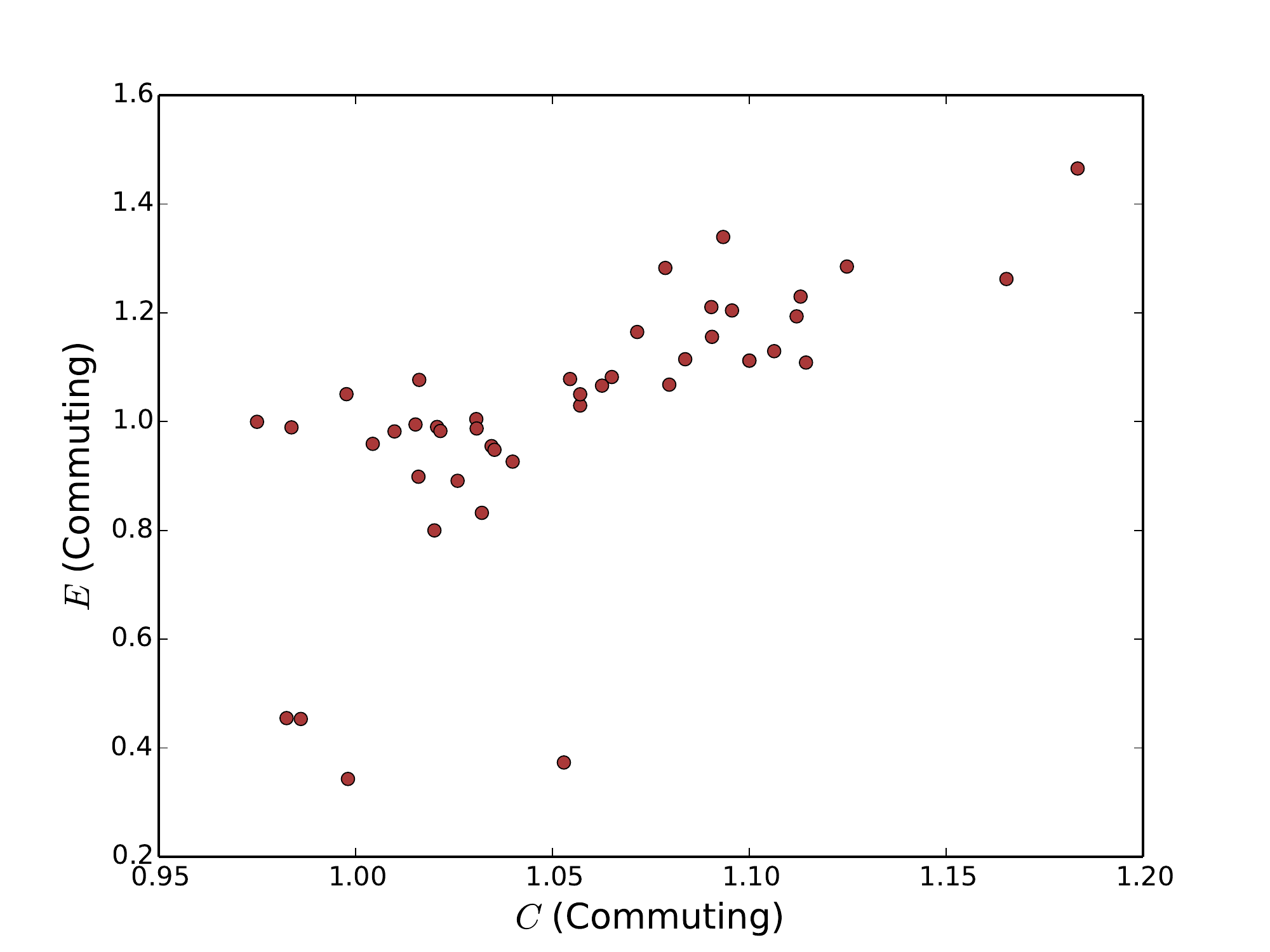}
d\includegraphics[width=7cm]{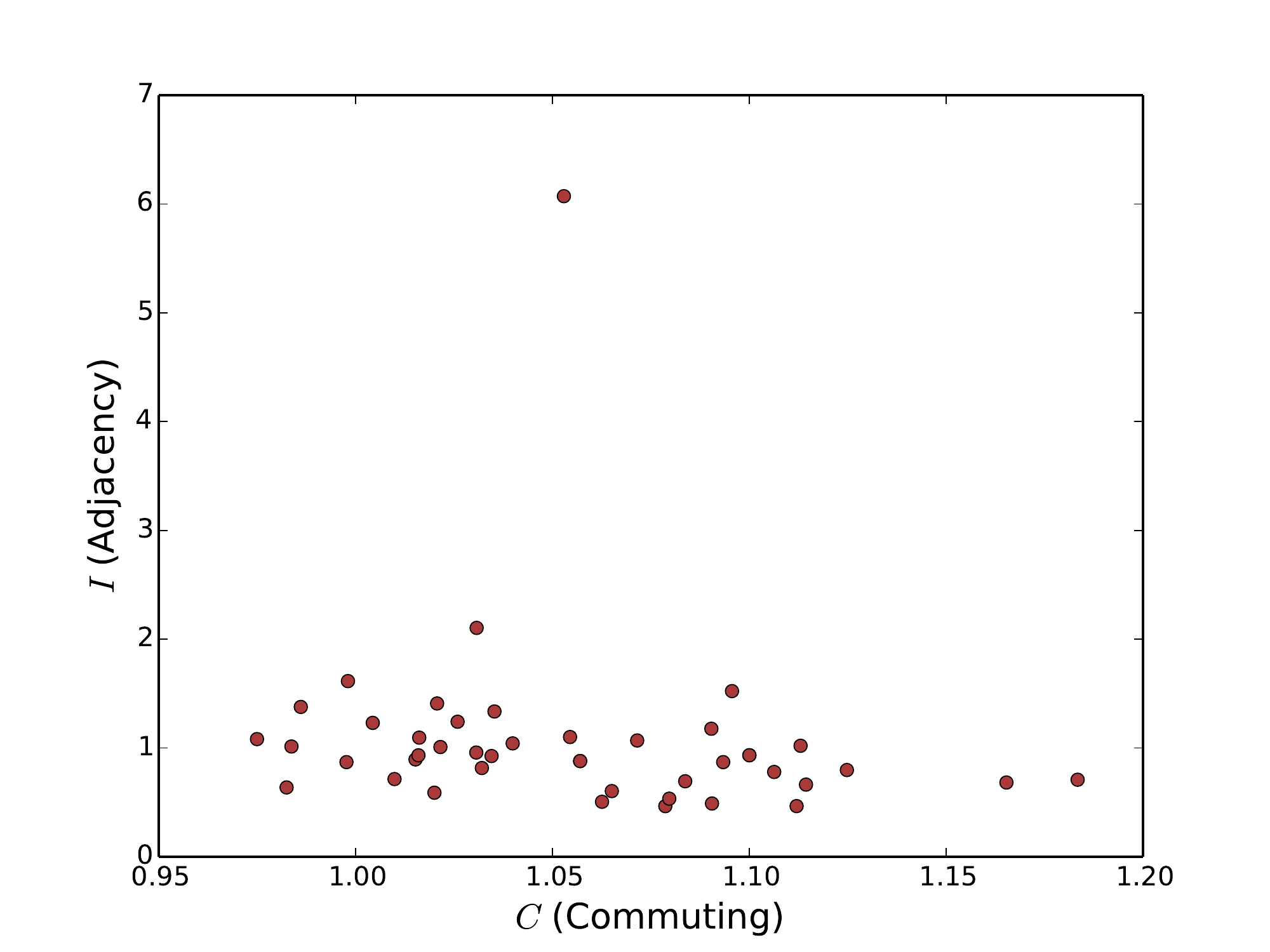}
e\includegraphics[width=7cm]{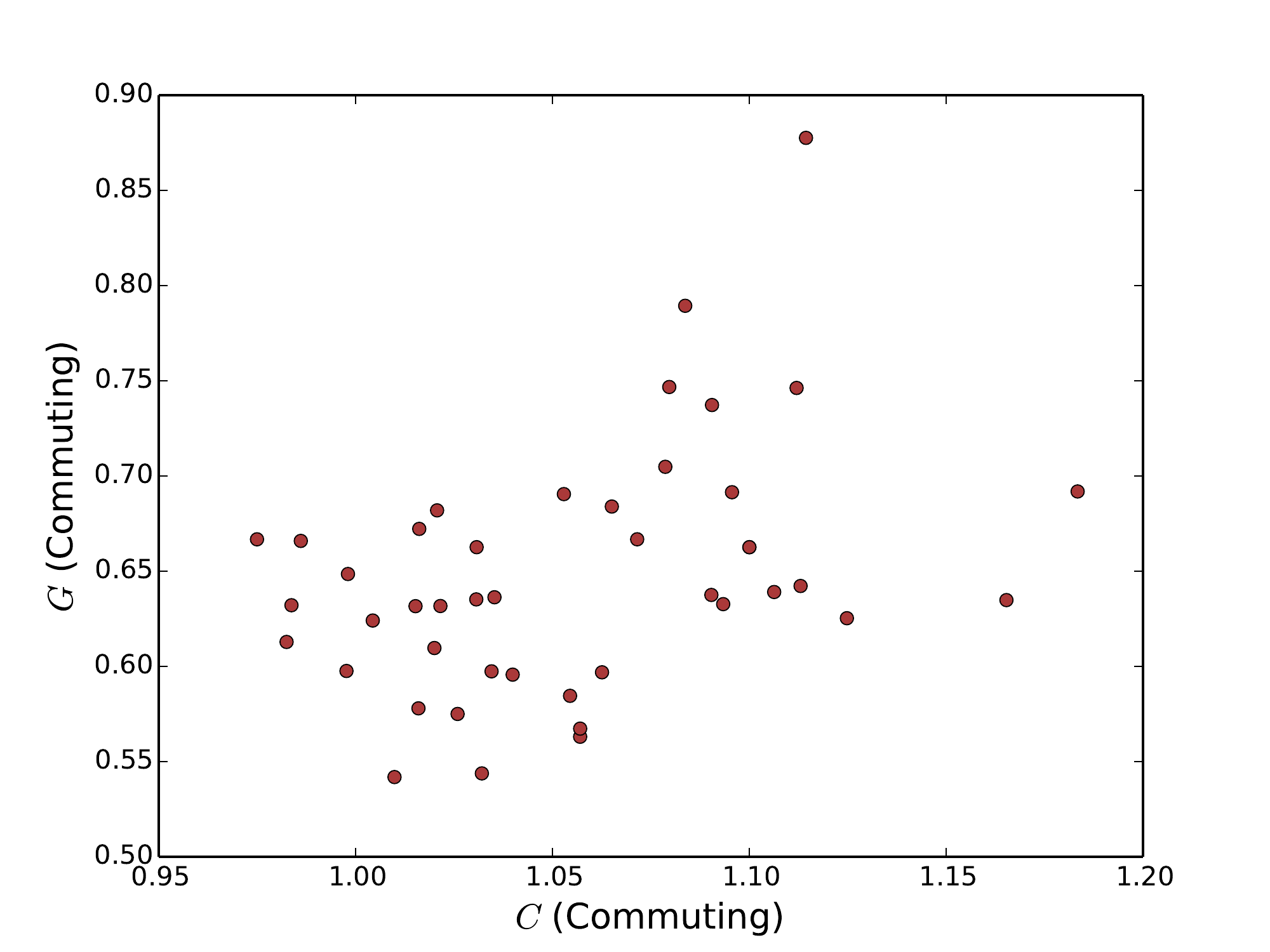}
f\includegraphics[width=7cm]{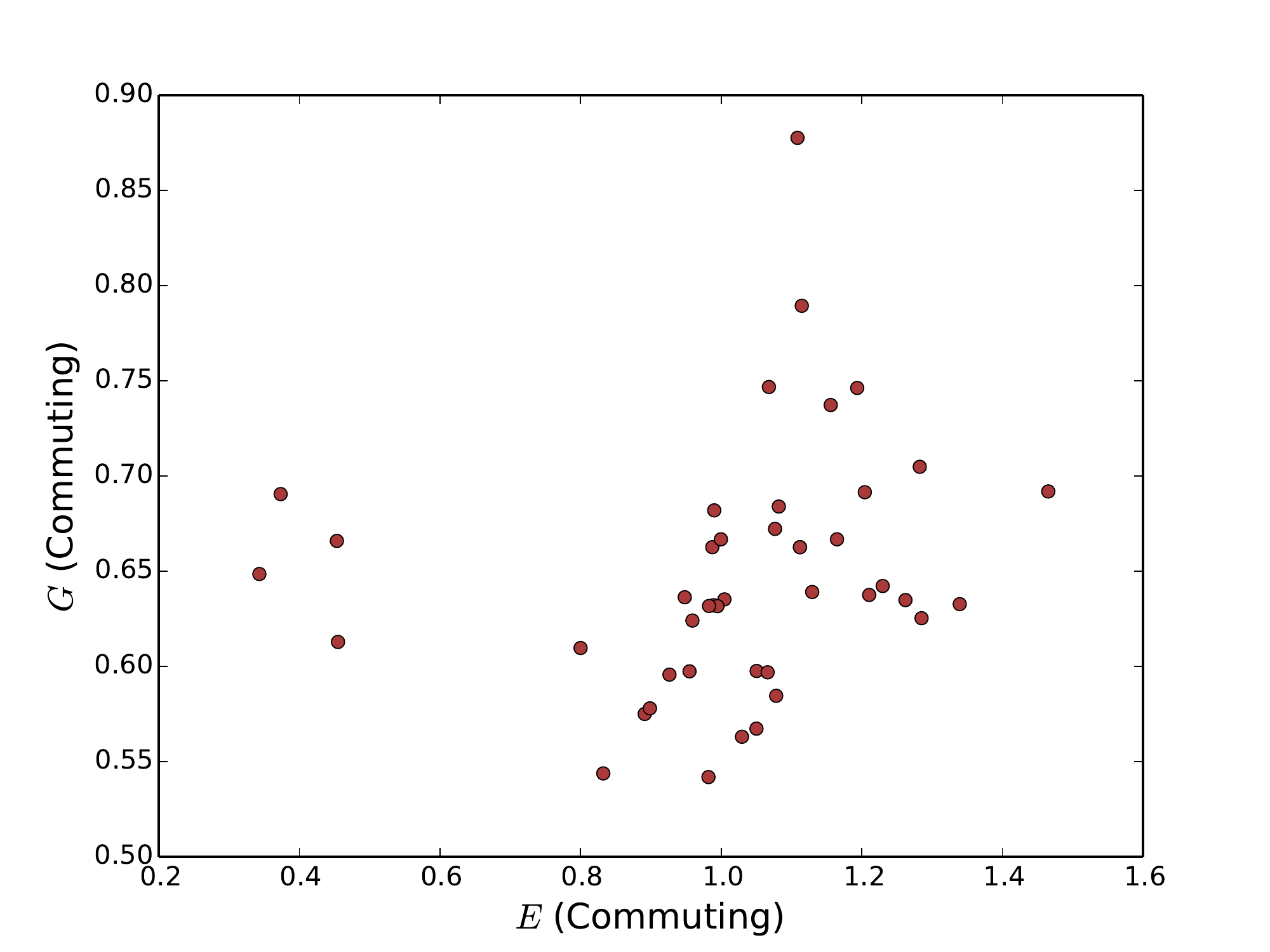}
g\includegraphics[width=7cm]{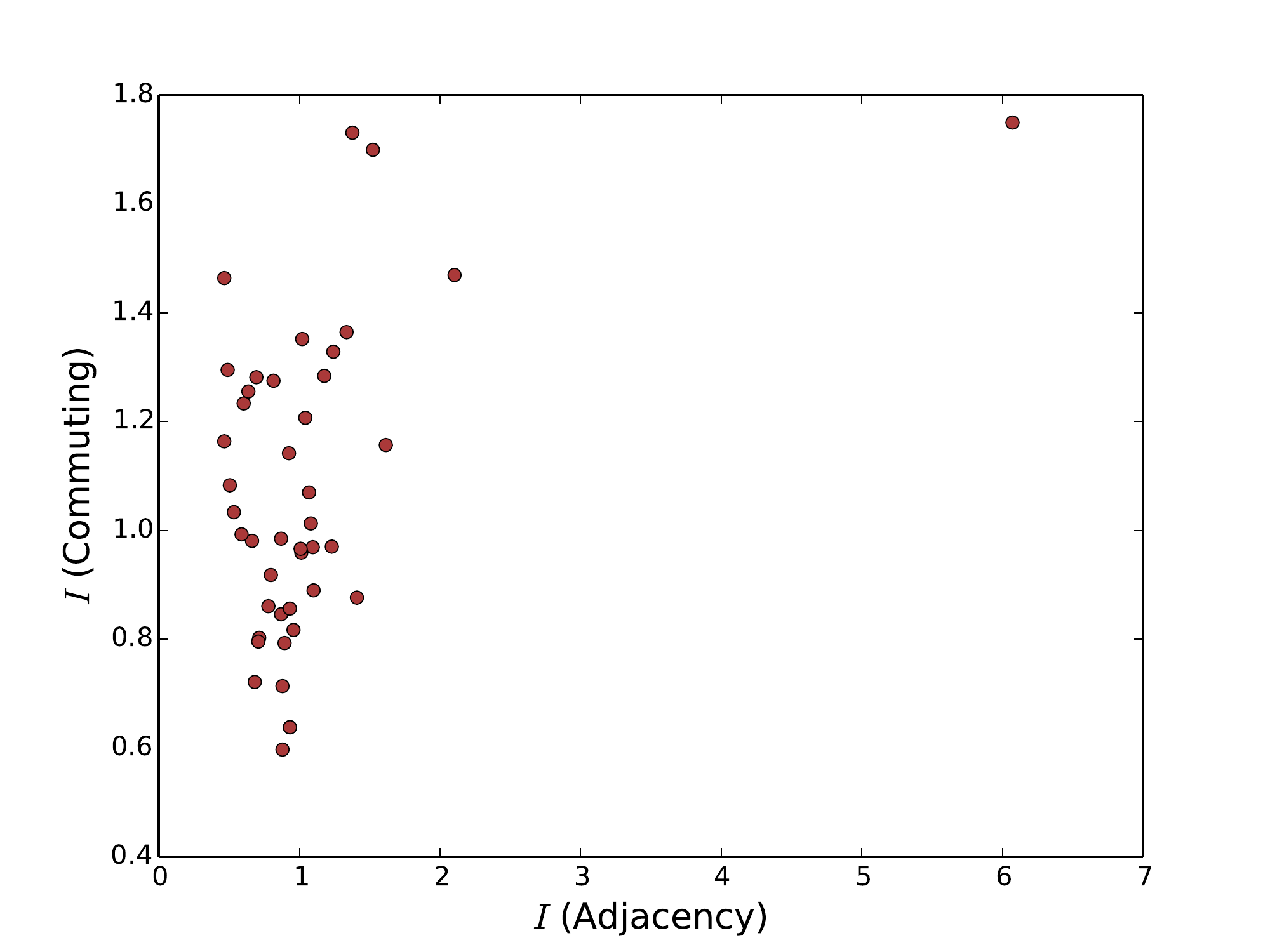}
h\includegraphics[width=7cm]{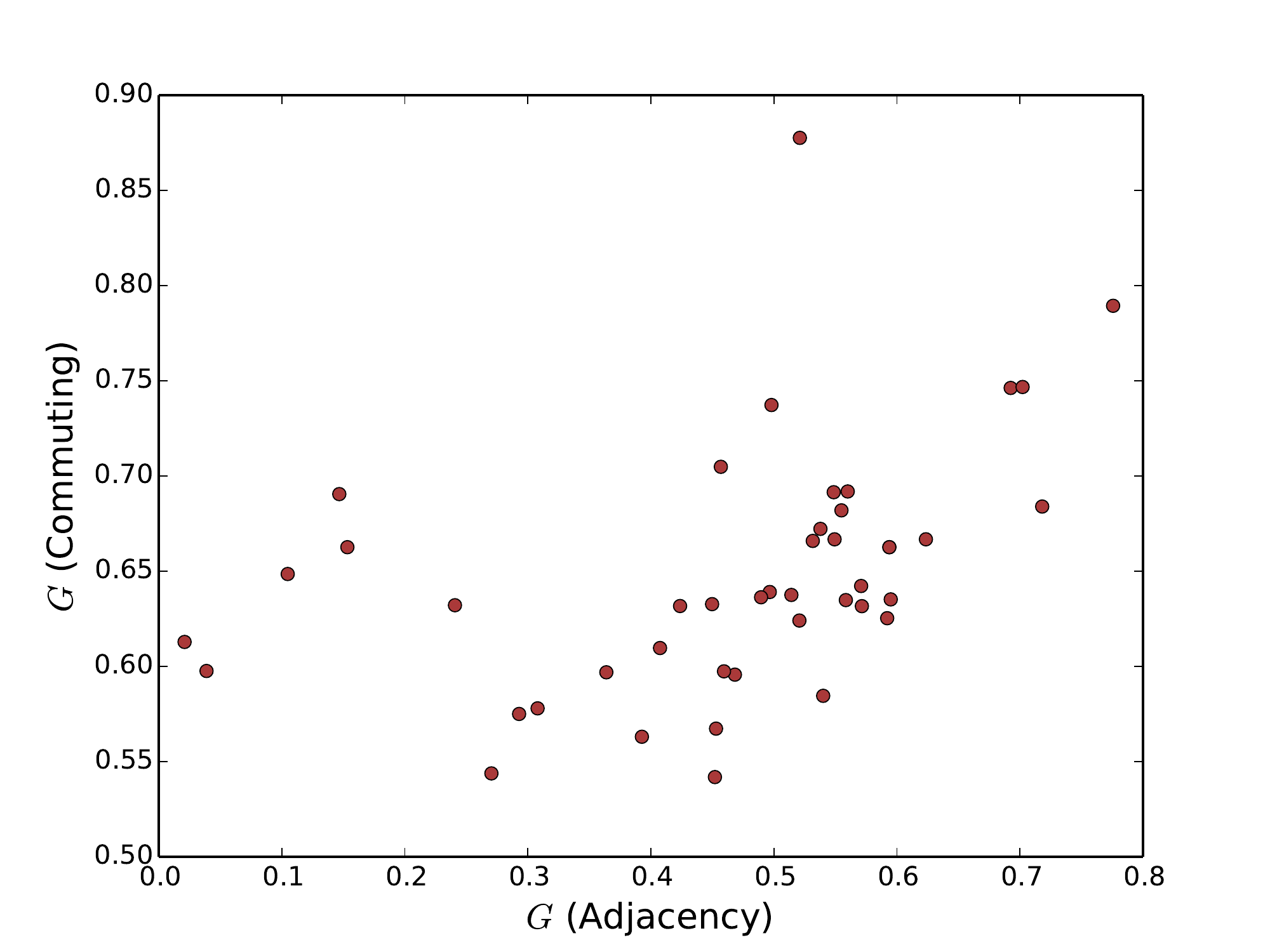}
\caption[Comparison between the indices studied in the main manuscript]{Comparison between the indices studied in the main manuscript. \textbf{a} $C$ (Adjacency) and $E$ (Commuting),   \textbf{b} $C$ (Commuting) and $C$ (Adjacency),
 \textbf{c} $C$ (Commuting) and $E$ (Commuting),
    \textbf{d} $C$ (Commuting) and $I$ (Adjacency),
 \textbf{e} $C$ (Commuting) and $G$ (Commuting),
   \textbf{f} $E$ (Commuting) and $G$ (Commuting),
   \textbf{g} $I$ (Adjacency) and $I$ (Commuting),
   \textbf{h} $G$ (Adjacency) and $G$ (Commuting)
\label{comparison}}
\end{center}
\end{figure}

In Supplementary Figure \ref{comparison}, we compare the indices studied in this work, showing that most of them are related to each other yet not necessarily linearly. We find that those indices capturing the clustering of African Americans are also related to those related to mixing and exposure. The more clustered together, more sensible and exposed to the rest of the population. When comparing the same indices in the commuting and adjacency graphs, we observe that they have a non-linear relation, highlight again the importance of considering both of them.

\subsection*{Correlations between the segregation indices and other measures of COVID-19 incidence}

The COVID-19 data used was obtained from \cite{Data1} and includes several temporal snapshots until mid-may.  The main variables we used are the difference on infected/deceased African Americans, where $0$ would mean that the percentage of African Americans in the population of a state is the same than the percentage of infected, and the ratio calculated as the percentage of infected African Americans divided by their percentage among the overall population, which will be one if they are equal and higher than one if there are more African Americans infected/deceased among the overall population. Supplementary Table \ref{tablefit} summarises the results obtained for the linear fit for the Figure 2 in the main manuscript.

\begin{table}
\begin{center}
\begin{tabular}{ |c|c|c|c|c| }
 \hline
Date & Index & Network type & slope & intercept \\
  \hline
12/04/2020 & $C$ & Adjacency & 8.26 & -3.47  \\
12/04/2020 & $E$ & Adjacency &-12.35 & 27.82  \\
12/04/2020 & $I$ & Adjacency & -19.66 & 30.01  \\
12/04/2020 & $G$ & Adjacency & 31.86 & -3.28 \\
12/04/2020 & $C$ & Commuting & 139.14 & -135.94  \\
12/04/2020 & $E$ & Commuting & 38.83 & -29.92  \\
12/04/2020 & $I$ & Commuting & -7.70 & 20.99  \\
12/04/2020 & $G$ & Commuting & 33.65 & -9.10\\
19/04/2020 & $C$ & Adjacency & 6.95 & -1.87 \\
19/04/2020 & $E$ & Adjacency &-6.45 & 19.04  \\
19/04/2020 & $I$ & Adjacency &-7.633 & 17.77 \\
19/04/2020 & $G$ & Adjacency & 20.20 & 1.65 \\
19/04/2020 & $C$ & Commuting & 117.98 & -114.11  \\
19/04/2020 & $E$ & Commuting & 25.93& -16.45  \\
19/04/2020 & $I$ & Commuting & -4.79 & 16.15 \\
19/04/2020 & $G$ & Commuting & 24.88 & -5.02\\
 \hline
\end{tabular}
\caption{Coefficients obtained from the linear fits in Figure 2 of the main manuscript}
\label{tablefit}
\end{center}
\end{table}

\begin{figure}[!htbp]
\begin{center}
\includegraphics[width=14cm]{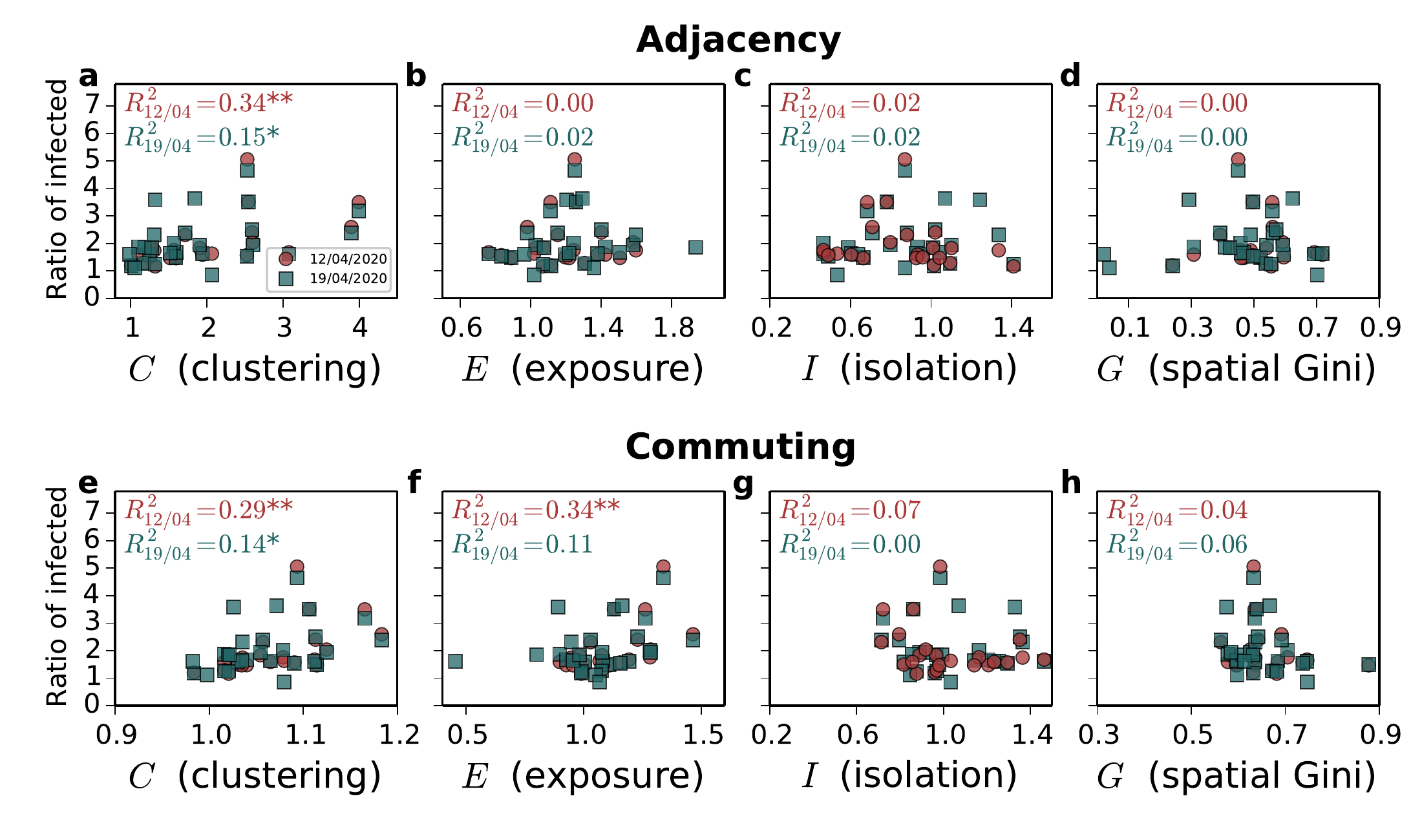}
\caption[Relation between the ratio of
    infected African Americans and the four indices considered]{\textbf{Relation between the ratio of
    infected African Americans and the four indices considered.} \textbf{a-g} Indices computed over the adjacency network: \textbf{a} $C$ (clustering), \textbf{b} $E$ (exposure), \textbf{c} $I$ (isolation),
  \textbf{d} $G$ (spatial Gini). \textbf{e-h} Indices computed over
  the commuting network:  \textbf{e} $C$ (clustering), \textbf{f} $E$ (exposure), \textbf{g} $I$ (isolation),
  \textbf{h} $G$ (spatial Gini).  Each of the colours
  corresponds to a temporal snapshot of the data set, red for
  $12/04/2020$ and blue for $19/04/2020$. The $R^2$ is computed as the
  square of the linear correlation coefficient.
\label{scatter_ratioinf}}
\end{center}
\end{figure}

\begin{figure}[!htbp]
\begin{center}
\includegraphics[width=14cm]{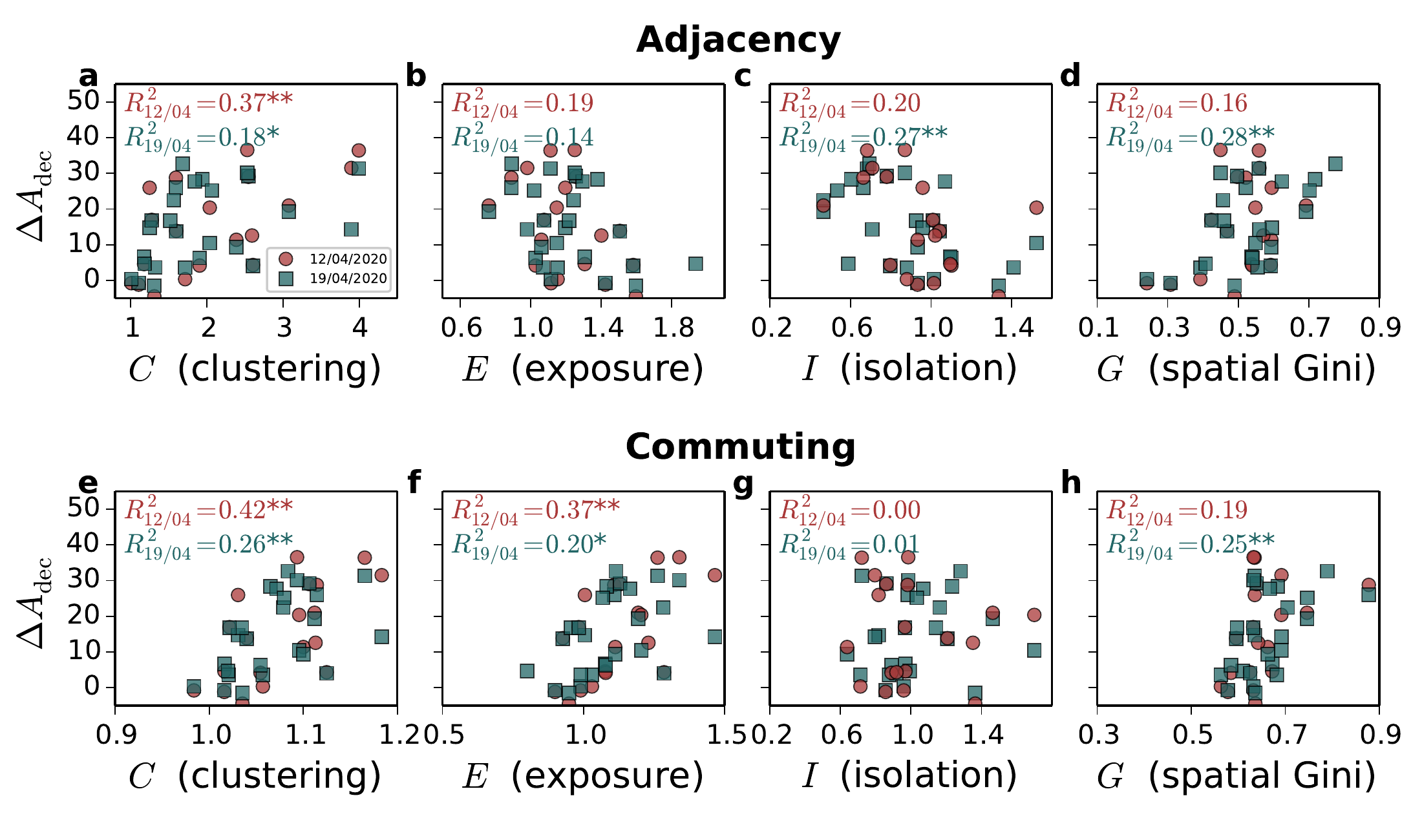}
\caption[Relation between the difference on the percentage of deceased African American as a function of the four indices considered]{\textbf{Relation between the difference on the percentage of deceased African American as a function of the four indices considered.} \textbf{a-g} Indices computed over the adjacency network: \textbf{a} $C$ (clustering), \textbf{b} $E$ (exposure), \textbf{c} $I$ (isolation),
  \textbf{d} $G$ (spatial Gini). \textbf{e-h} Indices computed over
  the commuting network:  \textbf{e} $C$ (clustering), \textbf{f} $E$ (exposure), \textbf{g} $I$ (isolation),
  \textbf{h} $G$ (spatial Gini).  Each of the colours
  corresponds to a temporal snapshot of the data set, red for
  $12/04/2020$ and blue for $19/04/2020$. The $R^2$ is computed as the
  square of the linear correlation coefficient.
\label{scatter_diffdec}}
\end{center}
\end{figure}

\begin{figure}[!htbp]
\begin{center}
\includegraphics[width=14cm]{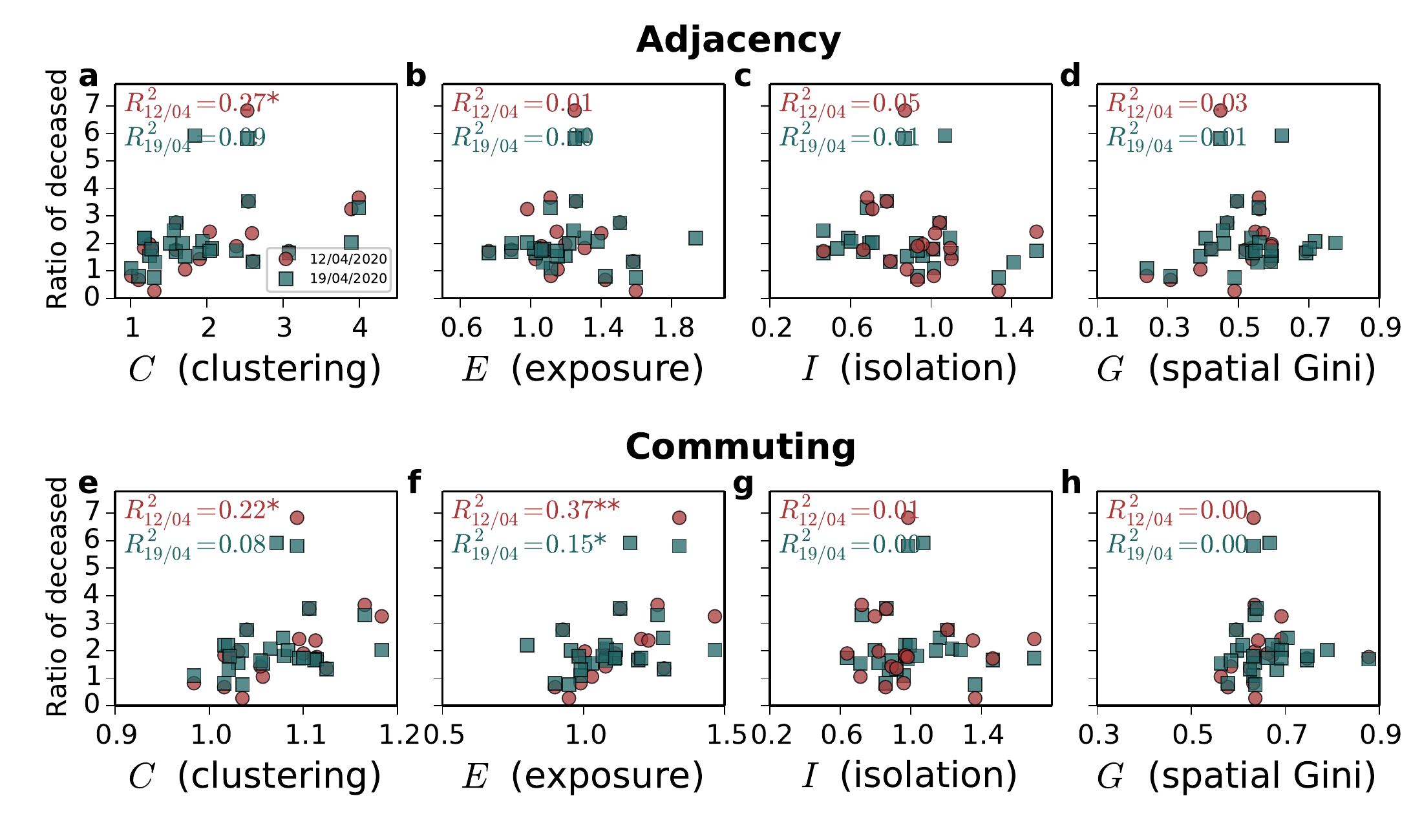}
\caption[Relation between the ratio of
    deceased African Americans and the four indices considered]{\textbf{Relation between the ratio of
    deceased African Americans and the four indices considered.} \textbf{a-g} Indices computed over the adjacency network: \textbf{a} $C$ (clustering), \textbf{b} $E$ (exposure), \textbf{c} $I$ (isolation),
  \textbf{d} $G$ (spatial Gini). \textbf{e-h} Indices computed over
  the commuting network:  \textbf{e} $C$ (clustering), \textbf{f} $E$ (exposure), \textbf{g} $I$ (isolation),
  \textbf{h} $G$ (spatial Gini).  Each of the colours
  corresponds to a temporal snapshot of the data set, red for
  $12/04/2020$ and blue for $19/04/2020$. The $R^2$ is computed as the
  square of the linear correlation coefficient.
\label{scatter_ratiodec}}
\end{center}
\end{figure}

Supplementary Figure \ref{scatter_ratioinf} summarises the results obtained in the case of the ratio of infected African Americans. As detailed in the main manuscript, there are two versions of each index depending on whether the walkers move upon the adjacency or the commuting network. Compared to the difference in percentage correlations are much lower for the ratio, likely as a consequence of the several outliers. While states with a low percentage of African Americans among the overall population might easily suffer a huge increase on the ratio, those with a higher percentage of African Americans among the population might have a lower increase.

We have also evaluated how our metrics relate to the ratio and difference among the deceased African Americans (See Supplementary Figures \ref{scatter_diffdec}  and \ref{scatter_ratiodec} ). Despite many more factors such as the age or underlying health conditions might influence the deceased individuals, still, most of the correlations remain significant to some extent, especially those related to their exposure. Moreover, those indices computed on the commuting network seem to be more informative than those based on the adjacency, which seems to point out that residential segregation provides only a partial picture of ethnic inequality. Mobility is also crucial to understand the mixing between different ethnicities, it is not only relevant where certain ethnicities live but also where they work and with whom they interact when they do so.

Overall, despite our metrics are informative in both cases, they seem to be more related to the difference in percentage more than the ratio. There are states in which the percentage of African Americans among the population is low and, therefore, the ratio can increase drastically.

Ideally, each ethnicity $\alpha$ should be compared with the corresponding
ratio to the overall population and the incidence of COVID-19 cases. As
this data was not available during the preparation of this work, we look
at the gap $\Delta A_{\rm inf}$ of African American and the relation with
the Isolation level of all other ethnicities in this study. Considering
the quantity for all other ethnicities defined as:
\begin{equation}
  \widetilde{\gamma}_O^{i} = \frac{1}{\Gamma-1} \sum_{\beta\neq\alpha}
  \widetilde{\gamma}_\beta^{i}
\end{equation}
the Isolation index for an ethnicity $\alpha$ is given by:
\begin{equation}
  I_\alpha = \frac{1}{N} \sum_{i=1}^{N} \frac{\widetilde{\gamma}_\alpha^{i}}
  {\widetilde{\gamma}_O^{i}}
\end{equation}

\begin{figure}[!htbp]
  \center
  \includegraphics[width=15cm]{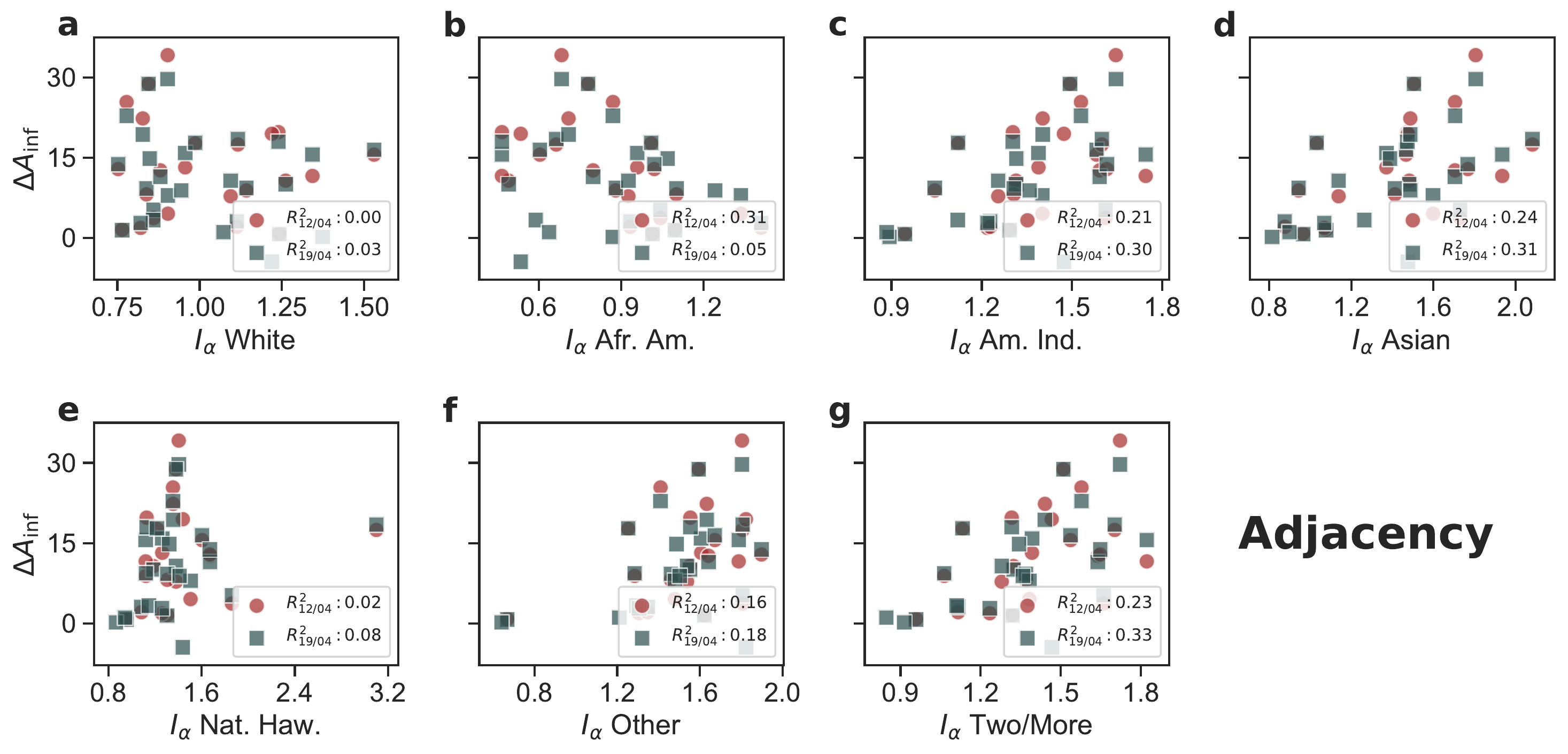}
  \caption{Isolation index of all ethnicities as a function of the infected
  rate gap of COVID-19 cases in the African American population on the
  adjacency network. African American is the only ethnicity to exhibit a
  negative correlation of isolation and $\Delta A_{\rm inf}$, suggesting that
  a higher infection rate can be related to lower isolation.}
  \label{cctall_inf_adj}
\end{figure}

\begin{figure}[!htbp]
  \center
  \includegraphics[width=15cm]{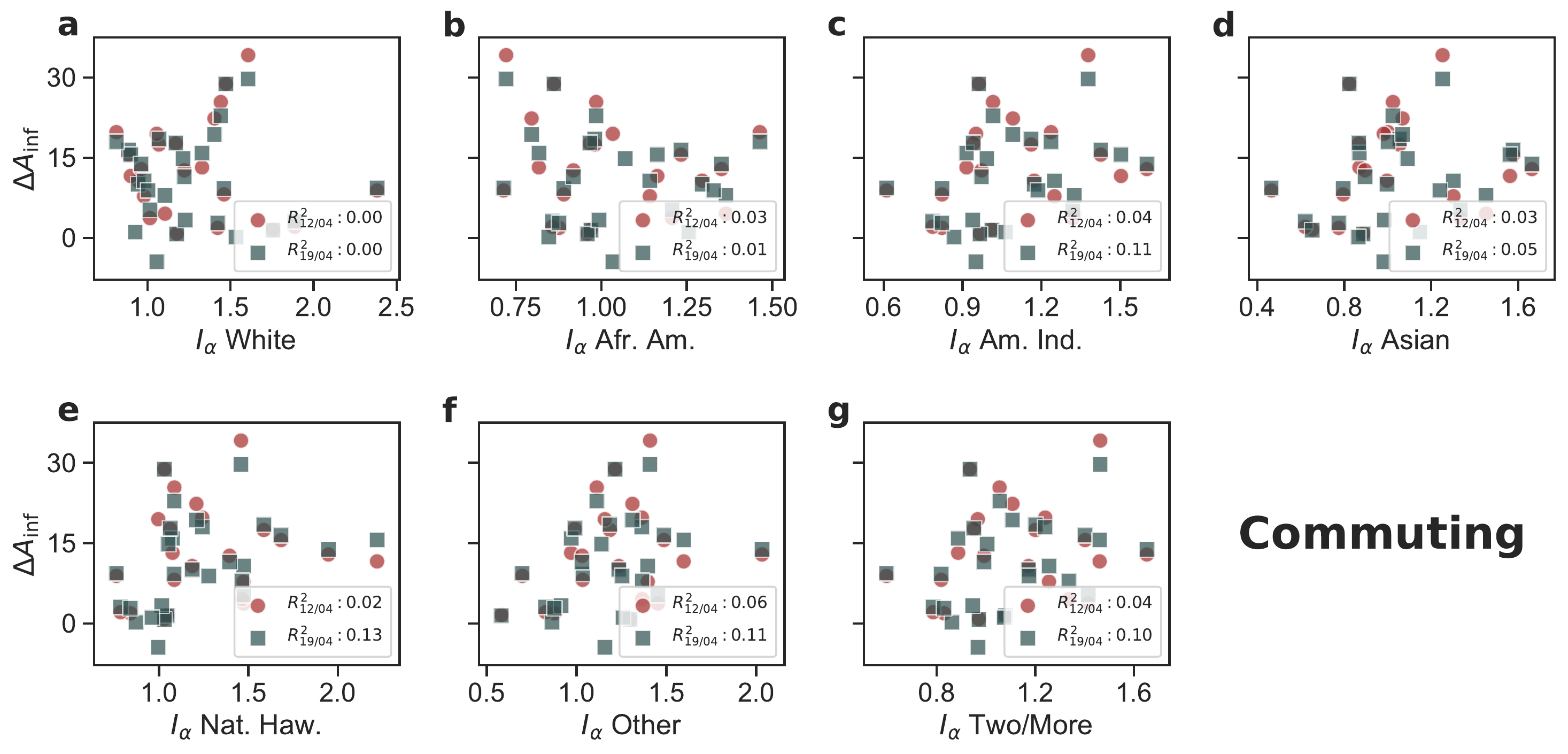}
  \caption{Isolation index of all ethnicities as a function of the infected
  rate gap of COVID-19 cases in the African American population considering
  the commute network. There is no significant correlation for any of the
  ethnicities.}
  \label{cctall_inf_comm}
\end{figure}

Correlations with the infection rate gap $\Delta A_{\rm inf}$ on the
adjacency and commute network are reported on Fig. \ref{cctall_inf_adj}
and \ref{cctall_inf_comm} for all ethnicities. Quantities are obtained from
the COVID-19 data set at 2 different periods, 12-04-2020 and 19-04-2020
respectively. The corresponding $R^2$ of the Pearson correlation is
reported in the inset of each panel. There is a negative correlation in
\textbf{b} which indicates that less isolated African Americans have a
higher incidence of infection cases. Whites \textbf{a} and Native
Hawaiians \textbf{e} exhibit no correlation while the remaining ethnicities
\textbf{c-d} and \textbf{f-g} have a positive $R^2$ which decreases over
time. We found no correlation for any ethnicity on the commuting network.

\begin{figure}[!htbp]
  \center
  \includegraphics[width=15cm]{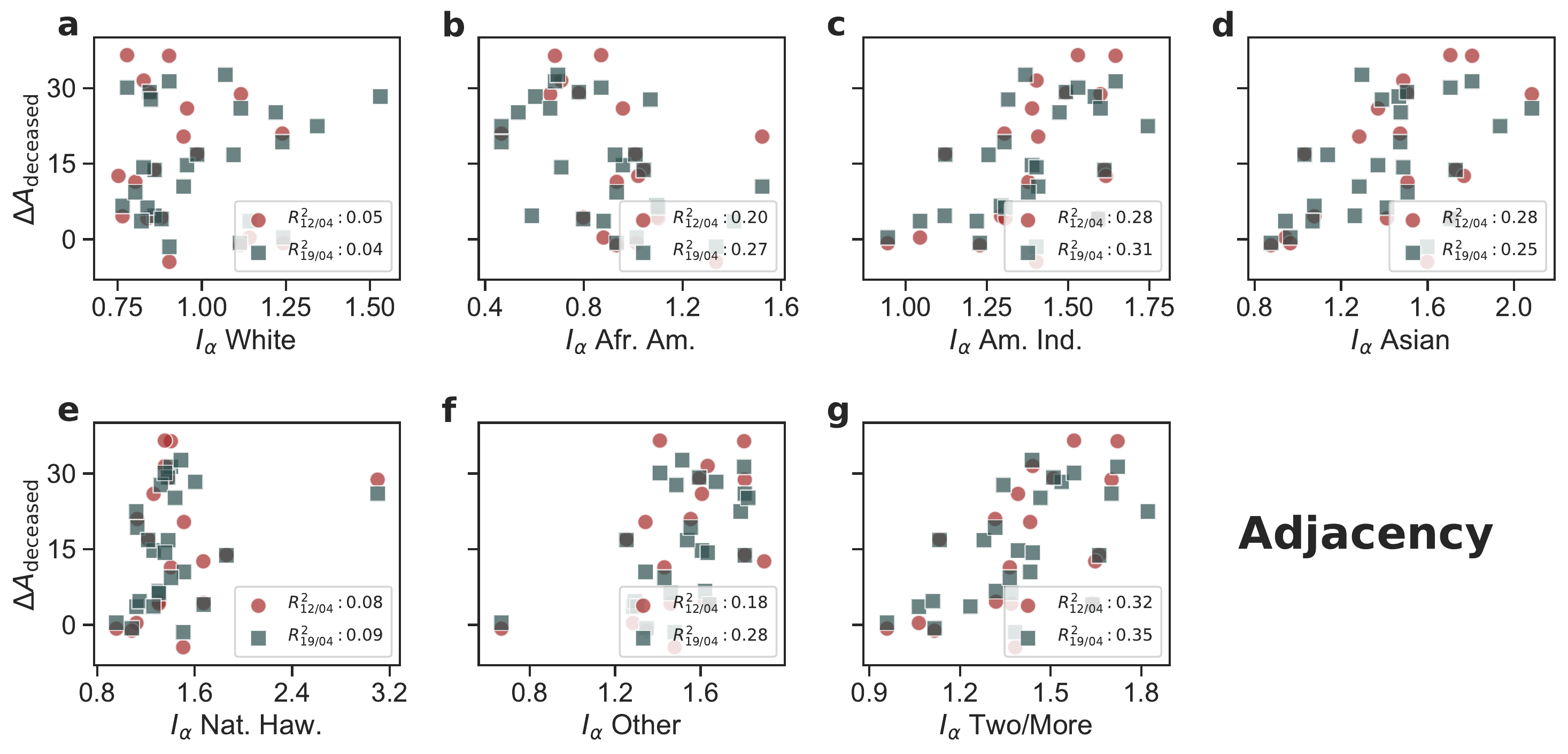}
  \caption{Isolation index of all ethnicities as a function of the deceased
  rate gap of COVID-19 cases in the African American population considering
  the adjacency network.}
  \label{cctall_death_adj}
\end{figure}

\begin{figure}[!htbp]
  \center
  \includegraphics[width=15cm]{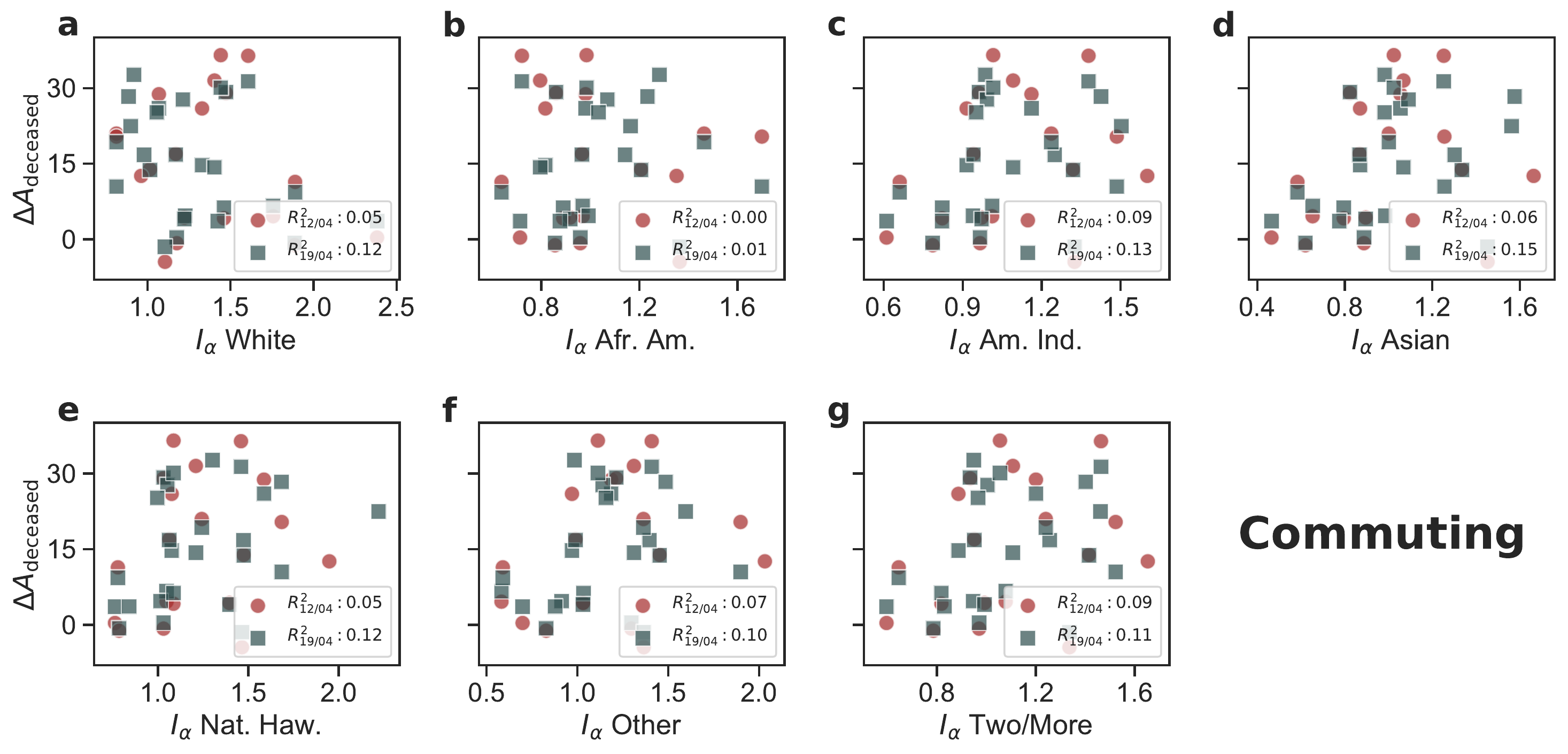}
  \caption{Isolation index of all ethnicities as a function of the deceased
  rate gap of COVID-19 cases in the African American population considering
  the commute network. There is no significant correlation for any of the
  ethnicities.}
  \label{cctall_death_comm}
\end{figure}

Similarly, correlations with $\Delta A_{\rm deceased}$ are computed for
the deceased data on the adjacency and commute networks (See Fig.
\ref{cctall_death_adj} and \ref{cctall_death_comm}). We can observe a
similar pattern with the results obtained from infected rates where there
is correlations for the same group of ethnicities and no significant
relationship on the commuting network.

\section{Local segregation maps through CMFPT and CCT and spatial correlation}

In the main manuscript, we show the values for the local segregation indices $\xi$ and$ \psi$ for Chicago and Los Angeles showing that there were significant differences on their spatial distribution as well as in their maximum values. Here we provide also results for Detroit and Houston to show that again there are significant differences. In this case, Detroit is the most populated city in Michigan, which is one of the states with highest values in most of the indices considered and Houston is the most populated city in Texas, which is a state with consistent low values in most segregation indices. Regarding the impact of COVID-19 among the African Americans of those states, in Michigan the gap is around 34$\%$ in early April and 24$\%$ in mid-may. In Texas, instead, the gap is around 2$\%$ at the beginning of April and 5$\%$ in mid-May.

\begin{figure}[!htbp]
\begin{center}
\includegraphics[width=1\textwidth]{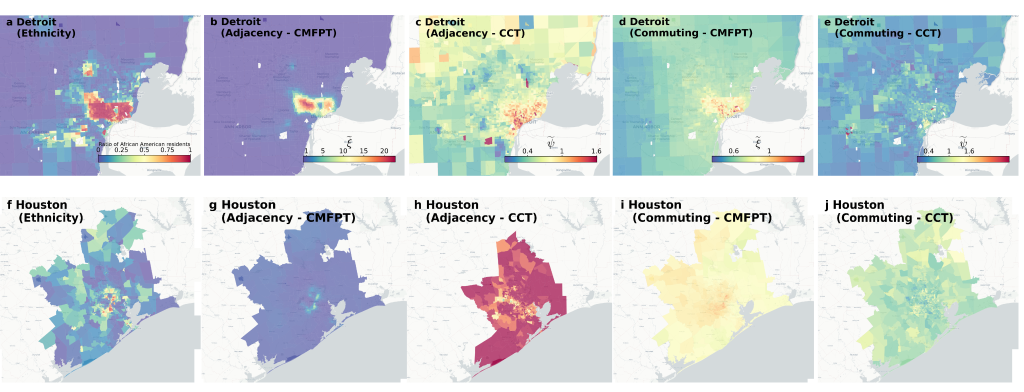}
\caption[Maps of local segregation in American Cities]{\textbf{Maps of local segregation in American cities.} Ratio of African American population and local segregation indices computed with CMFPT
and CCT in \textbf{a-e} Detroit and  \textbf{f-j} Houston. For Detroit: \textbf{a} Ratio of African American population, \textbf{b-c} $\widetilde{\xi}$ and $\widetilde{\psi}$ computed over the adjacency graph and \textbf{d-e} $\widetilde{\xi}$ and $\widetilde{\psi}$ computed over the commuting graph. For Houston: \textbf{f} Ratio of African American population, \textbf{g-h} $\widetilde{\xi}$ and $\widetilde{\psi}$ computed over the adjacency graph and \textbf{i-j} $\widetilde{\xi}$ and $\widetilde{\psi}$ computed over the commuting graph.
\label{maps}}
\end{center}
\end{figure}

In the main manuscript and Supplementary Figure \ref{maps} we plot the local measures of segregation in each of the census tracts of Chicago, Los Angeles, Detroit and Houston. Those maps display certain common patterns that we quantify in Supplementary Figure \ref{mapcorr}. Therein we have calculated the Kendall $\tau_k$ correlation coefficient performing pairwise comparisons of the values for each tract unit. Additionally to the segregation indices, we also compared the values for the ratio of African American population. It is relevant to note that while the value of $\tau_k$ for the ratio of African American population and $\widetilde{\xi}$ computed in the adjacency graph is around $0.8$ for all the four cities studied, there are stronger variations when $\widetilde{\xi}$ is computed in the commuting graph -- i.e., $0.81$ in Detroit and $0.55$ in Houston -- meaning that the effect of commuting in the segregation of African American population can display strong differences across cities and, therefore, mobility offers a different picture of urban segregation.

\begin{figure}[!htbp]
\begin{center}
\includegraphics[width=15cm]{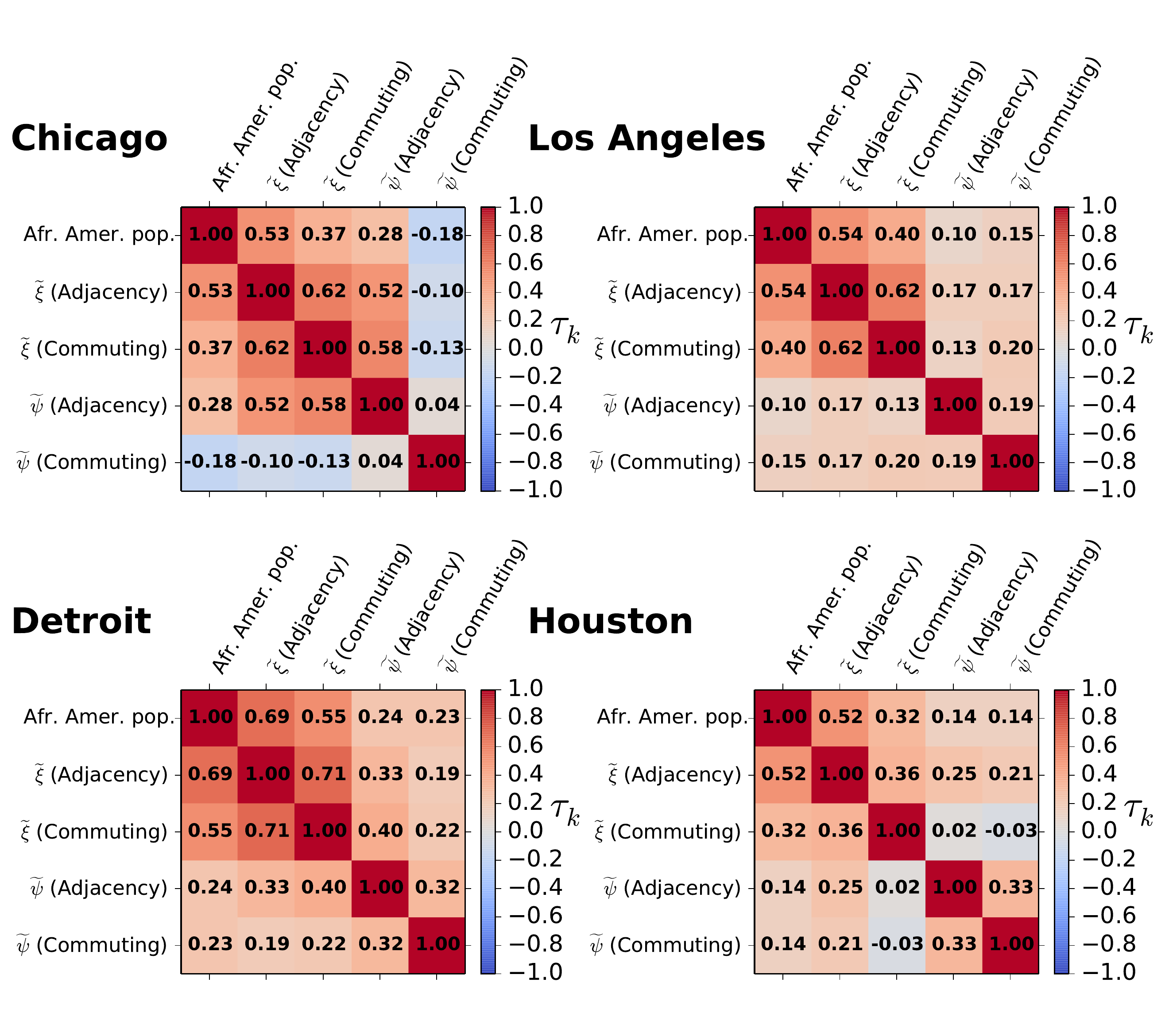}
\caption[Pearson correlation coefficient between the each of the local metrics of segregation $\widetilde{\xi}$ and $\widetilde{\psi}$ and the local ratio of African American population]{\textbf{Correlations between the each of the local metrics of segregation $\widetilde{\xi}$ and $\widetilde{\psi}$ and the local ratio of African American population.} Correlation between each of the
local indices of segregation and the ratio of African American population by census tract as well.
On the top row Chicago and Los Angeles and on the bottom row Detroit and Houston.
\label{mapcorr}}
\end{center}
\end{figure}

\section{Temporal analysis of correlations with segregation indices and other socioeconomic indicators}

\subsection*{Statistical analysis of the COVID-19 incidence data}

\begin{figure}[!htbp]
\begin{center}
\includegraphics[width=10cm]{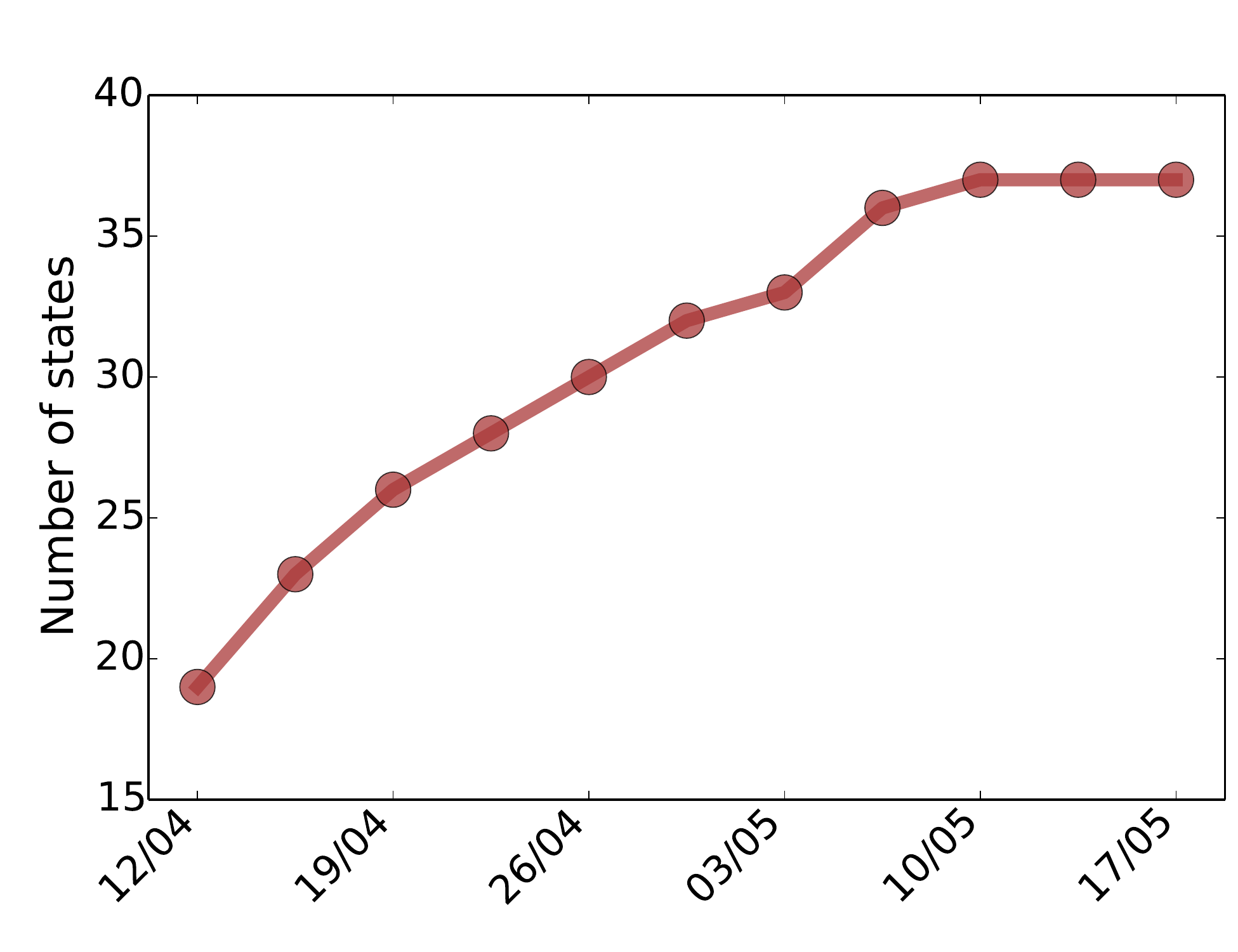}
\caption{Number of states included in the analysis for each temporal snapshot.
\label{numpoints}}
\end{center}
\end{figure}

In this section, we provide the temporal evolution of correlations
between the difference in the percentage of COVID-19 incidence among
African Americans and other segregation indices.  First of all,
Supplementary Figure \ref{numpoints} shows the number of states
included in each of the temporal snapshots. As can be seen, it
increases with time yet already in the first temporal snapshots there
are almost 20 of them. It is important to note that by mid-April, the
US reached the first peak of the pandemic.

Additionally to the number of states included in the analysis we also observe significant changes in the values across time for the different states analysed. We provide in Supplementary Figures \ref{evolinf} and \ref{evoldec} the evolution of the difference in percentage of infected and deceased African Americans. States are split in quartiles of the distribution of the percentage of African Americans among the overall population. While the average seems almost stable in most of the quartiles this is more a product of compensating changes than of stability in the values for a single state. For instance, in the first quartile there is a sharp increase in Minnesota compensated by a decrease in DC. On the third quartile, the sharp decrease in Illinois is compensated by the increase in Arkansas. It is also important to note that some states display strong discrepancies between the percentage on deceased and infected as, for instance, Minnesota.

\begin{figure}[!htbp]
\begin{center}
\includegraphics[width=15cm]{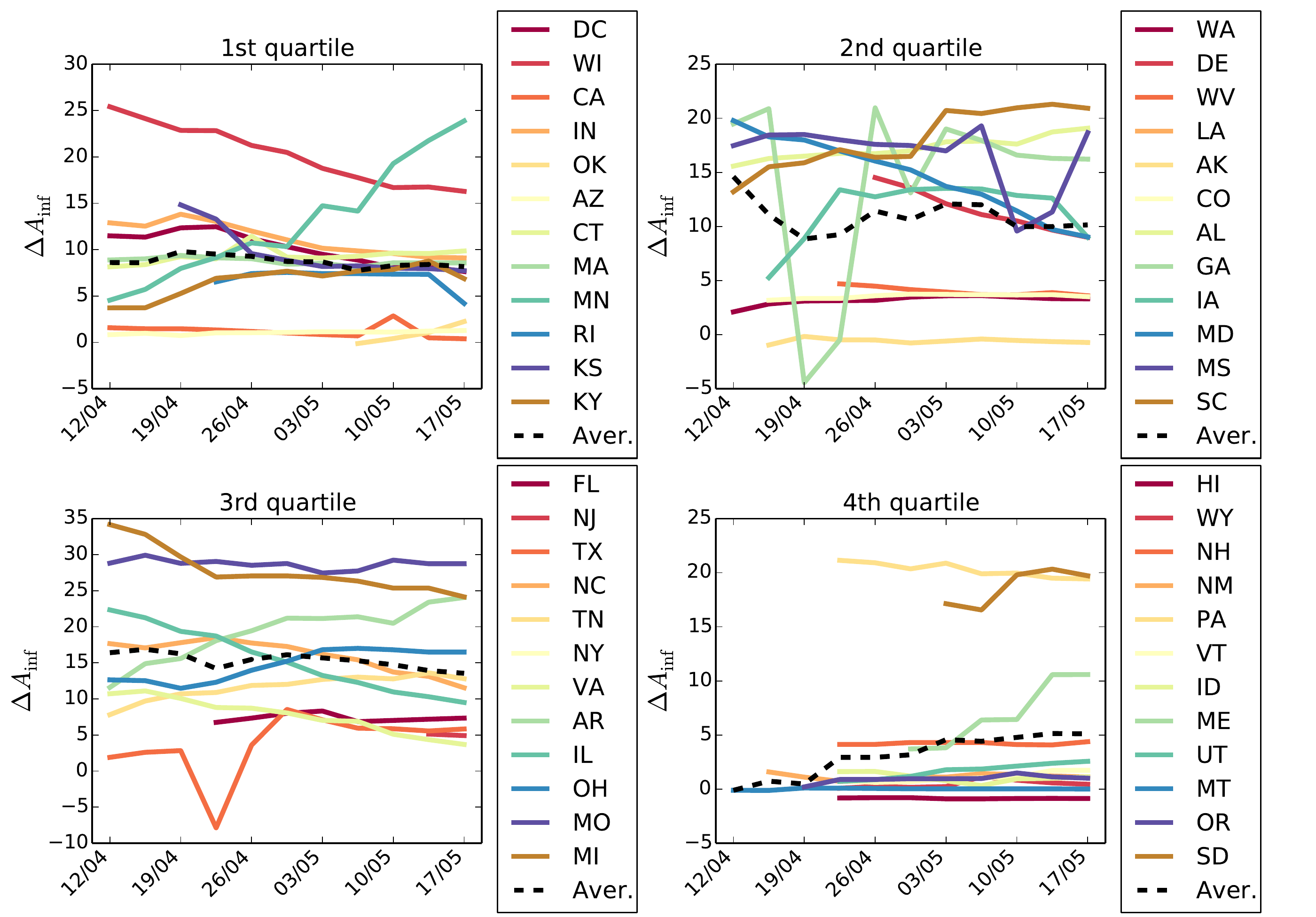}
\caption{The temporal evolution of the difference  in percentage on infected African Americans by state. Each plot represents a quartile of the distribution of percentage of African American population.
\label{evolinf}}
\end{center}
\end{figure}

\begin{figure}[!htbp]
\begin{center}
\includegraphics[width=15cm]{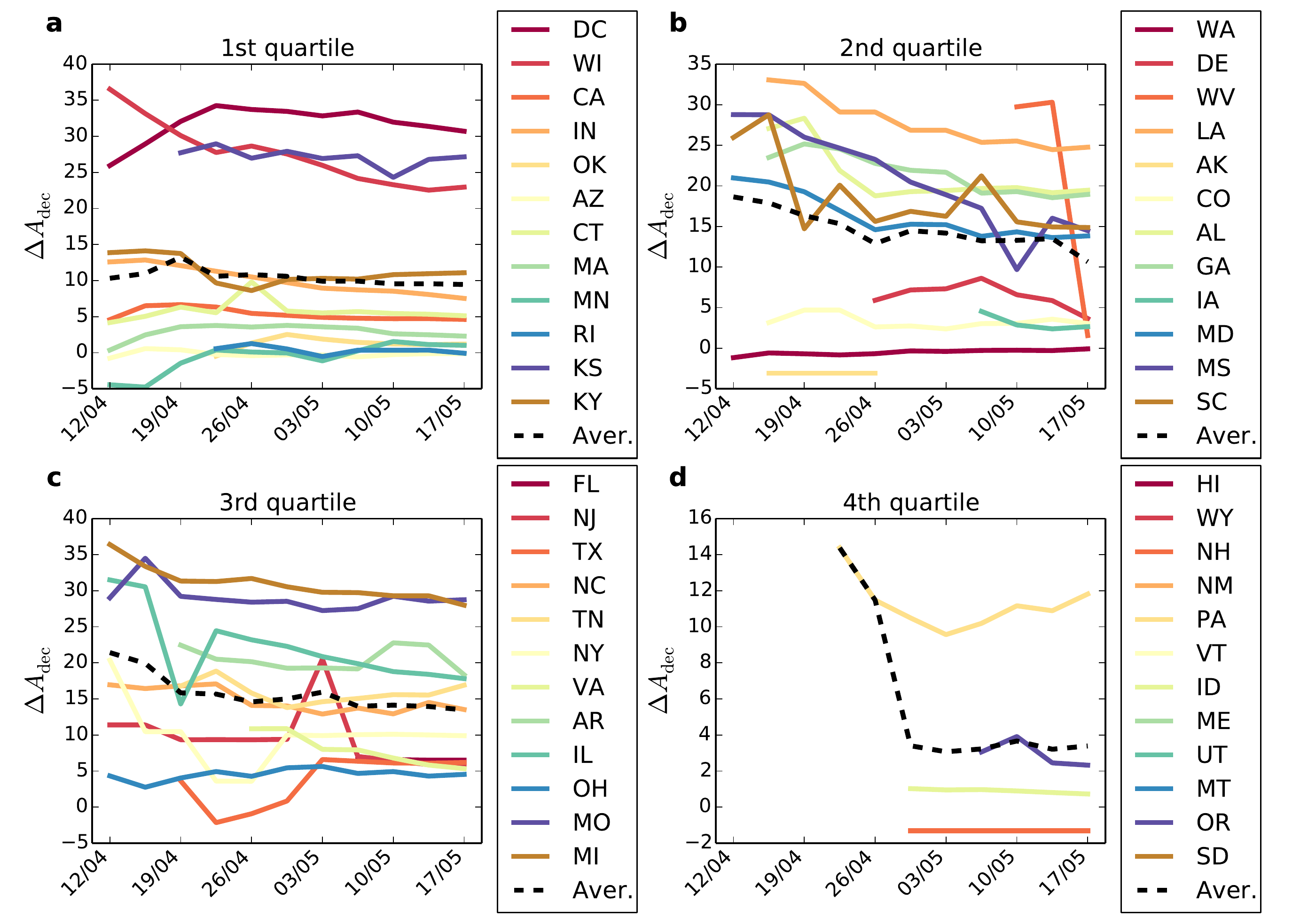}
\caption{The temporal evolution of the difference in percentage on deceased African Americans by state. Each plot represents a quartile of the distribution of percentage of African American population.
\label{evoldec}}
\end{center}
\end{figure}

\subsection*{Correlations with the ratio of infected African Americans and the difference in percentage and the ratio of deceased African Americans}

In the main manuscript, the main variable analysed is the difference in the percentage of African Americans since other factors might influence the deceased and the ratio that can lead to several outliers. In Supplementary Figures \ref{correvol_diff_deceased},  \ref{correvol_ratio_infected} and  \ref{correvol_ratio_deceased}, we show respectively the correlation with the difference in percentage of deceased African Americans, the ratio of infected and the ratio of deceased. In the case of the difference in percentage we can see that despite correlations are lower they are stable across time. It is important to note that in the case of deceased individuals other factors like the age or the underlying health conditions might play a significant role. In the case of both ratios, correlations are slightly high in the first snapshots suffer a steeper decrease. Again $C$ and $E$ computed in the commuting graph seem to outperform the rest of metrics.

\begin{figure}[!htbp]
\begin{center}
\includegraphics[width=14cm]{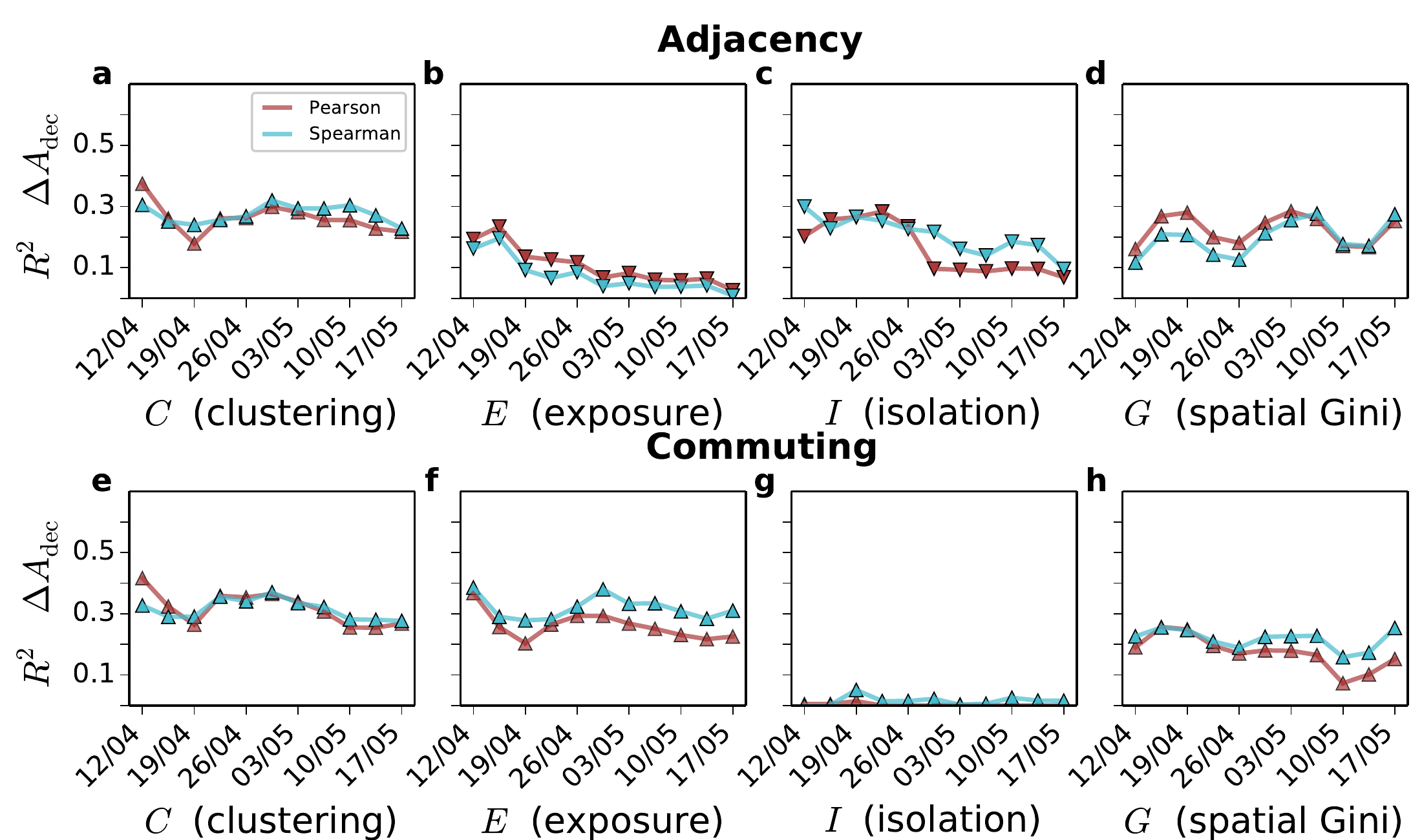}
\caption[Evolution of the Pearson and Spearman correlation ($R^2$) found between the difference in percentage of deceased African Americans and the four indices studied in the main manuscript]{Evolution of the Pearson and Spearman correlation ($R^2$) found between the difference in percentage of deceased African Americans and the four indices studied in the main manuscript. \textbf{a} $C$ (clustering), \textbf{b} $E$ (exposure), \textbf{c} $I$ (isolation),
  \textbf{d} $G$ (spatial Gini). \textbf{e-h} Indices computed over the commuting network: \textbf{e} $C$ (clustering), \textbf{f} $E$ (exposure), \textbf{g} $I$ (isolation),
  \textbf{h} $G$ (spatial Gini).
\label{correvol_diff_deceased}}
\end{center}
\end{figure}

\begin{figure}[!htbp]
\begin{center}
\includegraphics[width=14cm]{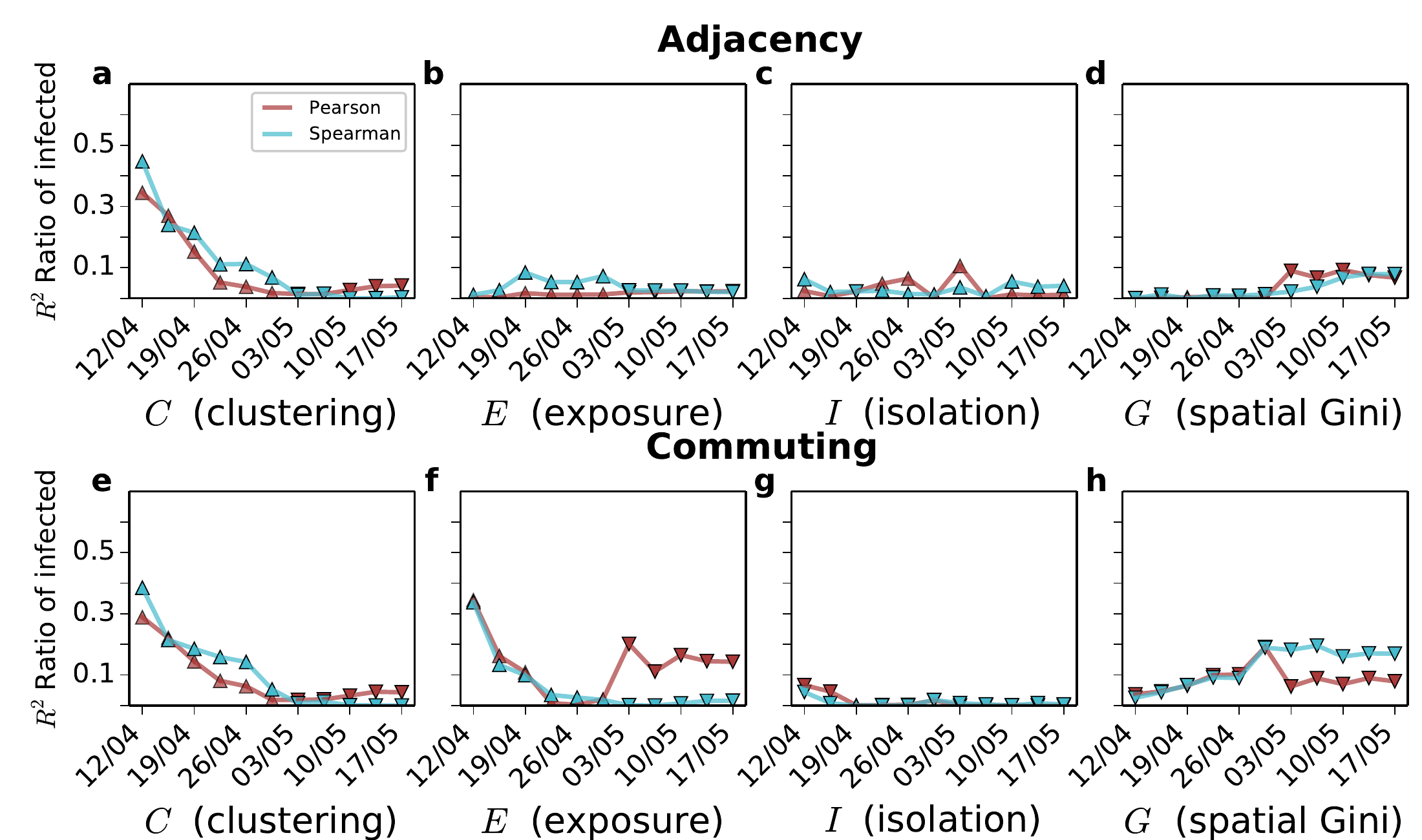}
\caption[Evolution of the Pearson and Spearman correlation ($R^2$) found between the ratio of infected African Americans and the four indices studied in the main manuscript]{Evolution of the Pearson and Spearman correlation ($R^2$) found between the ratio of infected African American and each of the indices studied in this work. \textbf{a-g} Indices computed over the adjacency network: \textbf{a} $C$ (clustering), \textbf{b} $E$ (exposure), \textbf{c} $I$ (isolation),
  \textbf{d} $G$ (spatial Gini). \textbf{e-h} Indices computed over the commuting network: \textbf{e} $C$ (clustering), \textbf{f} $E$ (exposure), \textbf{g} $I$ (isolation),   \textbf{h} $G$ (spatial Gini).
\label{correvol_ratio_infected}}
\end{center}
\end{figure}

\begin{figure}[!htbp]
\begin{center}
\includegraphics[width=14cm]{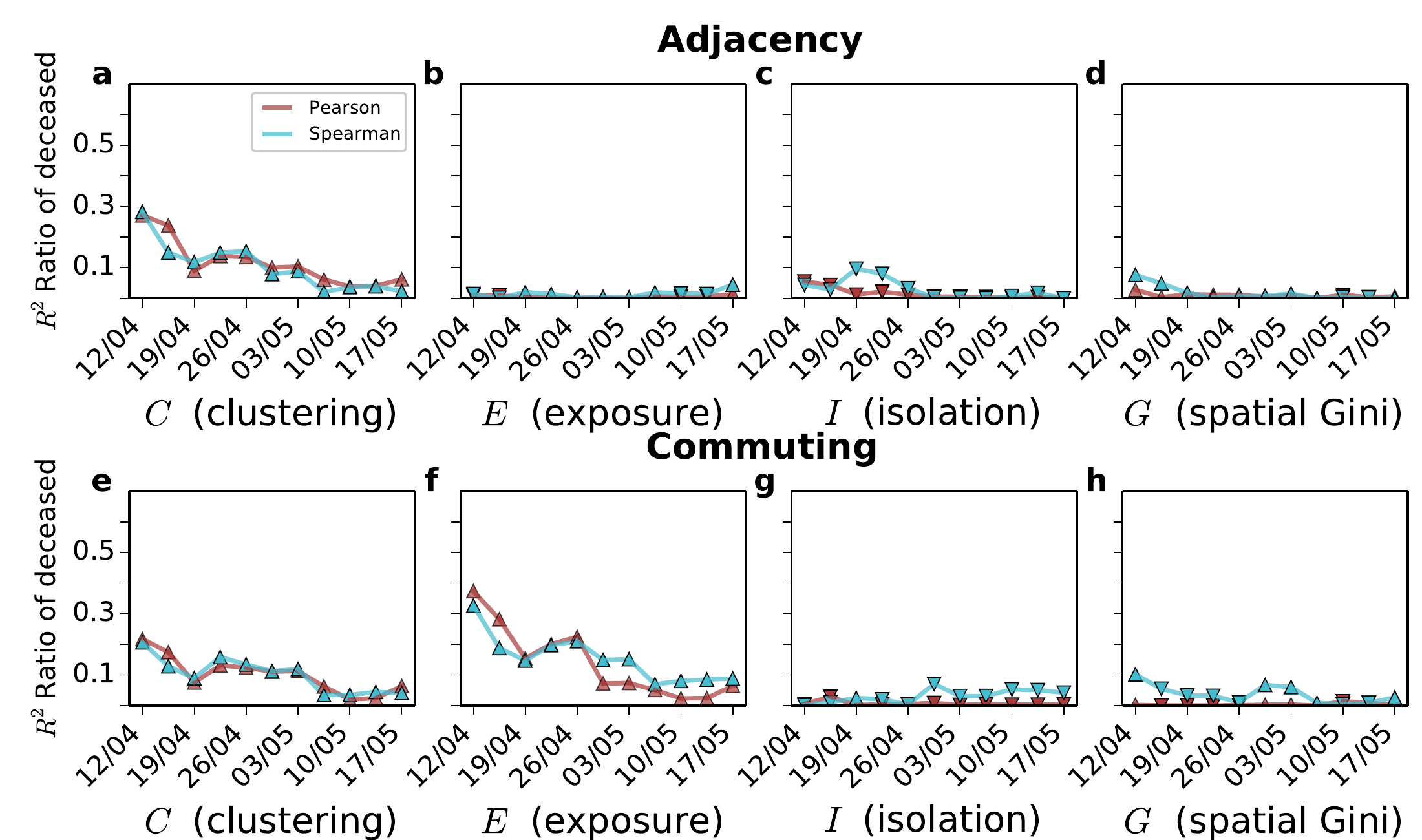}
\caption[Evolution of the Pearson and Spearman correlation ($R^2$) found between the ratio of deceased African Americans and the four indices studied in the main manuscript]{Evolution of the Pearson and Spearman correlation ($R^2$) found between the ratio of deceased African Americans and the four indices studied in the main manuscript.  \textbf{a-g} Indices computed over the adjacency network: \textbf{a} $C$ (clustering), \textbf{b} $E$ (exposure), \textbf{c} $I$ (isolation),
  \textbf{d} $G$ (spatial Gini). \textbf{e-h} Indices computed over the commuting network: \textbf{e} $C$ (clustering), \textbf{f} $E$ (exposure), \textbf{g} $I$ (isolation),  \textbf{h} $G$ (spatial Gini).
\label{correvol_ratio_deceased}}
\end{center}
\end{figure}

\subsection*{Temporal evolution of correlations with another data set}

We also had access to another project that aggregates data on the ethnicity of both infected and deceased African Americans by COVID-19 through three different temporal snapshots $22/04/2020$, $04/05/2020$ and $15/05/2020$ \cite{Data2}. In Supplementary Figure \ref{correvol_diff_infected_data2}, we report the correlations in each of the three snapshots for the difference in the percentage of infected African Americans, which are in line with the results obtained for the other data set. Correlations are considerably high and significant for the first stages of the pandemic and decrease with time as the different lock-downs take place. As in the results provided in the main manuscript, those indices computed on the commuting graph seem to provide a better correlation than those computed on the adjacency one. Again those indices connected to the exposure of African Americans only yield significant correlation on the commuting network. The results obtained for the ratio of infected and deceased African Americans as well as the difference on the percentage of deceased are also compatible with those obtained with the previous data set (See Supplementary Figures \ref{correvol_ratio_infected_data2}, \ref{correvol_diff_deceased_data2} and \ref{correvol_ratio_deceased_data2}).

\begin{figure}[!htbp]
\begin{center}
\includegraphics[width=14cm]{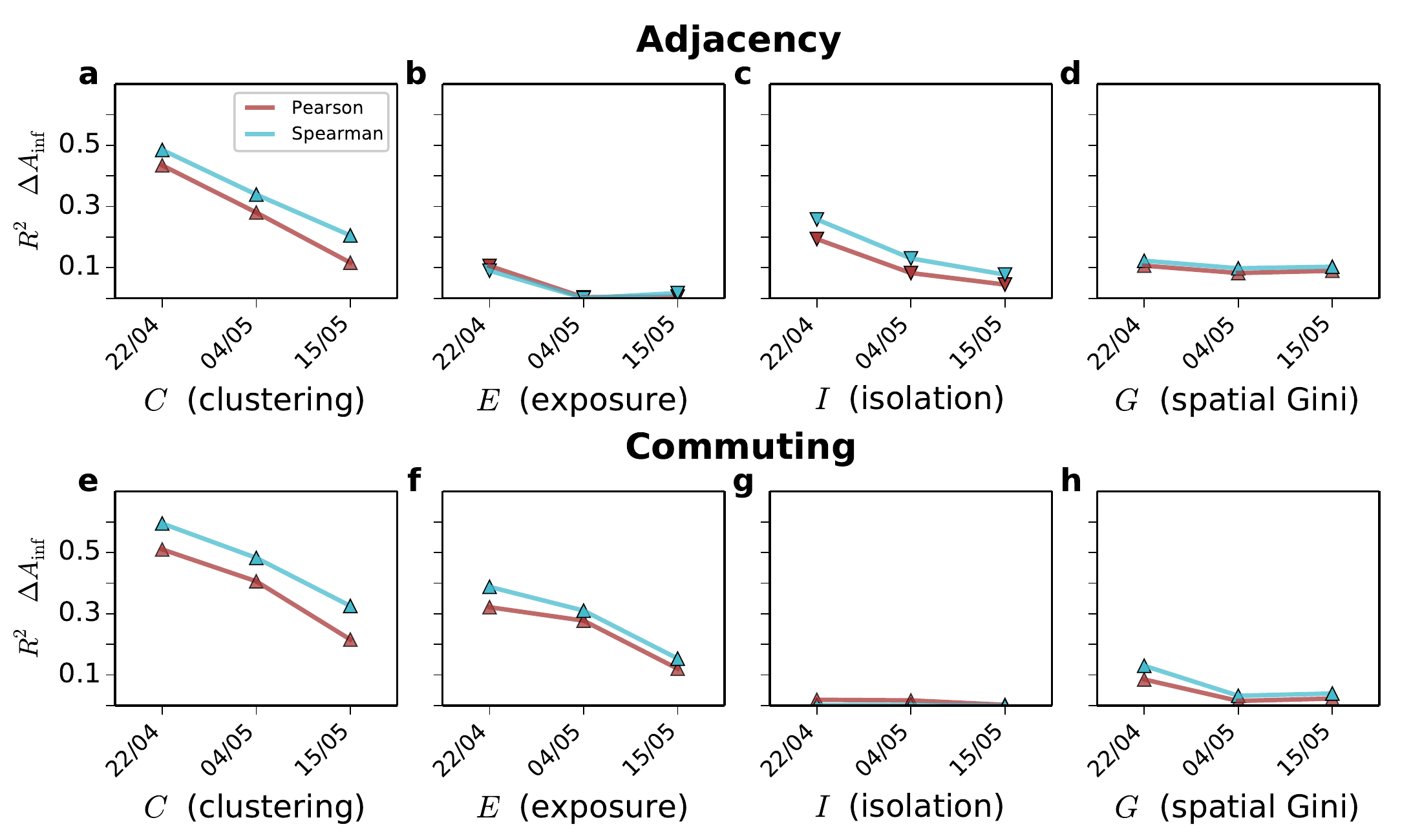}
\caption[Evolution of the Pearson and Spearman correlation ($R^2$) found between the difference on the deceased African Americans and the four indices studied in the main manuscript using another data source]{Evolution of the Pearson and Spearman correlation ($R^2$) found between the difference on the deceased African Americans and the four indices studied in the main manuscript using another data source. \textbf{a-g} Indices computed over the adjacency network:\textbf{a} $C$ (clustering), \textbf{b} $E$ (exposure), \textbf{c} $I$ (isolation),
  \textbf{d} $G$ (spatial Gini). \textbf{e-h} Indices computed over the commuting network: \textbf{e} $C$ (clustering), \textbf{f} $E$ (exposure), \textbf{g} $I$ (isolation),   \textbf{h} $G$ (spatial Gini). The markers indicate the sign of the relation, positive for triangles pointing up and negative for triangles
  pointing down.
\label{correvol_diff_infected_data2}}
\end{center}
\end{figure}

\begin{figure}[!htbp]
\begin{center}
\includegraphics[width=14cm]{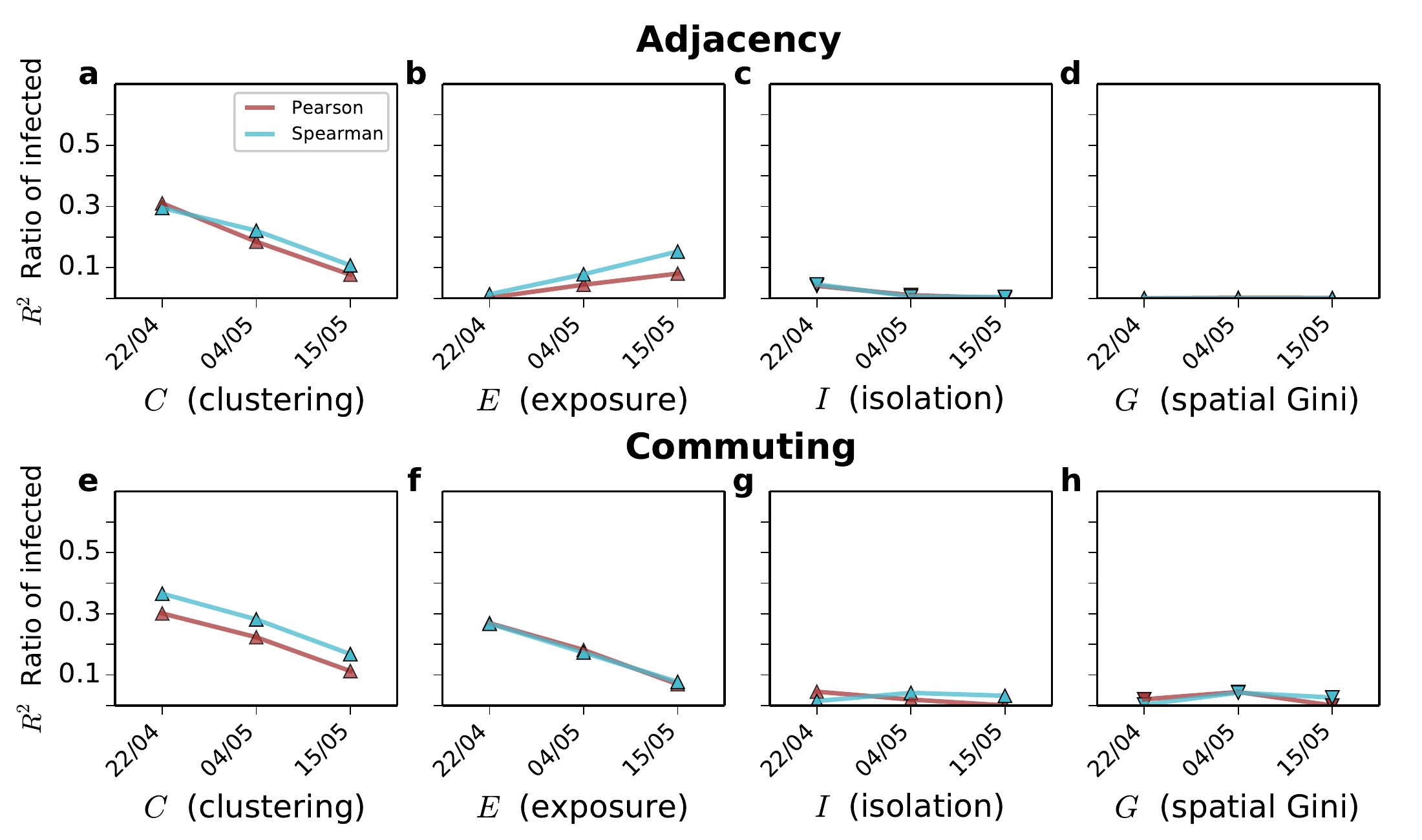}
\caption[Evolution of the Pearson and Spearman correlation ($R^2$) found between the difference on the deceased African Americans and the four indices studied in the main manuscript using another data source]{Evolution of the Pearson and Spearman correlation ($R^2$) found between the difference on the deceased African Americans and the four indices studied in the main manuscript using another data source.  \textbf{a-g} Indices computed over the adjacency network: \textbf{a} $C$ (clustering), \textbf{b} $E$ (exposure), \textbf{c} $I$ (isolation),
  \textbf{d} $G$ (spatial Gini). \textbf{e-h} Indices computed over the commuting network: \textbf{e} $C$ (clustering), \textbf{f} $E$ (exposure), \textbf{g} $I$ (isolation), \textbf{h} $G$ (spatial Gini). The markers indicate the sign of the relation, positive for triangles pointing up and negative for triangles
  pointing down.
\label{correvol_diff_deceased_data2}}
\end{center}
\end{figure}

\begin{figure}[!htbp]
\begin{center}
\includegraphics[width=14cm]{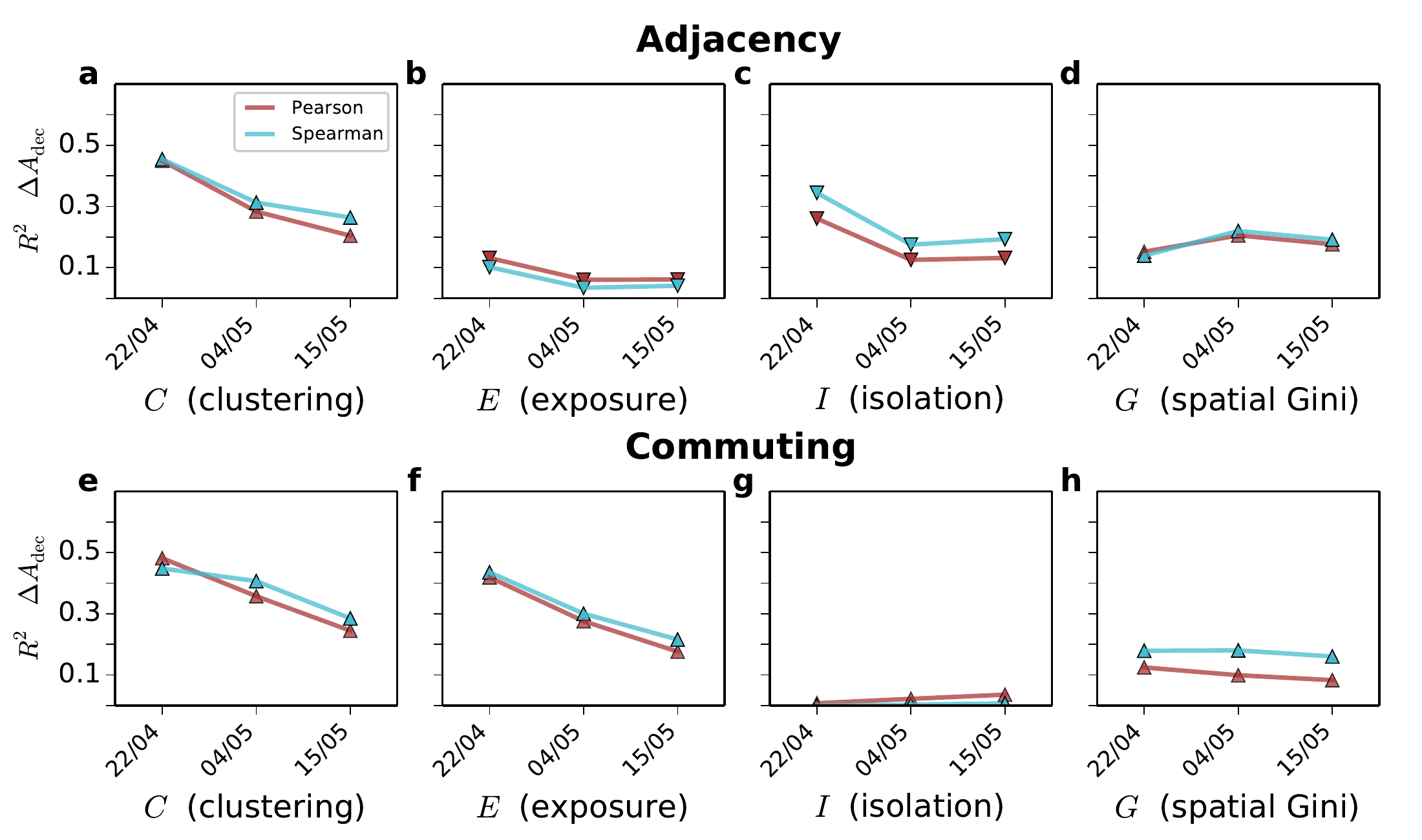}
\caption[Evolution of the Pearson and Spearman correlation ($R^2$) found between the ratio of deceased African Americans and the four indices studied in the main manuscript using another data source]{Evolution of the Pearson and Spearman correlation ($R^2$) found between the ratio of deceased African Americans and the four indices studied in the main manuscript using another data source.  \textbf{a-g} Indices computed over the adjacency network: \textbf{a} $C$ (clustering), \textbf{b} $E$ (exposure), \textbf{c} $I$ (isolation),
  \textbf{d} $G$ (spatial Gini). \textbf{e-h} Indices computed over the commuting network: \textbf{e} $C$ (clustering), \textbf{f} $E$ (exposure), \textbf{g} $I$ (isolation), \textbf{h} $G$ (spatial Gini). The markers indicate the sign of the relation, positive for triangles pointing up and negative for triangles
  pointing down.
\label{correvol_ratio_deceased_data2}}
\end{center}
\end{figure}

\begin{figure}[!htbp]
\begin{center}
\includegraphics[width=14cm]{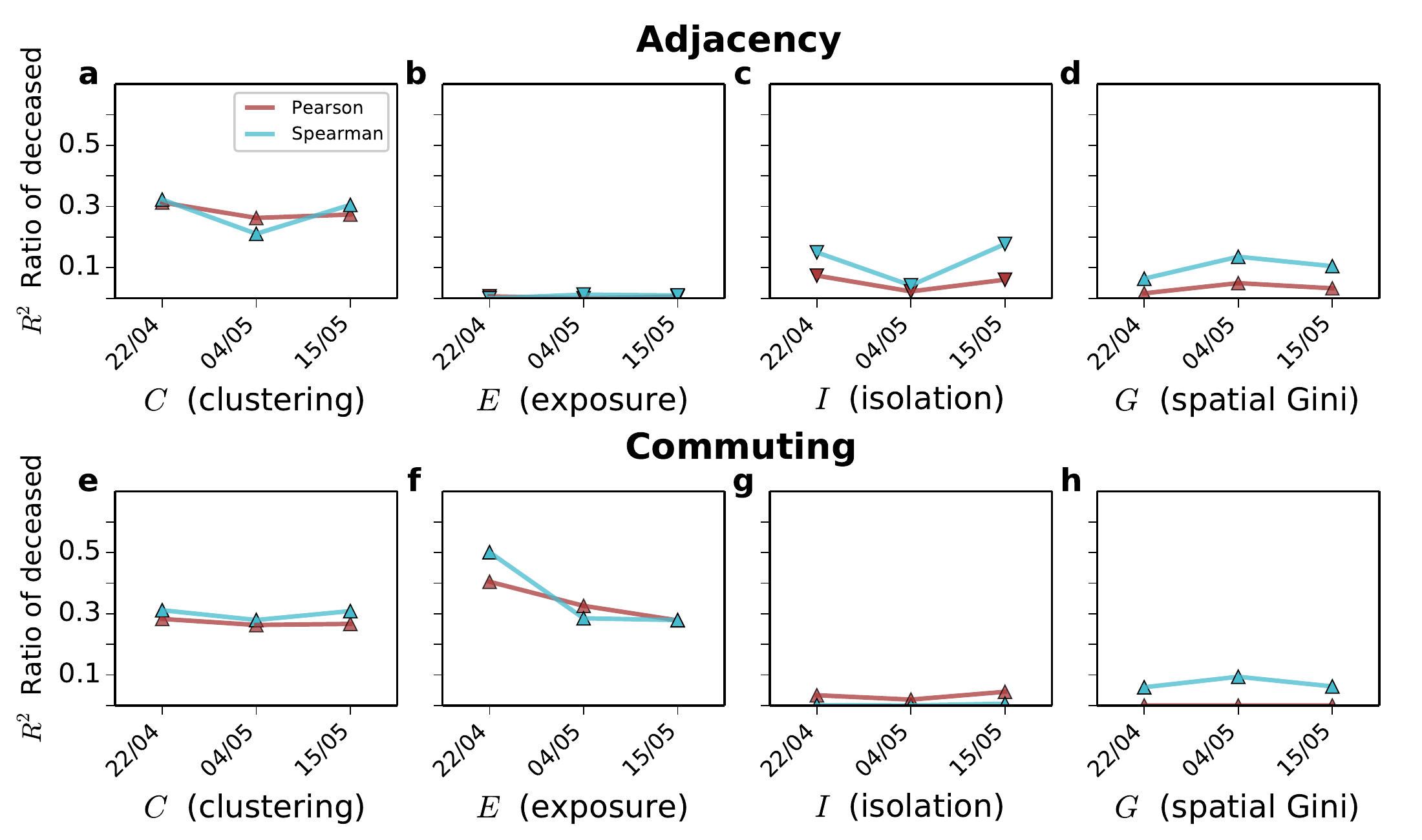}
\caption[Evolution of the Pearson and Spearman correlation ($R^2$) found between the ratio of infected African Americans and the four indices studied in the main manuscript using another data source ]{Evolution of the Pearson and Spearman correlation ($R^2$) found between the ratio of infected African Americans and the four indices studied in the main manuscript using another data source.  \textbf{a-g} Indices computed over the adjacency network:\textbf{a} $C$ (clustering), \textbf{b} $E$ (exposure), \textbf{c} $I$ (isolation),
  \textbf{d} $G$ (spatial Gini). \textbf{e-h} Indices computed over the commuting network: \textbf{e} $C$ (clustering), \textbf{f} $E$ (exposure), \textbf{g} $I$ (isolation), \textbf{h} $G$ (spatial Gini). The markers indicate the sign of the relation, positive for triangles pointing up and negative for triangles
  pointing down.
\label{correvol_ratio_infected_data2}}
\end{center}
\end{figure}

\subsection*{Formulation of the alternative indices $C'$ and $E'$}

In the main manuscript we have studied the metrics $C$ and $E$ which are computed from the elements of the normalised CMFPT $\widetilde{\tau}_{\alpha,\beta}$. However, there are more potential ways to capture the clustering and exposure of an ethnicity by doing the other calculations from that matrix. Here we propose the two alternative formulations for those two metrics
\begin{align*}
C'=\frac{\overline{\tau}_{OA}}{\overline{\tau}_{OO}}\\
E'=\frac{\overline{\tau}_{AA}}{\overline{\tau}_{AO}}
\end{align*}
The first quantity comes from the ratio between the time from other ethnicities to African Americans and the time from other ethnicities to others, where higher values correspond more isolated African Americans compared to other ethnicities.
The second quantity instead, is the ratio between the time separating African Americans and the time between African Americans and any other ethnicity, where higher values correspond to African Americans more exposed to others than to themselves. The correlation between our alternative proposals and the difference in the percentage of African Americans infected is shown in Supplementary Figure \ref{Figure4_alternative}. While correlations are slightly lower, they are still significant. One interesting finding is that $E'$ changes the sign of the correlation when computed over the adjacency graph and the commuting network. Highlighting once again the need of considering mobility to understand the segregation and exposure of ethnicities in urbanscapes.

\begin{figure}[!htbp]
\begin{center}
\includegraphics[width=14cm]{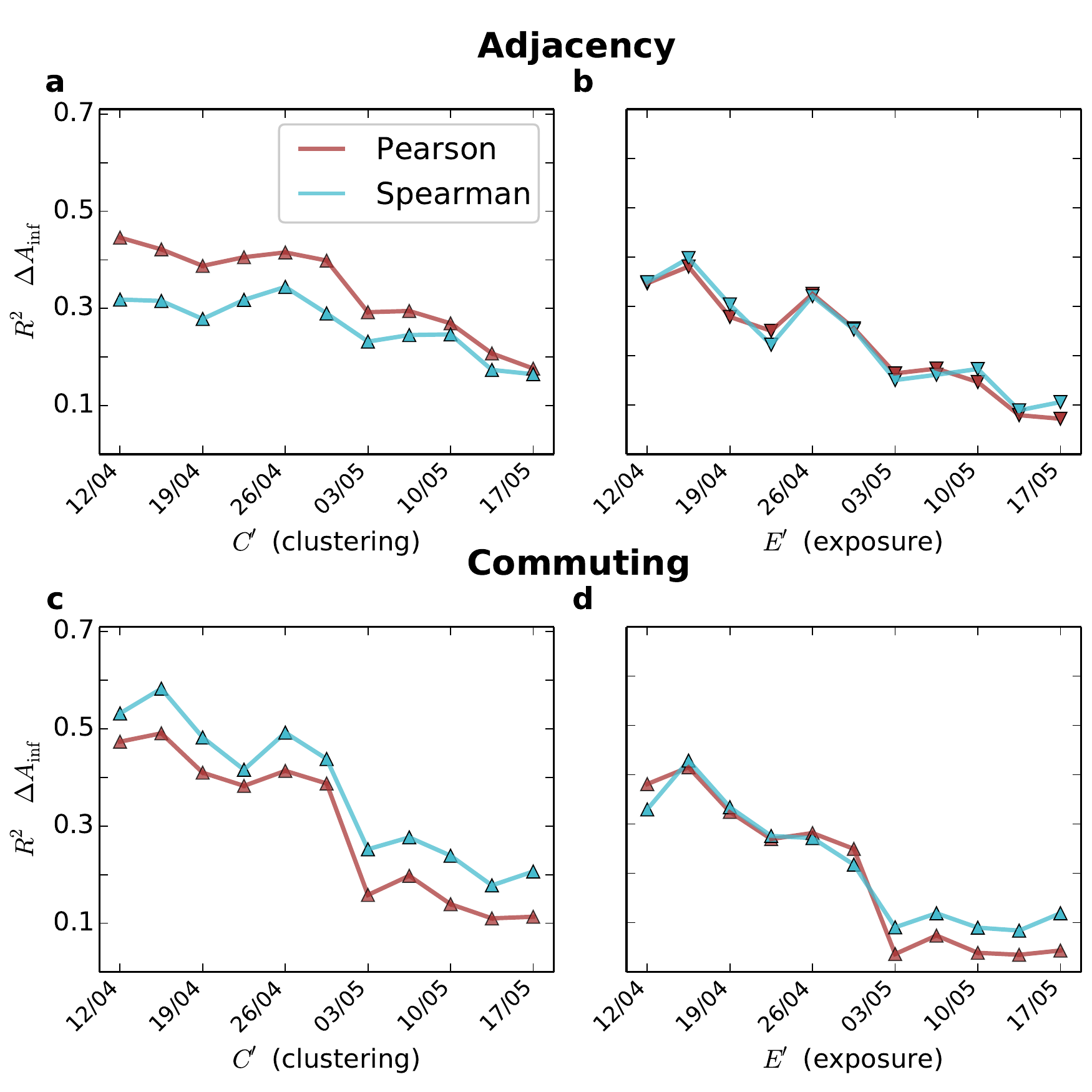}
\caption[Evolution of the Pearson and Spearman correlation ($R^2$) found between the difference on the percentage of infected African Americans and the alternative indices proposed $C'$ and $E'$]{Evolution of the Pearson and Spearman correlation ($R^2$) found between the difference of infected African Americans and the alternative indices proposed $C'$ and $E'$. \textbf{a} $C'$ (clustering) and \textbf{b} $E'$ (exposure) calculated upon the adjacency network. \textbf{c} $C'$ (clustering) and \textbf{d} $E'$ (exposure) calculated upon the commuting network. The markers indicate the sign of the relation, positive for triangles pointing up and negative for triangles
  pointing down.
\label{Figure4_alternative}}
\end{center}
\end{figure}

\subsection*{Temporal  evolution of correlations with other segregation indices from the literature}

We have also studied the correlations between the difference in the percentage of COVID-19 incidence among African Americans and other segregation indices from the literature. First of all, we obtained the segregation index $\sigma_\alpha$ proposed in \cite{Ballester2014}, which is also based on the movement of random walks is spatial systems and captures the probability that a randomly chosen individual of group $\alpha$ meets another individual of the same group, or in this case, ethnicity. Additionally, we also computed Moran's I, which is a measure of spatial auto-correlation and compares the ethnic composition of neighbourhoods \cite{Cliff1981}. The correlation of both metrics with the difference in the percentage of infected among African Americans.
The evolution of the correlations is shown in Supplementary Figure
\ref{balmoran}, where only the Moran index calculated over the adjacency graph seems to display a significant correlation.

\begin{figure}[!htbp]
\begin{center}
\includegraphics[width=14cm]{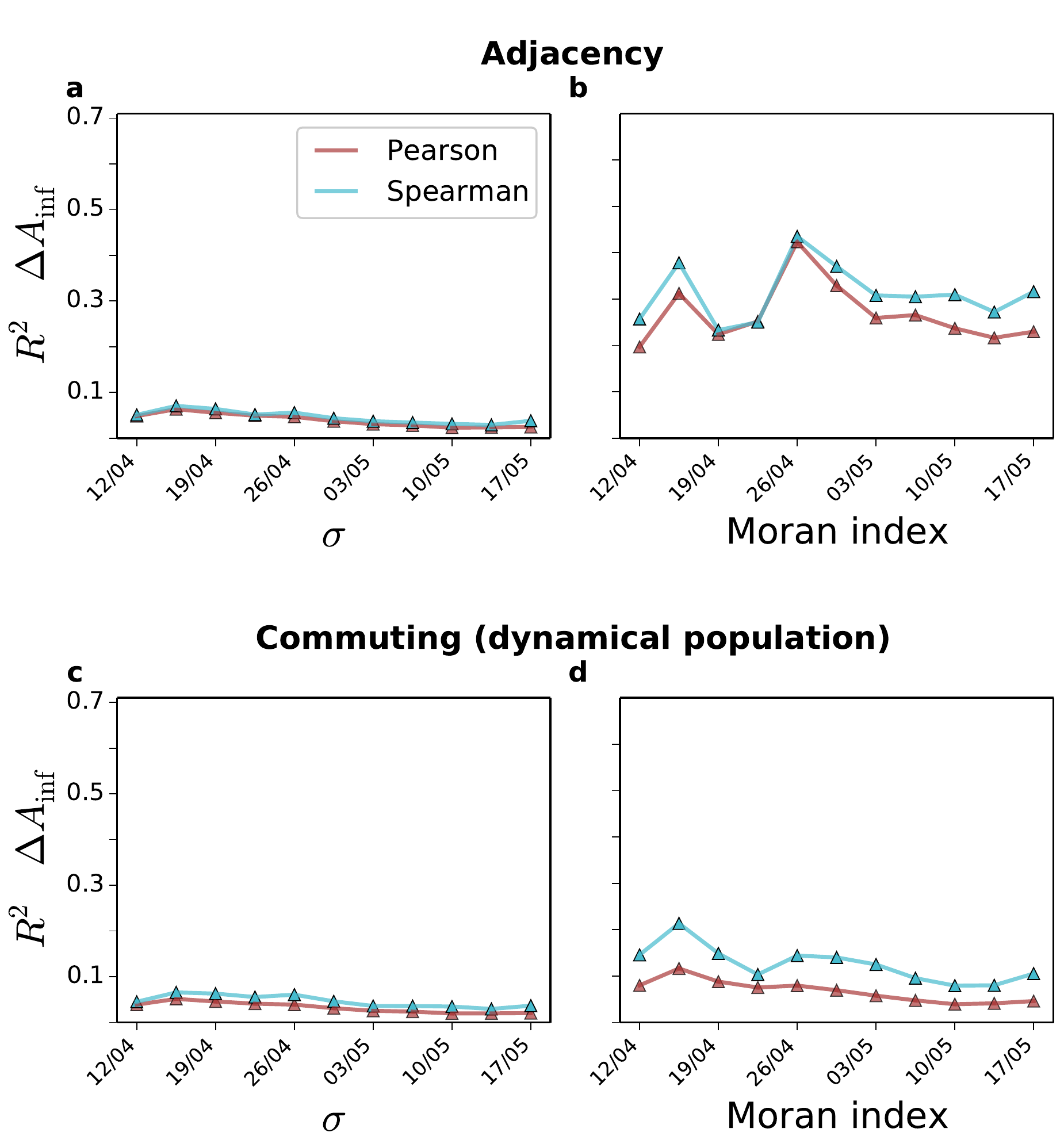}
\caption[Correlations between the incidence of COVID-19 in African American population and the segregation indices $\sigma$ and Moran's I]{\textbf{The temporal evolution of the correlation ($R^2$)  between the incidence of COVID-19 in African American population and the segregation indices $\sigma$ and Moran's I.} \textbf{a} $\sigma$ and  \textbf{b} Moran's I computed over the adjacency graph.  \textbf{c} $\sigma$ and  \textbf{d} Moran's I computed over the commuting graph. The markers indicate the sign of the relation, positive for triangles pointing up and negative for triangles
  pointing down.
\label{balmoran}}
\end{center}
\end{figure}

For the second metric we have build a matrix of distance between ethnicities similar to the one obtained for $\widetilde{\tau}_{\alpha,\beta}$ using a measure proposed by \cite{Farber2012}. Inspired by the Getis and Ord statistic \cite{Getis2010}, the metric proposed  \cite{Farber2012} quantifies for each location $i$ the exposure of ethnicity $\alpha$ to ethnicity $\beta$ as
\begin{equation}
{}^\beta_\alpha G^*_i=\frac{\sum_{j=1}^n w_{ij}(\hat{d}_{pi})m_{j,\beta}}{\sum_{j=1}^n  m_{j,\beta}},
\end{equation}

where each $n$ is the total number of location in a city, $j$ corresponds to each of those locations and  $m_{j,\beta}$ is the population of ethnicity $\beta$ in location $j$, $\hat{d}_{pi}$ is an estimate of trip length and $w_{ij}(\hat{d}_{pi})$ is a function of the distance that is equal to $1$ when $d_{ij}<d_{pj}$  and $0$ otherwise. In the case of the adjacency graph only adjacent pair of tracts were considered whereas in the case of the commuting network only pairs connected by commuting trips were considered. Overall ${}^\beta_\alpha G^*_i$ quantify the ratio of population of ethnicity $\beta$ to which the individuals residing in $i$ are exposed.
 In our case we set the threshold $d_{pj}$ equal to the average commuting distance in each of the cities. Succinctly, ${}^\beta_\alpha G^*_i$ is a value between $0$ and $1$ that encapsulates the fraction of the population of ethnicity. We average the value of ${}^\beta_\alpha G^*_i$ to obtain a distance matrix between ethnicities in each of the cities as

\begin{equation}
{}^\beta_\alpha G^*_i ={\sum_{i=1}^n m_{i,{\alpha}} {}^\beta_\alpha G^*_i}{\sum_{i=1}^n m_{i,{\alpha}}},
\end{equation}

so that we take into account the fraction of population of ethnicity $\alpha$ in location $i$. Finally from the matrix ${}^\beta_\alpha <G^*>$ we compute the same exposure and clustering indices computed from $\widetilde{\tau}_{\alpha,\beta}$ in the main text. Calculating first

\begin{align*}
\overline{<G^*>}_{AO}=\frac{\sum_{\forall \beta\ne A }M^{\beta} {}^A_\alpha <G^*>}{\sum_{\forall \beta\ne A }e^\beta} \\
\overline{<G^*>}_{OA}=\frac{\sum_{\forall \beta\ne A }M^\beta {}^\beta_A <G^*>}{\sum_{\forall \beta\ne A }M^\beta}  \\
\overline{<G^*>}_{OO}=\frac{\sum_{\forall \alpha,\beta\ne A }{}^\beta_\alpha <G^*>M^{\alpha}M^{\beta}}{\sum_{\forall \alpha,\beta\ne A }M^{\alpha}M^\beta},
\label{Eq2}
\end{align*}

to finally obtain

\begin{align*}
C^f=\frac{\overline{<G^*>}_{AO}}{\overline{<G^*>}_{OO}}, \\
E^f=\frac{\overline{<G^*>}_{OO}}{\overline{<G^*>}_{OO}}.
\end{align*}

Additionally to the calculation of the indices in the adjacency and the commuting network with dynamical population we also computed it with the residential population and the commuting network to investigate the role played by the dynamical population. As can be seen in Supplementary Figure \ref{farber}, significant correlations appear with all indices yet the higher ones are with the exposure index especially when computed on the commuting network with dynamical population. Correlations are, however, lower and less stable than those obtained in the main manuscript.

\begin{figure}[!htbp]
\begin{center}
\includegraphics[width=12cm]{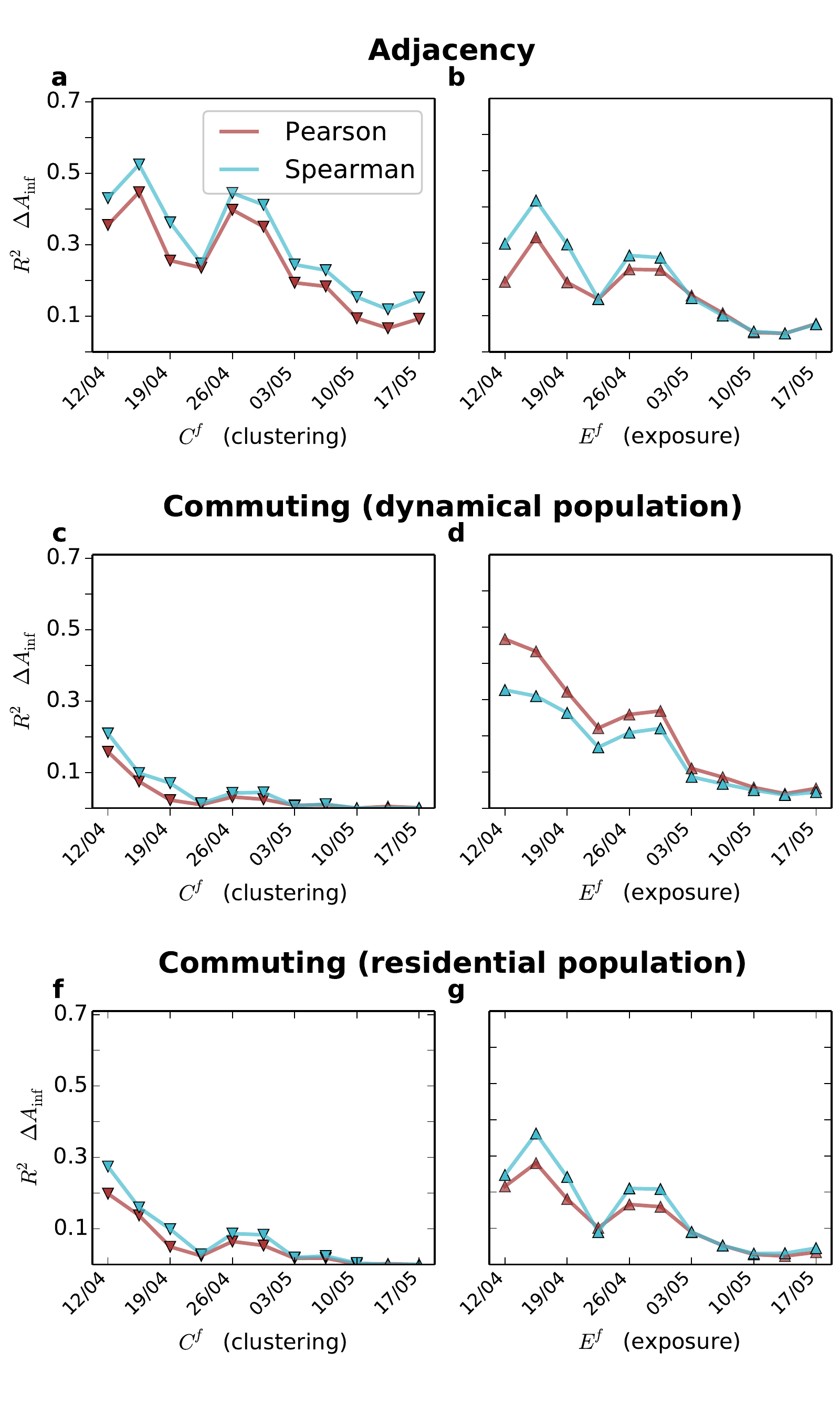}
\caption[Correlations between the incidence of COVID-19 in African American population and the clustering and exposure indices computed from the $G^*$ statistic proposed in \cite{Farber2012} ]{\textbf{Correlations between the incidence of COVID-19 in African American population and the clustering and exposure indices computed from the $G^*$ statistic proposed in \cite{Farber2012}.} \textbf{a,b} Correlation with the clustering $C^f$ and exposure $E^f$ indices computed over the adjacency graph. \textbf{c,d} Correlation with the clustering $C^f$ and exposure $E^f$ indices computed over the commuting graph when the dynamical population is incorporated. \textbf{e,f} Correlation with the clustering $C^f$ and exposure  $E^f$ indices computed over the commuting graph when only the residential population is incorporated. The markers indicate the sign of the relation, positive for triangles pointing up and negative for triangles
  pointing down.
\label{farber}}
\end{center}
\end{figure}

Overall, it is important to note that none of the additional segregation metrics we have studied in this section is more informative than the ones we proposed on the main manuscript based on CMFPT and CCT. Moreover, the use of the dynamical population together with the commuting network seems to improve some of the indices.

\subsection*{The relation between the incidence of COVID-19 in African Americans and socio-economic indicators}

We present in this section the correlations between a set of
socio-economic indicators and the incidence of COVID-19 in African
Americans. For the sake of brevity, we focus here only on the data set
used in the main manuscript as well as in the difference in the
percentage of infections which is the case where correlations are
higher. The set of indicators we have studied are the median household
income, the percentage of the population below the poverty level, the
percentage of insured and uninsured African Americans, the usage of
public transportation by both African Americans and the overall
population, the percentage of African American population in a state,
the average commuting distance and the ratio between the average
commuting distance of African Americans and the overall
population. All of the metrics are provided at the level of the
African American population and the results are shown in Supplementary
Figure \ref{socioeconomic}.  The median household income, the
percentage of the population below the poverty level, the percentage
of insured and uninsured African Americans were obtained from the 2018
American Community Survey elaborated by the U.S. Census Bureau
\cite{Ethnicity}. Most of the variables yield low or very low
correlations except for the usage of public transportation by African
Americans. Economic indicators such as median income or percentage of
poverty seem to slightly correlate with the incidence of COVID-19,
which could because because a more deprived African American community
puts them in a more risky situation. Regarding the health indicators
related to the degree of insurance of African Americans, it seems
there is no direct relation with the number of infected. Not so
surprising results since we are analysing the percentage of infected
and, therefore, the fact of having insurance might not change
significantly the risk of getting the illness. Finally, the usage of
public transportation seems to play a crucial role in the spread of
the disease, especially if we compare the use done by the African
American population and the overall population where no correlation
appears. The fact that African Americans use more public
transportation might put them on a more dangerous position as well as
might a reflection of their economic status. Moreover, it could happen
that in those cities in which African Americans are more segregated
they also have to use more the public transportation.

\begin{figure}[!htbp]
\begin{center}
\includegraphics[width=5.9cm]{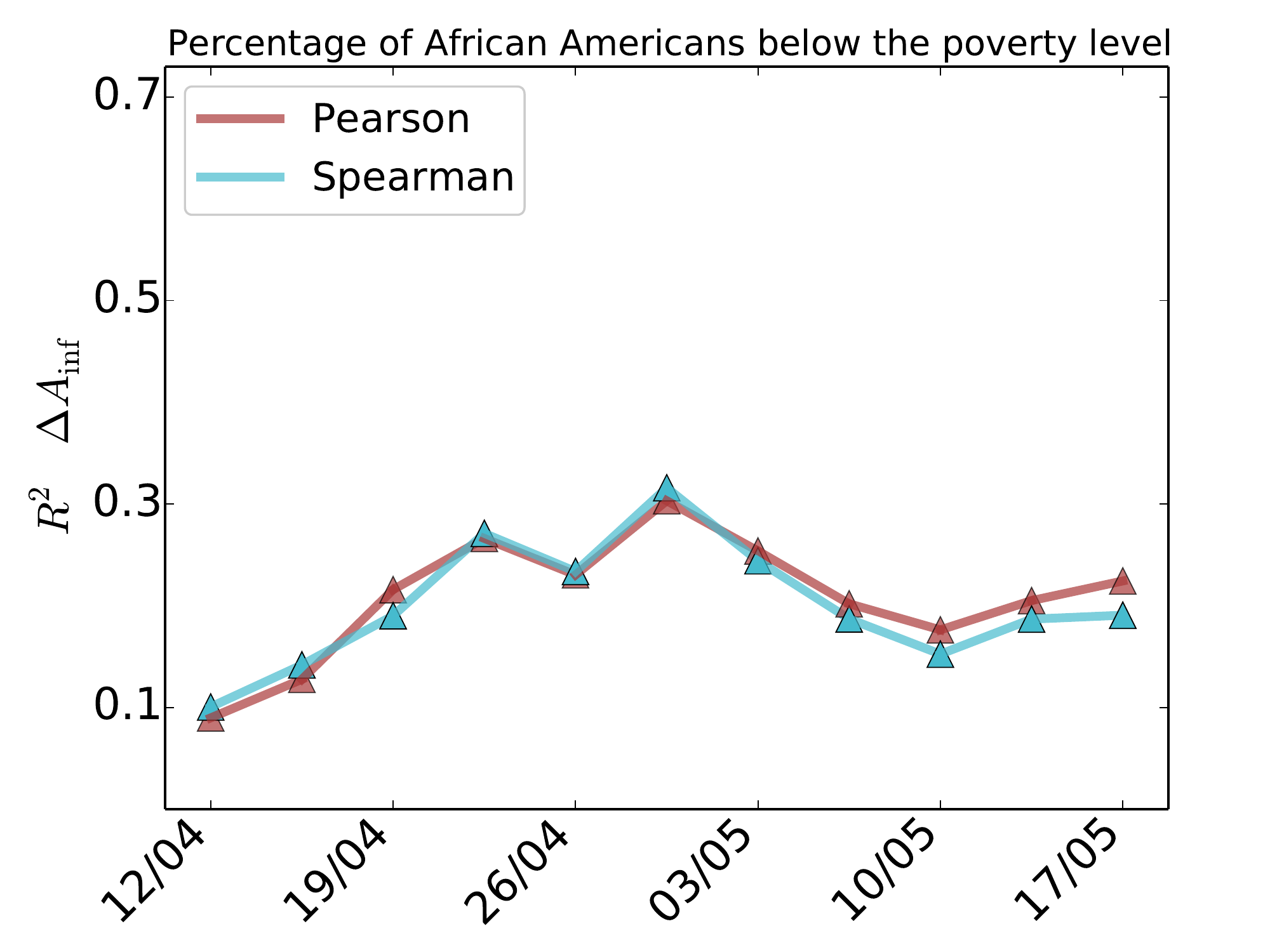}
\includegraphics[width=5.9cm]{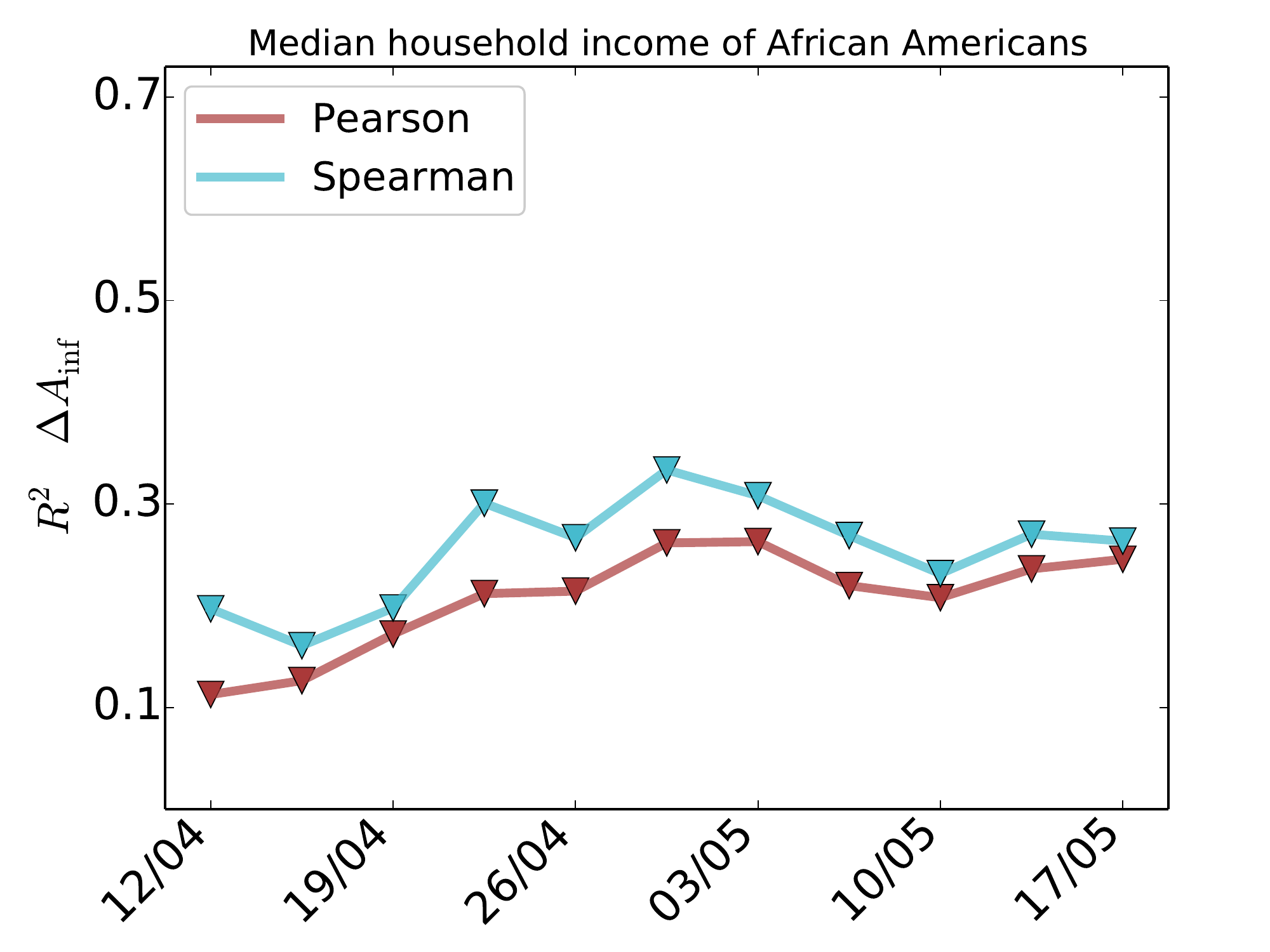}
\includegraphics[width=5.9cm]{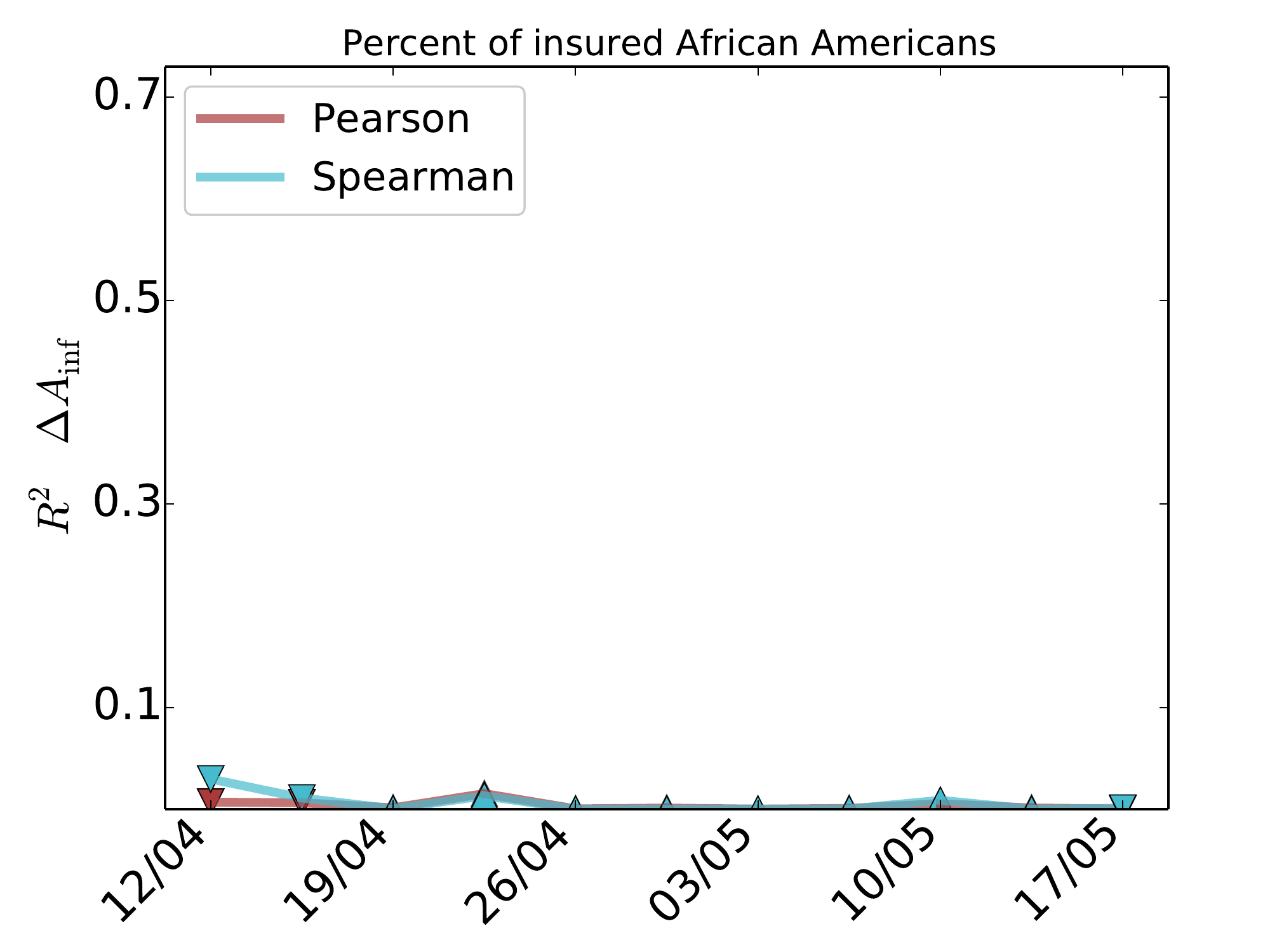}
\includegraphics[width=5.9cm]{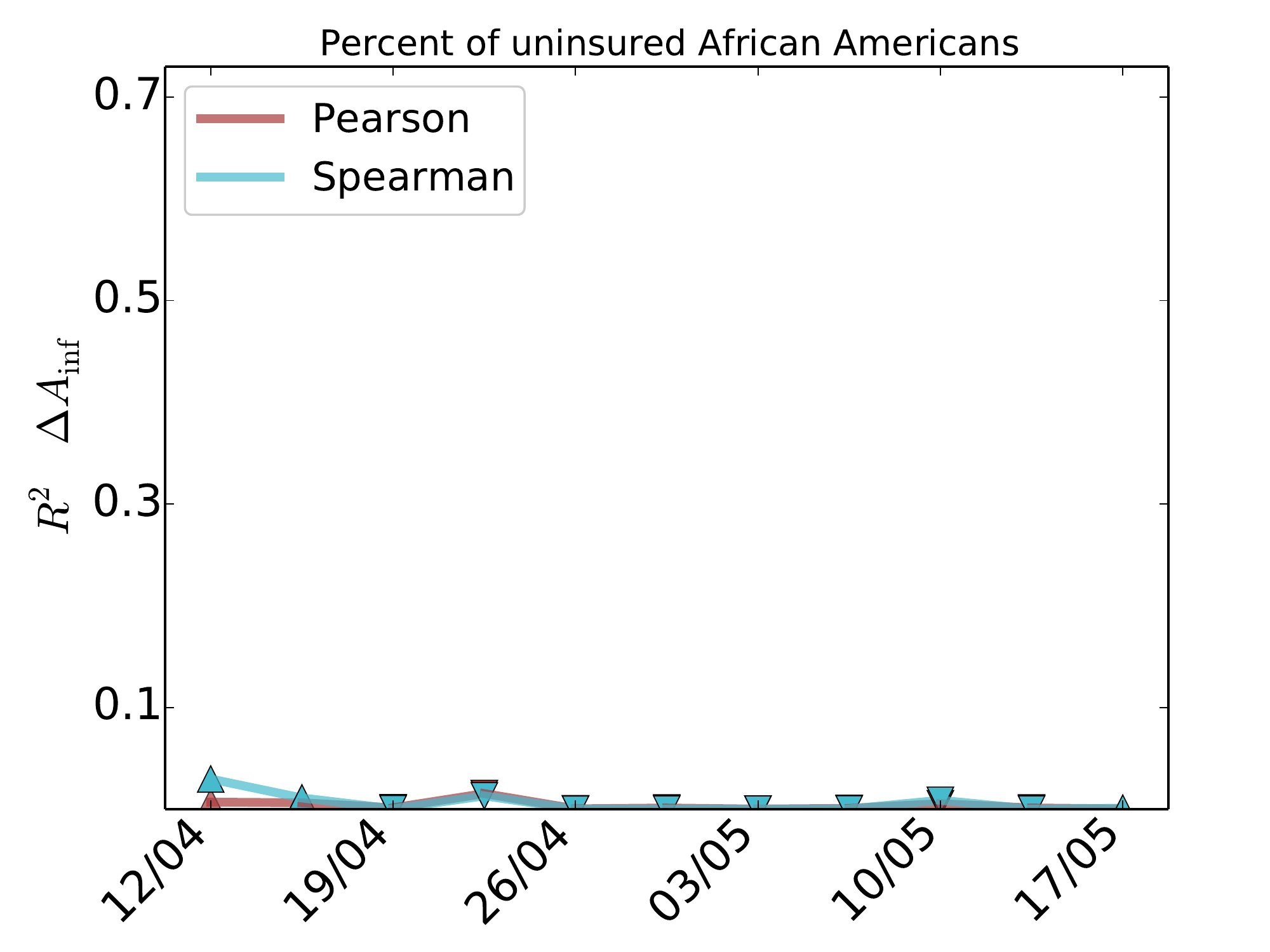}
\includegraphics[width=5.9cm]{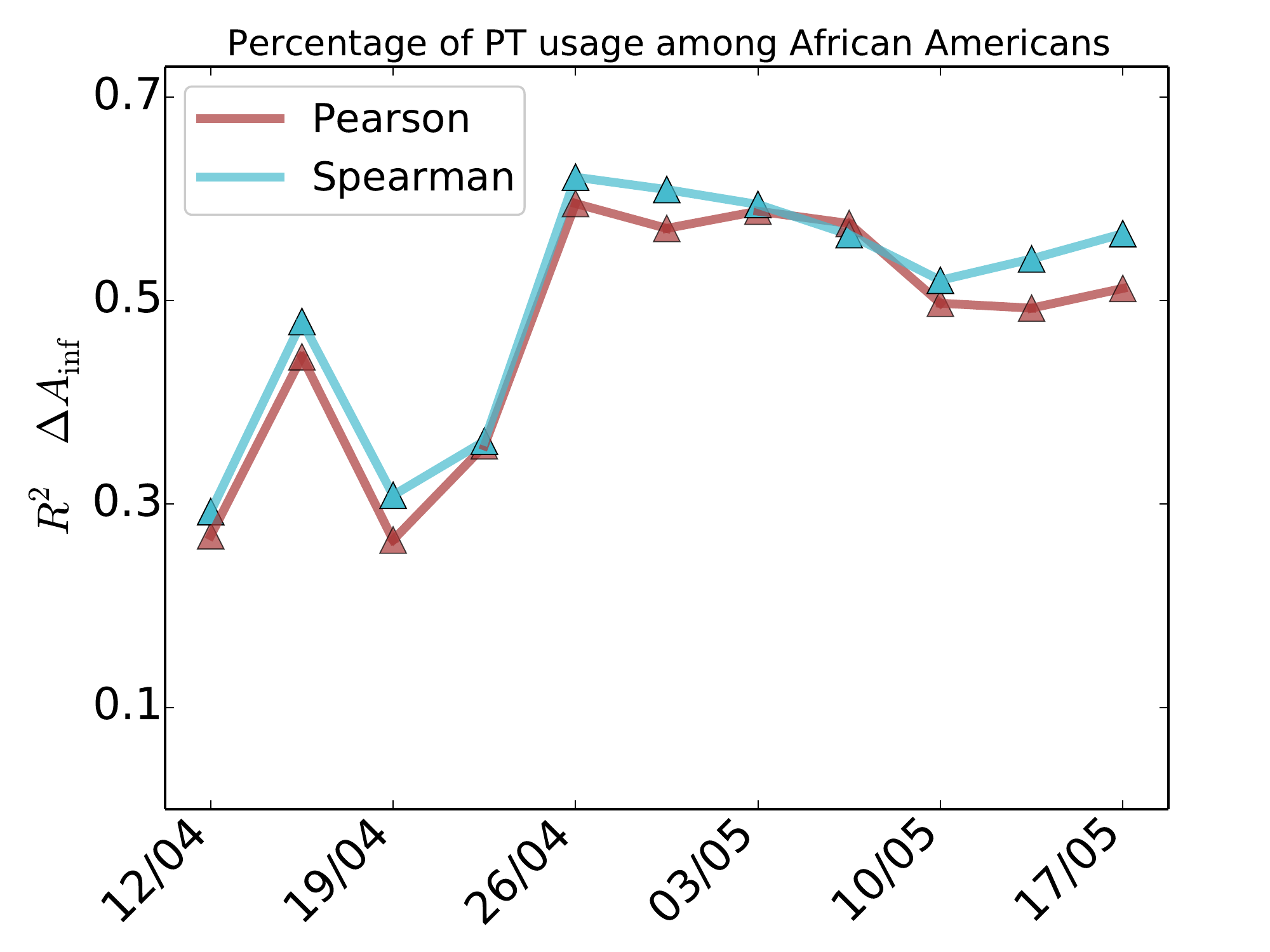}
\includegraphics[width=5.9cm]{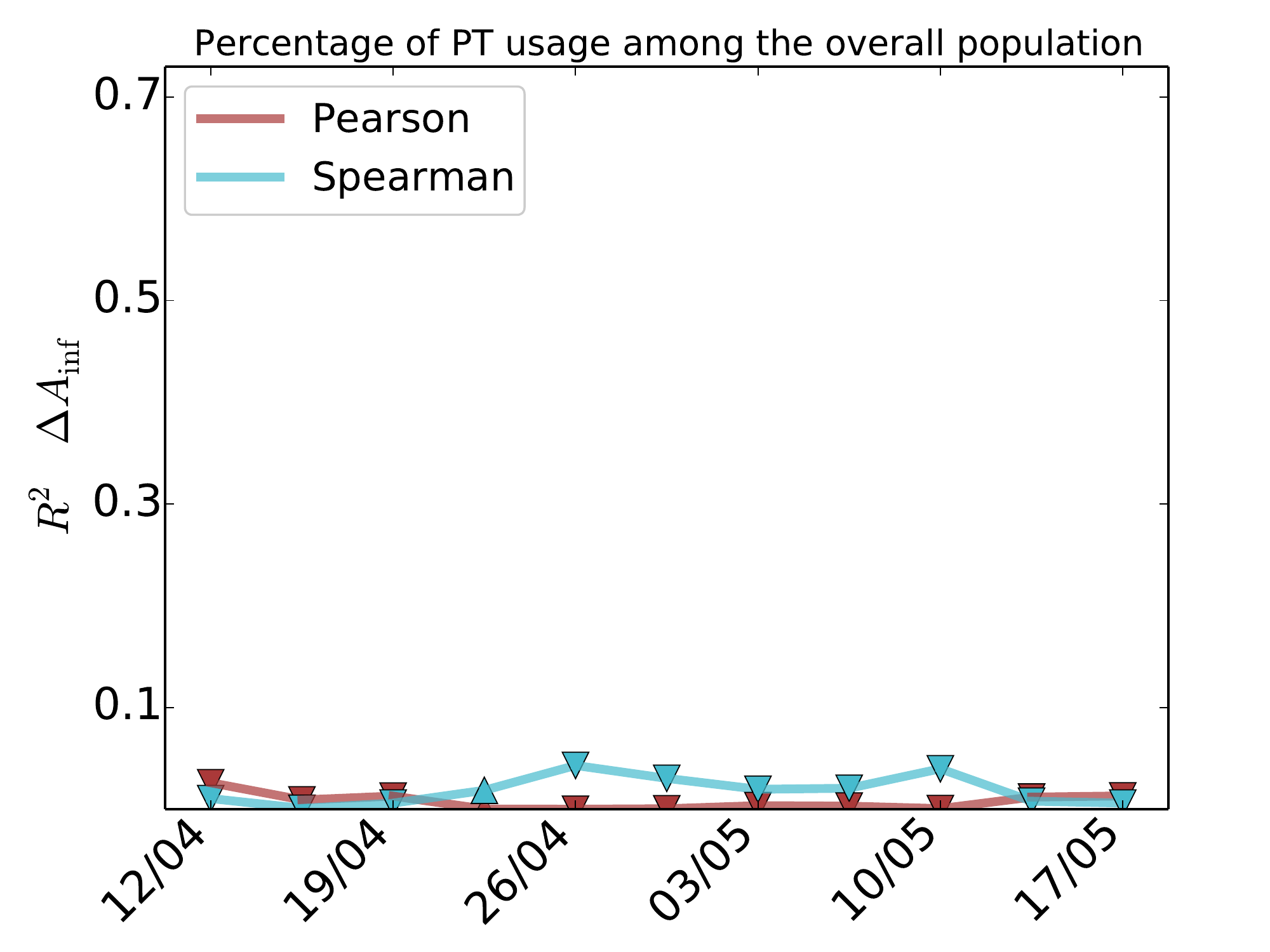}
\caption[Correlations between the incidence of COVID-19 in African American population and a set of socio-economic indicators]{\textbf{The temporal evolution of Pearson and Spearman correlations ($R^2$)  between the incidence of COVID-19 in African American population and a set of socio-economic indicators.} On the top row and from left to right we have the median household income of African Americans, the percentage of African Americans below the poverty level and the percent of insured African Americans. On the bottom row and from left to right there is the percent of uninsured African Americans, the percentage of use of public transportation among African Americans and the percentage of use among the overall population. The markers indicate the sign of the relation, positive for triangles pointing up and negative for triangles
  pointing down.
\label{socioeconomic}}
\end{center}
\end{figure}

\begin{figure}[!htbp]
\begin{center}
\includegraphics[width=5.9cm]{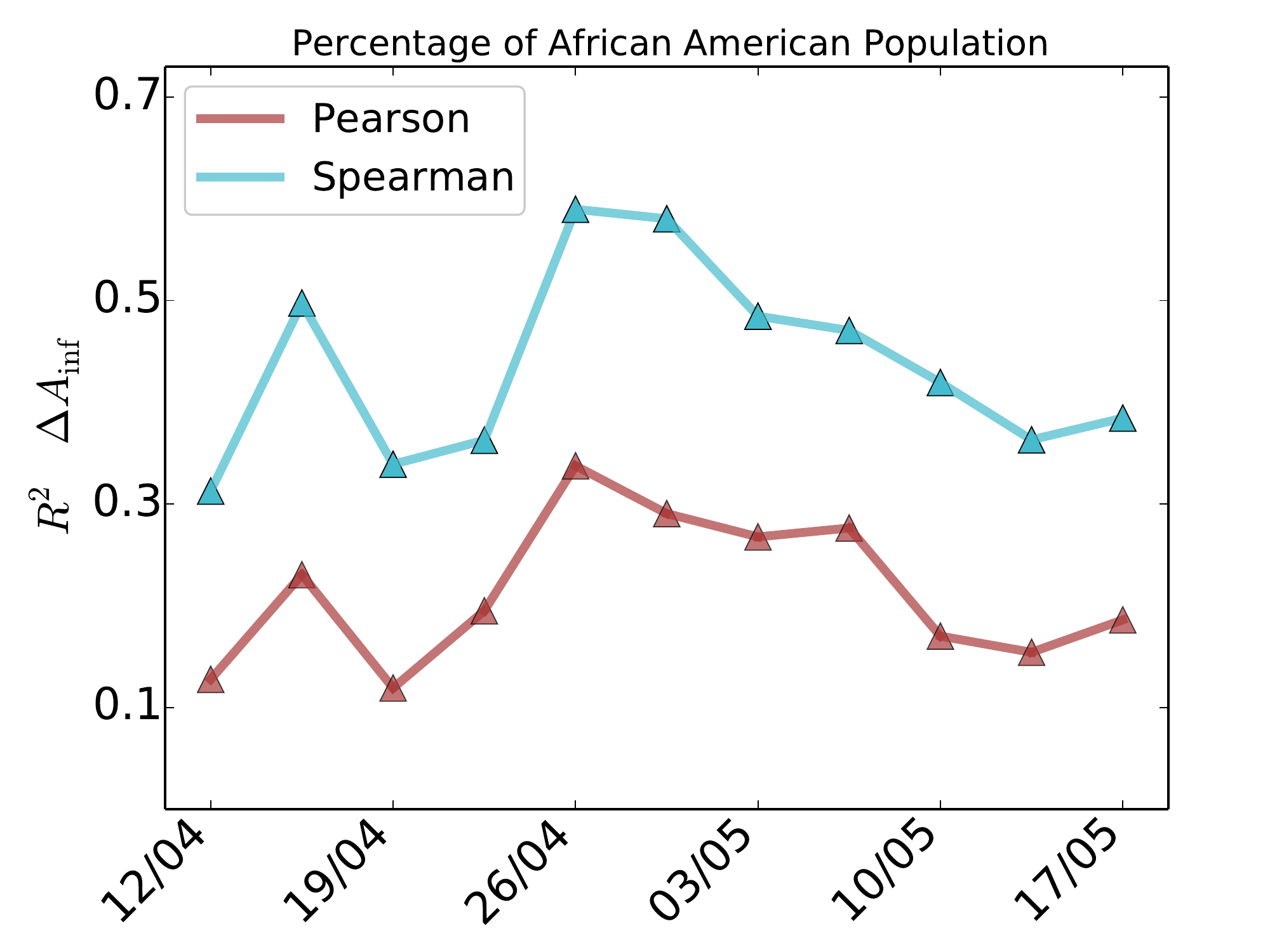}
\includegraphics[width=5.9cm]{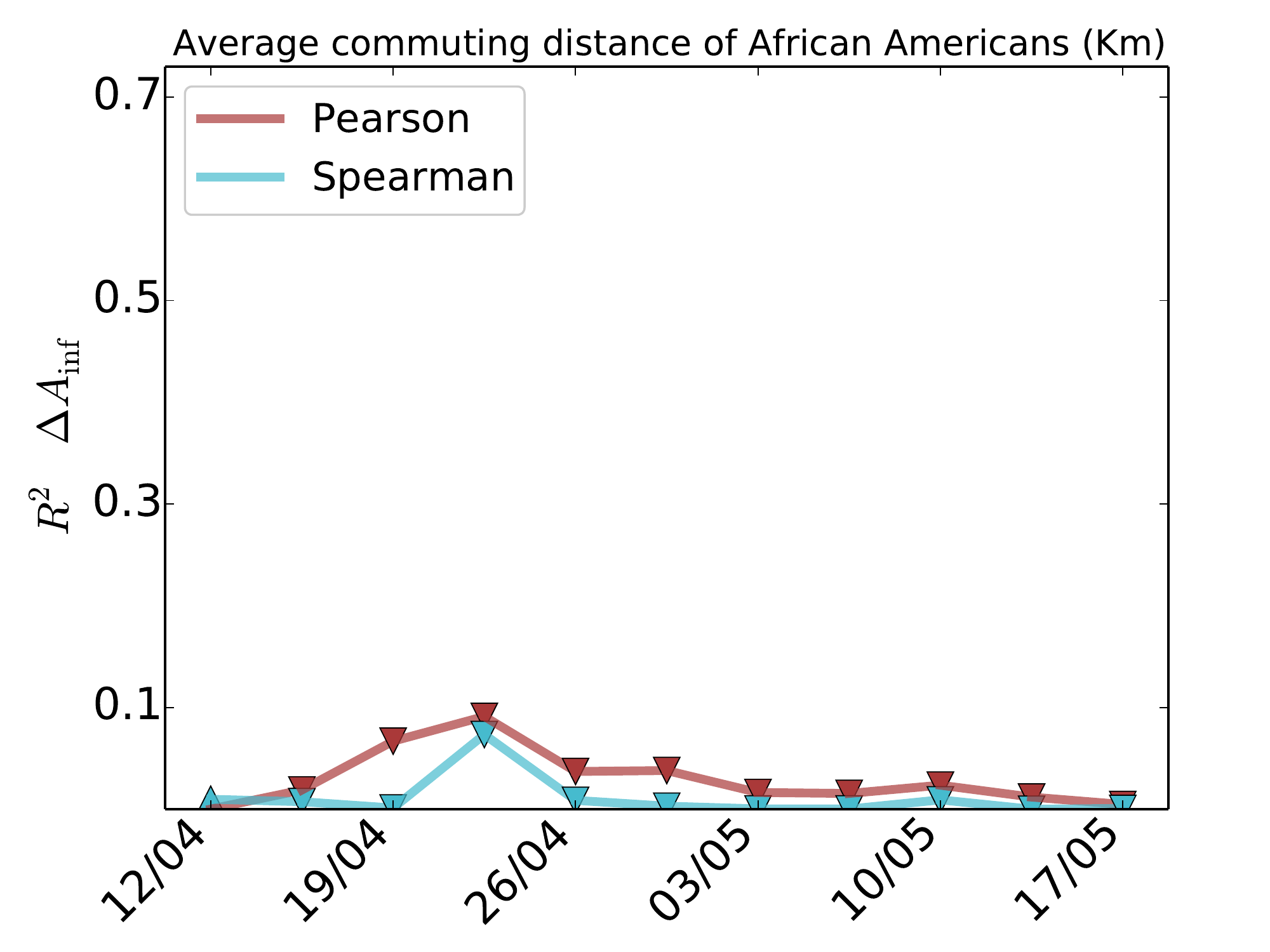}
\includegraphics[width=5.9cm]{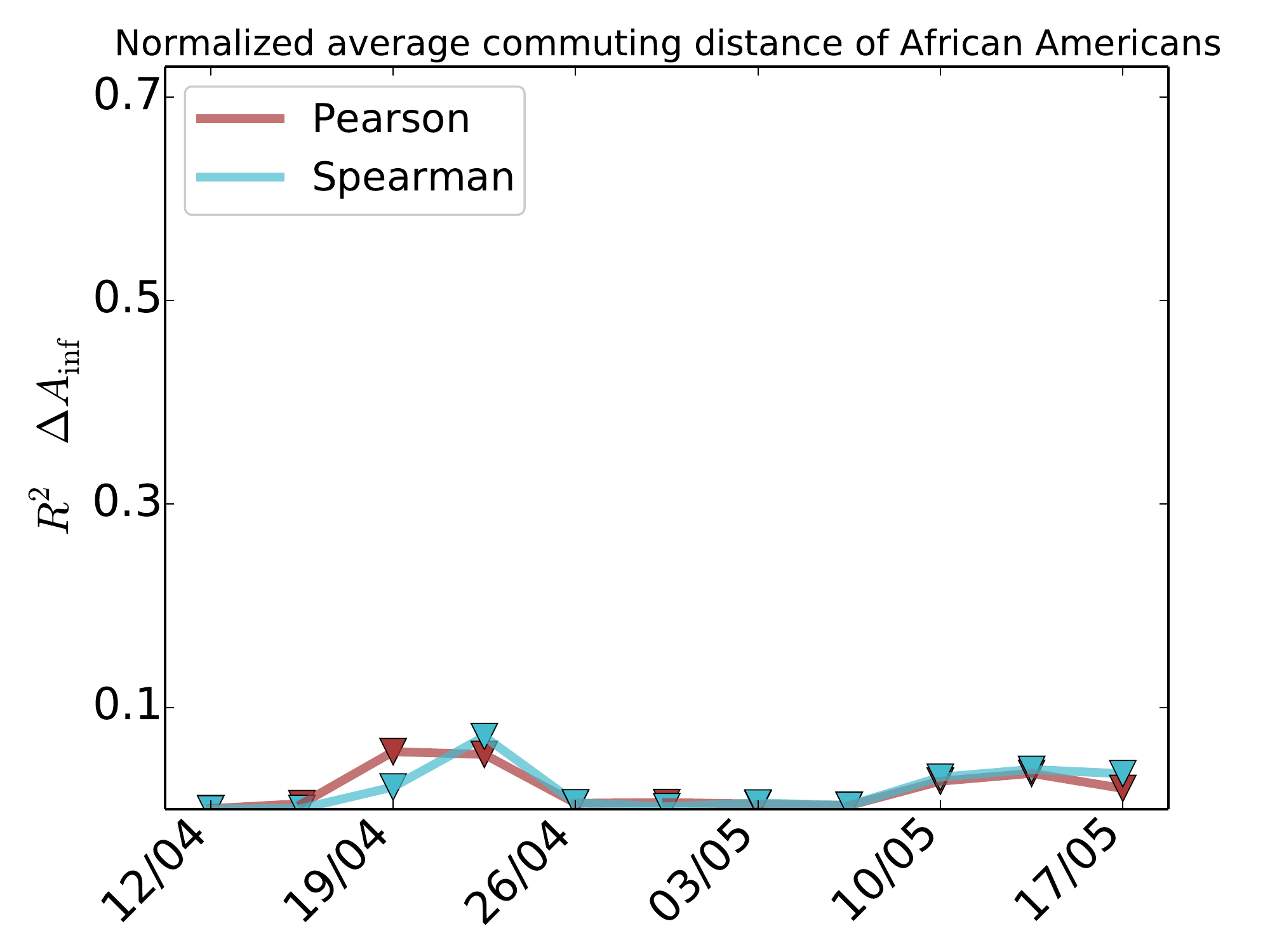}
\caption[Correlations between the incidence of COVID-19 in African American population and a set of population and mobility indicators]{\textbf{Correlations between the incidence of COVID-19 in African American population and a set of population and mobility indicators.} From left to right the temporal evolution of the Pearson and Spearman $R^2$ for the percentage of African Americans among the population, the average commuting distance of African Americans and its ratio with the commuting distance of the overall population. The markers indicate the sign of the relation, positive for triangles pointing up and negative for triangles
  pointing down.
\label{characteristics}}
\end{center}
\end{figure}

Additionally to those socio-economic variables we also tested if the overall African American population can also be used as a proxy for the difference in percentage. We also computed on our commuting networks the average commuting distance of the African American population as well as the ratio with the commuting distance of the overall population. As displayed in Supplementary Figure \ref{characteristics}, the overall percentage of African American population seems to be related to the difference in the percentage of infected. However, there is a striking difference between the Pearson and the Spearman correlation coefficients, which means that the rank is more or less conserved yet there are strong outliers. In other words, a state with more percentage of African American population will more easily have a higher difference on the infected yet the population does not align the points in a straight trend. Regarding the mobility indicators, none of them yields a significant correlation, meaning that it is not so relevant how far African Americans travel and where they travel and whom they meet.


\clearpage

\begin{thebibliography}{49}%
\makeatletter
\providecommand \@ifxundefined [1]{%
 \@ifx{#1\undefined}
}%
\providecommand \@ifnum [1]{%
 \ifnum #1\expandafter \@firstoftwo
 \else \expandafter \@secondoftwo
 \fi
}%
\providecommand \@ifx [1]{%
 \ifx #1\expandafter \@firstoftwo
 \else \expandafter \@secondoftwo
 \fi
}%
\providecommand \natexlab [1]{#1}%
\providecommand \enquote  [1]{``#1''}%
\providecommand \bibnamefont  [1]{#1}%
\providecommand \bibfnamefont [1]{#1}%
\providecommand \citenamefont [1]{#1}%
\providecommand \href@noop [0]{\@secondoftwo}%
\providecommand \href [0]{\begingroup \@sanitize@url \@href}%
\providecommand \@href[1]{\@@startlink{#1}\@@href}%
\providecommand \@@href[1]{\endgroup#1\@@endlink}%
\providecommand \@sanitize@url [0]{\catcode `\\12\catcode `\$12\catcode
  `\&12\catcode `\#12\catcode `\^12\catcode `\_12\catcode `\%12\relax}%
\providecommand \@@startlink[1]{}%
\providecommand \@@endlink[0]{}%
\providecommand \url  [0]{\begingroup\@sanitize@url \@url }%
\providecommand \@url [1]{\endgroup\@href {#1}{\urlprefix }}%
\providecommand \urlprefix  [0]{URL }%
\providecommand \Eprint [0]{\href }%
\providecommand \doibase [0]{http://dx.doi.org/}%
\providecommand \selectlanguage [0]{\@gobble}%
\providecommand \bibinfo  [0]{\@secondoftwo}%
\providecommand \bibfield  [0]{\@secondoftwo}%
\providecommand \translation [1]{[#1]}%
\providecommand \BibitemOpen [0]{}%
\providecommand \bibitemStop [0]{}%
\providecommand \bibitemNoStop [0]{.\EOS\space}%
\providecommand \EOS [0]{\spacefactor3000\relax}%
\providecommand \BibitemShut  [1]{\csname bibitem#1\endcsname}%
\let\auto@bib@innerbib\@empty
\bibitem [{\citenamefont {CDC}\ and\ \citenamefont {Team}(2020)}]{Covid2020}%
  \BibitemOpen
  \bibfield  {author} {\bibinfo {author} {\bibfnamefont {C.}~\bibnamefont
  {CDC}}\ and\ \bibinfo {author} {\bibfnamefont {R.}~\bibnamefont {Team}},\
  }\href@noop {} {\bibfield  {journal} {\bibinfo  {journal} {MMWR Morb Mortal
  Wkly Rep}\ }\textbf {\bibinfo {volume} {69}},\ \bibinfo {pages} {343}
  (\bibinfo {year} {2020})}\BibitemShut {NoStop}%
\bibitem [{\citenamefont {Millett}\ \emph {et~al.}(2020)\citenamefont
  {Millett}, \citenamefont {Jones}, \citenamefont {Benkeser}, \citenamefont
  {Baral}, \citenamefont {Mercer}, \citenamefont {Beyrer}, \citenamefont
  {Honermann}, \citenamefont {Lankiewicz}, \citenamefont {Mena}, \citenamefont
  {Crowley}, \citenamefont {Sherwood},\ and\ \citenamefont
  {Sullivan}}]{Millett2020}%
  \BibitemOpen
  \bibfield  {author} {\bibinfo {author} {\bibfnamefont {G.~A.}\ \bibnamefont
  {Millett}}, \bibinfo {author} {\bibfnamefont {A.~T.}\ \bibnamefont {Jones}},
  \bibinfo {author} {\bibfnamefont {D.}~\bibnamefont {Benkeser}}, \bibinfo
  {author} {\bibfnamefont {S.}~\bibnamefont {Baral}}, \bibinfo {author}
  {\bibfnamefont {L.}~\bibnamefont {Mercer}}, \bibinfo {author} {\bibfnamefont
  {C.}~\bibnamefont {Beyrer}}, \bibinfo {author} {\bibfnamefont
  {B.}~\bibnamefont {Honermann}}, \bibinfo {author} {\bibfnamefont
  {E.}~\bibnamefont {Lankiewicz}}, \bibinfo {author} {\bibfnamefont
  {L.}~\bibnamefont {Mena}}, \bibinfo {author} {\bibfnamefont {J.~S.}\
  \bibnamefont {Crowley}}, \bibinfo {author} {\bibfnamefont {J.}~\bibnamefont
  {Sherwood}}, \ and\ \bibinfo {author} {\bibfnamefont {P.}~\bibnamefont
  {Sullivan}},\ }\href {\doibase
  https://doi.org/10.1016/j.annepidem.2020.05.003} {\bibfield  {journal}
  {\bibinfo  {journal} {Annals of Epidemiology}\ } (\bibinfo {year} {2020}),\
  https://doi.org/10.1016/j.annepidem.2020.05.003}\BibitemShut {NoStop}%
\bibitem [{\citenamefont {DiMaggio}\ \emph {et~al.}(2020)\citenamefont
  {DiMaggio}, \citenamefont {Klein}, \citenamefont {Berry},\ and\ \citenamefont
  {Frangos}}]{DiMaggio2020}%
  \BibitemOpen
  \bibfield  {author} {\bibinfo {author} {\bibfnamefont {C.}~\bibnamefont
  {DiMaggio}}, \bibinfo {author} {\bibfnamefont {M.}~\bibnamefont {Klein}},
  \bibinfo {author} {\bibfnamefont {C.}~\bibnamefont {Berry}}, \ and\ \bibinfo
  {author} {\bibfnamefont {S.}~\bibnamefont {Frangos}},\ }\href {\doibase
  10.1101/2020.05.14.20101691} {\bibfield  {journal} {\bibinfo  {journal}
  {medRxiv}\ } (\bibinfo {year} {2020}),\ 10.1101/2020.05.14.20101691},\
  \BibitemShut {NoStop}%
\bibitem [{\citenamefont {Goldstein}\ and\ \citenamefont
  {Atherwood}(2020)}]{Goldstein2020}%
  \BibitemOpen
  \bibfield  {author} {\bibinfo {author} {\bibfnamefont {J.~R.}\ \bibnamefont
  {Goldstein}}\ and\ \bibinfo {author} {\bibfnamefont {S.}~\bibnamefont
  {Atherwood}},\ }\href {\doibase 10.1101/2020.05.21.20109116} {\bibfield
  {journal} {\bibinfo  {journal} {medRxiv}\ } (\bibinfo {year} {2020}),\
  10.1101/2020.05.21.20109116},\
  \BibitemShut {NoStop}%
\bibitem [{\citenamefont {Cyrus}\ \emph {et~al.}(2020)\citenamefont {Cyrus},
  \citenamefont {Clarke}, \citenamefont {Hadley}, \citenamefont {Bursac},
  \citenamefont {Trepka}, \citenamefont {Devieux}, \citenamefont {Bagci},
  \citenamefont {Furr-Holden}, \citenamefont {Coudray}, \citenamefont
  {Mariano}, \citenamefont {Kiplagat}, \citenamefont {Noel}, \citenamefont
  {Ravelo}, \citenamefont {Paley},\ and\ \citenamefont {Wagner}}]{Cyrus2020}%
  \BibitemOpen
  \bibfield  {author} {\bibinfo {author} {\bibfnamefont {E.}~\bibnamefont
  {Cyrus}}, \bibinfo {author} {\bibfnamefont {R.}~\bibnamefont {Clarke}},
  \bibinfo {author} {\bibfnamefont {D.}~\bibnamefont {Hadley}}, \bibinfo
  {author} {\bibfnamefont {Z.}~\bibnamefont {Bursac}}, \bibinfo {author}
  {\bibfnamefont {M.~J.}\ \bibnamefont {Trepka}}, \bibinfo {author}
  {\bibfnamefont {J.~G.}\ \bibnamefont {Devieux}}, \bibinfo {author}
  {\bibfnamefont {U.}~\bibnamefont {Bagci}}, \bibinfo {author} {\bibfnamefont
  {D.}~\bibnamefont {Furr-Holden}}, \bibinfo {author} {\bibfnamefont {M.~S.}\
  \bibnamefont {Coudray}}, \bibinfo {author} {\bibfnamefont {Y.}~\bibnamefont
  {Mariano}}, \bibinfo {author} {\bibfnamefont {S.}~\bibnamefont {Kiplagat}},
  \bibinfo {author} {\bibfnamefont {I.}~\bibnamefont {Noel}}, \bibinfo {author}
  {\bibfnamefont {G.}~\bibnamefont {Ravelo}}, \bibinfo {author} {\bibfnamefont
  {M.}~\bibnamefont {Paley}}, \ and\ \bibinfo {author} {\bibfnamefont {E.~F.}\
  \bibnamefont {Wagner}},\ }\href {\doibase 10.1101/2020.05.15.20096552}
  {\bibfield  {journal} {\bibinfo  {journal} {medRxiv}\ } (\bibinfo {year}
  {2020}),\ 10.1101/2020.05.15.20096552},\
  \BibitemShut {NoStop}%
\bibitem [{\citenamefont {England}(2020)}]{UKCovidDisparity2020}%
  \BibitemOpen
  \bibfield  {author} {\bibinfo {author} {\bibfnamefont {P.~H.}\ \bibnamefont
  {England}},\ }\href
  {https://www.gov.uk/government/publications/covid-19-review-of-disparities-in-risks-and-outcomes}
  {\emph {\bibinfo {title} {COVID-19: review of disparities in risks and
  outcomes}}},\ \bibinfo {type} {Tech. Rep.}\ (\bibinfo  {institution} {UK
  Government},\ \bibinfo {year} {2020})\BibitemShut {NoStop}%
\bibitem [{\citenamefont {Raisi-Estabragh}\ \emph {et~al.}(2020)\citenamefont
  {Raisi-Estabragh}, \citenamefont {McCracken}, \citenamefont {Bethell},
  \citenamefont {Cooper}, \citenamefont {Cooper}, \citenamefont {Caulfield},
  \citenamefont {Munroe}, \citenamefont {Harvey},\ and\ \citenamefont
  {Petersen}}]{Raisi-Estabragh2020}%
  \BibitemOpen
  \bibfield  {author} {\bibinfo {author} {\bibfnamefont {Z.}~\bibnamefont
  {Raisi-Estabragh}}, \bibinfo {author} {\bibfnamefont {C.}~\bibnamefont
  {McCracken}}, \bibinfo {author} {\bibfnamefont {M.~S.}\ \bibnamefont
  {Bethell}}, \bibinfo {author} {\bibfnamefont {J.}~\bibnamefont {Cooper}},
  \bibinfo {author} {\bibfnamefont {C.}~\bibnamefont {Cooper}}, \bibinfo
  {author} {\bibfnamefont {M.~J.}\ \bibnamefont {Caulfield}}, \bibinfo {author}
  {\bibfnamefont {P.~B.}\ \bibnamefont {Munroe}}, \bibinfo {author}
  {\bibfnamefont {N.~C.}\ \bibnamefont {Harvey}}, \ and\ \bibinfo {author}
  {\bibfnamefont {S.~E.}\ \bibnamefont {Petersen}},\ }\href {\doibase
  10.1101/2020.06.01.20118943} {\bibfield  {journal} {\bibinfo  {journal}
  {medRxiv}\ } (\bibinfo {year} {2020}),\ 10.1101/2020.06.01.20118943},\
  \BibitemShut {NoStop}%
\bibitem [{\citenamefont {Li}\ \emph {et~al.}(2020)\citenamefont {Li},
  \citenamefont {Hannah}, \citenamefont {Durbin}, \citenamefont {Dreher},
  \citenamefont {McAuley}, \citenamefont {Marayati}, \citenamefont {Spiera},
  \citenamefont {Ali}, \citenamefont {Gometz}, \citenamefont {Kostman},\ and\
  \citenamefont {Choudhri}}]{Li2020}%
  \BibitemOpen
  \bibfield  {author} {\bibinfo {author} {\bibfnamefont {A.~Y.}\ \bibnamefont
  {Li}}, \bibinfo {author} {\bibfnamefont {T.~C.}\ \bibnamefont {Hannah}},
  \bibinfo {author} {\bibfnamefont {J.}~\bibnamefont {Durbin}}, \bibinfo
  {author} {\bibfnamefont {N.}~\bibnamefont {Dreher}}, \bibinfo {author}
  {\bibfnamefont {F.~M.}\ \bibnamefont {McAuley}}, \bibinfo {author}
  {\bibfnamefont {N.~F.}\ \bibnamefont {Marayati}}, \bibinfo {author}
  {\bibfnamefont {Z.}~\bibnamefont {Spiera}}, \bibinfo {author} {\bibfnamefont
  {M.}~\bibnamefont {Ali}}, \bibinfo {author} {\bibfnamefont {A.}~\bibnamefont
  {Gometz}}, \bibinfo {author} {\bibfnamefont {J.}~\bibnamefont {Kostman}}, \
  and\ \bibinfo {author} {\bibfnamefont {T.~F.}\ \bibnamefont {Choudhri}},\
  }\href {\doibase 10.1101/2020.04.17.20069708} {\bibfield  {journal} {\bibinfo
   {journal} {medRxiv}\ } (\bibinfo {year} {2020}),\
  10.1101/2020.04.17.20069708},\ 
  \BibitemShut {NoStop}%
\bibitem [{\citenamefont {Prats-Uribe}\ \emph {et~al.}(2020)\citenamefont
  {Prats-Uribe}, \citenamefont {Paredes},\ and\ \citenamefont
  {PRIETO-ALHAMBRA}}]{Prats-Uribe2020}%
  \BibitemOpen
  \bibfield  {author} {\bibinfo {author} {\bibfnamefont {A.}~\bibnamefont
  {Prats-Uribe}}, \bibinfo {author} {\bibfnamefont {R.}~\bibnamefont
  {Paredes}}, \ and\ \bibinfo {author} {\bibfnamefont {D.}~\bibnamefont
  {PRIETO-ALHAMBRA}},\ }\href {\doibase 10.1101/2020.05.06.20092676} {\bibfield
   {journal} {\bibinfo  {journal} {medRxiv}\ } (\bibinfo {year} {2020}),\
  10.1101/2020.05.06.20092676},\ 
  \BibitemShut {NoStop}%
\bibitem [{\citenamefont {Takagi}\ \emph {et~al.}(2020)\citenamefont {Takagi},
  \citenamefont {Kuno}, \citenamefont {Yokoyama}, \citenamefont {Ueyama},
  \citenamefont {Matsushiro}, \citenamefont {Hari},\ and\ \citenamefont
  {Ando}}]{Takagi2020}%
  \BibitemOpen
  \bibfield  {author} {\bibinfo {author} {\bibfnamefont {H.}~\bibnamefont
  {Takagi}}, \bibinfo {author} {\bibfnamefont {T.}~\bibnamefont {Kuno}},
  \bibinfo {author} {\bibfnamefont {Y.}~\bibnamefont {Yokoyama}}, \bibinfo
  {author} {\bibfnamefont {H.}~\bibnamefont {Ueyama}}, \bibinfo {author}
  {\bibfnamefont {T.}~\bibnamefont {Matsushiro}}, \bibinfo {author}
  {\bibfnamefont {Y.}~\bibnamefont {Hari}}, \ and\ \bibinfo {author}
  {\bibfnamefont {T.}~\bibnamefont {Ando}},\ }\href {\doibase
  10.1101/2020.05.25.20112599} {\bibfield  {journal} {\bibinfo  {journal}
  {medRxiv}\ } (\bibinfo {year} {2020}),\ 10.1101/2020.05.25.20112599},\
  \BibitemShut {NoStop}%
\bibitem [{\citenamefont {Khanijahani}(2020)}]{Khanijahani2020}%
  \BibitemOpen
  \bibfield  {author} {\bibinfo {author} {\bibfnamefont {A.}~\bibnamefont
  {Khanijahani}},\ }\href {\doibase 10.1101/2020.06.03.20120667} {\bibfield
  {journal} {\bibinfo  {journal} {medRxiv}\ } (\bibinfo {year} {2020}),\
  10.1101/2020.06.03.20120667},\ 
  \BibitemShut {NoStop}%
\bibitem [{\citenamefont {Bilal}\ \emph {et~al.}(2020)\citenamefont {Bilal},
  \citenamefont {Barber},\ and\ \citenamefont {Diez-Roux}}]{Bilal2020}%
  \BibitemOpen
  \bibfield  {author} {\bibinfo {author} {\bibfnamefont {U.}~\bibnamefont
  {Bilal}}, \bibinfo {author} {\bibfnamefont {S.}~\bibnamefont {Barber}}, \
  and\ \bibinfo {author} {\bibfnamefont {A.~V.}\ \bibnamefont {Diez-Roux}},\
  }\href {\doibase 10.1101/2020.05.01.20087833} {\bibfield  {journal} {\bibinfo
   {journal} {medRxiv}\ } (\bibinfo {year} {2020}),\
  10.1101/2020.05.01.20087833},\ 
  \BibitemShut {NoStop}%
\bibitem [{\citenamefont {Chinazzi}\ \emph {et~al.}(2020)\citenamefont
  {Chinazzi}, \citenamefont {Davis}, \citenamefont {Ajelli}, \citenamefont
  {Gioannini}, \citenamefont {Litvinova}, \citenamefont {Merler}, \citenamefont
  {y~Piontti}, \citenamefont {Mu}, \citenamefont {Rossi}, \citenamefont {Sun}
  \emph {et~al.}}]{Chinazzi2020}%
  \BibitemOpen
  \bibfield  {author} {\bibinfo {author} {\bibfnamefont {M.}~\bibnamefont
  {Chinazzi}}, \bibinfo {author} {\bibfnamefont {J.~T.}\ \bibnamefont {Davis}},
  \bibinfo {author} {\bibfnamefont {M.}~\bibnamefont {Ajelli}}, \bibinfo
  {author} {\bibfnamefont {C.}~\bibnamefont {Gioannini}}, \bibinfo {author}
  {\bibfnamefont {M.}~\bibnamefont {Litvinova}}, \bibinfo {author}
  {\bibfnamefont {S.}~\bibnamefont {Merler}}, \bibinfo {author} {\bibfnamefont
  {A.~P.}\ \bibnamefont {y~Piontti}}, \bibinfo {author} {\bibfnamefont
  {K.}~\bibnamefont {Mu}}, \bibinfo {author} {\bibfnamefont {L.}~\bibnamefont
  {Rossi}}, \bibinfo {author} {\bibfnamefont {K.}~\bibnamefont {Sun}},  \emph
  {et~al.},\ }\href@noop {} {\bibfield  {journal} {\bibinfo  {journal}
  {Science}\ }\textbf {\bibinfo {volume} {368}},\ \bibinfo {pages} {395}
  (\bibinfo {year} {2020})}\BibitemShut {NoStop}%
\bibitem [{\citenamefont {Buckee}\ \emph {et~al.}(2020)\citenamefont {Buckee},
  \citenamefont {Balsari}, \citenamefont {Chan}, \citenamefont {Crosas},
  \citenamefont {Dominici}, \citenamefont {Gasser}, \citenamefont {Grad},
  \citenamefont {Grenfell}, \citenamefont {Halloran}, \citenamefont {Kraemer}
  \emph {et~al.}}]{Buckee2020}%
  \BibitemOpen
  \bibfield  {author} {\bibinfo {author} {\bibfnamefont {C.~O.}\ \bibnamefont
  {Buckee}}, \bibinfo {author} {\bibfnamefont {S.}~\bibnamefont {Balsari}},
  \bibinfo {author} {\bibfnamefont {J.}~\bibnamefont {Chan}}, \bibinfo {author}
  {\bibfnamefont {M.}~\bibnamefont {Crosas}}, \bibinfo {author} {\bibfnamefont
  {F.}~\bibnamefont {Dominici}}, \bibinfo {author} {\bibfnamefont
  {U.}~\bibnamefont {Gasser}}, \bibinfo {author} {\bibfnamefont {Y.~H.}\
  \bibnamefont {Grad}}, \bibinfo {author} {\bibfnamefont {B.}~\bibnamefont
  {Grenfell}}, \bibinfo {author} {\bibfnamefont {M.~E.}\ \bibnamefont
  {Halloran}}, \bibinfo {author} {\bibfnamefont {M.~U.}\ \bibnamefont
  {Kraemer}},  \emph {et~al.},\ }\href@noop {} {\bibfield  {journal} {\bibinfo
  {journal} {Science (New York, NY)}\ }\textbf {\bibinfo {volume} {368}},\
  \bibinfo {pages} {145} (\bibinfo {year} {2020})}\BibitemShut {NoStop}%
\bibitem [{\citenamefont {Kraemer}\ \emph {et~al.}(2020)\citenamefont
  {Kraemer}, \citenamefont {Yang}, \citenamefont {Gutierrez}, \citenamefont
  {Wu}, \citenamefont {Klein}, \citenamefont {Pigott}, \citenamefont
  {Du~Plessis}, \citenamefont {Faria}, \citenamefont {Li}, \citenamefont
  {Hanage} \emph {et~al.}}]{Kraemer2020}%
  \BibitemOpen
  \bibfield  {author} {\bibinfo {author} {\bibfnamefont {M.~U.}\ \bibnamefont
  {Kraemer}}, \bibinfo {author} {\bibfnamefont {C.-H.}\ \bibnamefont {Yang}},
  \bibinfo {author} {\bibfnamefont {B.}~\bibnamefont {Gutierrez}}, \bibinfo
  {author} {\bibfnamefont {C.-H.}\ \bibnamefont {Wu}}, \bibinfo {author}
  {\bibfnamefont {B.}~\bibnamefont {Klein}}, \bibinfo {author} {\bibfnamefont
  {D.~M.}\ \bibnamefont {Pigott}}, \bibinfo {author} {\bibfnamefont
  {L.}~\bibnamefont {Du~Plessis}}, \bibinfo {author} {\bibfnamefont {N.~R.}\
  \bibnamefont {Faria}}, \bibinfo {author} {\bibfnamefont {R.}~\bibnamefont
  {Li}}, \bibinfo {author} {\bibfnamefont {W.~P.}\ \bibnamefont {Hanage}},
  \emph {et~al.},\ }\href@noop {} {\bibfield  {journal} {\bibinfo  {journal}
  {Science}\ }\textbf {\bibinfo {volume} {368}},\ \bibinfo {pages} {493}
  (\bibinfo {year} {2020})}\BibitemShut {NoStop}%
\bibitem [{\citenamefont {Aleta}\ \emph {et~al.}(2020)\citenamefont {Aleta},
  \citenamefont {Martin-Corral}, \citenamefont {y~Piontti}, \citenamefont
  {Ajelli}, \citenamefont {Litvinova}, \citenamefont {Chinazzi}, \citenamefont
  {Dean}, \citenamefont {Halloran}, \citenamefont {Longini~Jr}, \citenamefont
  {Merler} \emph {et~al.}}]{Aleta2020}%
  \BibitemOpen
  \bibfield  {author} {\bibinfo {author} {\bibfnamefont {A.}~\bibnamefont
  {Aleta}}, \bibinfo {author} {\bibfnamefont {D.}~\bibnamefont
  {Martin-Corral}}, \bibinfo {author} {\bibfnamefont {A.~P.}\ \bibnamefont
  {y~Piontti}}, \bibinfo {author} {\bibfnamefont {M.}~\bibnamefont {Ajelli}},
  \bibinfo {author} {\bibfnamefont {M.}~\bibnamefont {Litvinova}}, \bibinfo
  {author} {\bibfnamefont {M.}~\bibnamefont {Chinazzi}}, \bibinfo {author}
  {\bibfnamefont {N.~E.}\ \bibnamefont {Dean}}, \bibinfo {author}
  {\bibfnamefont {M.~E.}\ \bibnamefont {Halloran}}, \bibinfo {author}
  {\bibfnamefont {I.~M.}\ \bibnamefont {Longini~Jr}}, \bibinfo {author}
  {\bibfnamefont {S.}~\bibnamefont {Merler}},  \emph {et~al.},\ }\href@noop {}
  {\bibfield  {journal} {\bibinfo  {journal} {medRxiv}\ } (\bibinfo {year}
  {2020})}\BibitemShut {NoStop}%
\bibitem [{\citenamefont {Brown}\ \emph {et~al.}(2013)\citenamefont {Brown},
  \citenamefont {Nicosia}, \citenamefont {Scellato}, \citenamefont {Noulas},\
  and\ \citenamefont {Mascolo}}]{Brown2013}%
  \BibitemOpen
  \bibfield  {author} {\bibinfo {author} {\bibfnamefont {C.}~\bibnamefont
  {Brown}}, \bibinfo {author} {\bibfnamefont {V.}~\bibnamefont {Nicosia}},
  \bibinfo {author} {\bibfnamefont {S.}~\bibnamefont {Scellato}}, \bibinfo
  {author} {\bibfnamefont {A.}~\bibnamefont {Noulas}}, \ and\ \bibinfo {author}
  {\bibfnamefont {C.}~\bibnamefont {Mascolo}},\ }\href {\doibase
  10.1140/epjb/e2013-40253-6} {\bibfield  {journal} {\bibinfo  {journal} {The
  European Physical Journal B}\ }\textbf {\bibinfo {volume} {86}},\ \bibinfo
  {pages} {290} (\bibinfo {year} {2013})}\BibitemShut {NoStop}%
\bibitem [{\citenamefont {Galvani}\ and\ \citenamefont
  {May}(2005)}]{Galvani2005}%
  \BibitemOpen
  \bibfield  {author} {\bibinfo {author} {\bibfnamefont {A.~P.}\ \bibnamefont
  {Galvani}}\ and\ \bibinfo {author} {\bibfnamefont {R.~M.}\ \bibnamefont
  {May}},\ }\href {\doibase 10.1038/438293a} {\bibfield  {journal} {\bibinfo
  {journal} {Nature}\ }\textbf {\bibinfo {volume} {438}},\ \bibinfo {pages}
  {293} (\bibinfo {year} {2005})}\BibitemShut {NoStop}%
\bibitem [{\citenamefont {Stein}(2011)}]{Stein2011}%
  \BibitemOpen
  \bibfield  {author} {\bibinfo {author} {\bibfnamefont {R.~A.}\ \bibnamefont
  {Stein}},\ }\href {\doibase 10.1016/j.ijid.2010.06.020} {\bibfield  {journal}
  {\bibinfo  {journal} {International Journal of Infectious Diseases}\ }\textbf
  {\bibinfo {volume} {15}},\ \bibinfo {pages} {E510} (\bibinfo {year}
  {2011})}\BibitemShut {NoStop}%
\bibitem [{\citenamefont {Pareek}\ \emph {et~al.}(2020)\citenamefont {Pareek},
  \citenamefont {Bangash}, \citenamefont {Pareek}, \citenamefont {Pan},
  \citenamefont {Sze}, \citenamefont {Minhas}, \citenamefont {Hanif},\ and\
  \citenamefont {Khunti}}]{Pareek2020}%
  \BibitemOpen
  \bibfield  {author} {\bibinfo {author} {\bibfnamefont {M.}~\bibnamefont
  {Pareek}}, \bibinfo {author} {\bibfnamefont {M.~N.}\ \bibnamefont {Bangash}},
  \bibinfo {author} {\bibfnamefont {N.}~\bibnamefont {Pareek}}, \bibinfo
  {author} {\bibfnamefont {D.}~\bibnamefont {Pan}}, \bibinfo {author}
  {\bibfnamefont {S.}~\bibnamefont {Sze}}, \bibinfo {author} {\bibfnamefont
  {J.~S.}\ \bibnamefont {Minhas}}, \bibinfo {author} {\bibfnamefont
  {W.}~\bibnamefont {Hanif}}, \ and\ \bibinfo {author} {\bibfnamefont
  {K.}~\bibnamefont {Khunti}},\ }\href@noop {} {\bibfield  {journal} {\bibinfo
  {journal} {The Lancet}\ } (\bibinfo {year} {2020})}\BibitemShut {NoStop}%
\bibitem [{\citenamefont {Laurencin}\ and\ \citenamefont
  {McClinton}(2020)}]{Laurencin2020}%
  \BibitemOpen
  \bibfield  {author} {\bibinfo {author} {\bibfnamefont {C.~T.}\ \bibnamefont
  {Laurencin}}\ and\ \bibinfo {author} {\bibfnamefont {A.}~\bibnamefont
  {McClinton}},\ }\href@noop {} {\bibfield  {journal} {\bibinfo  {journal}
  {Journal of Racial and Ethnic Health Disparities}\ ,\ \bibinfo {pages} {1}}
  (\bibinfo {year} {2020})}\BibitemShut {NoStop}%
\bibitem [{\citenamefont {Yancy}(2020)}]{Yancy2020}%
  \BibitemOpen
  \bibfield  {author} {\bibinfo {author} {\bibfnamefont {C.~W.}\ \bibnamefont
  {Yancy}},\ }\href@noop {} {\bibfield  {journal} {\bibinfo  {journal} {Jama}\
  } (\bibinfo {year} {2020})}\BibitemShut {NoStop}%
\bibitem [{\citenamefont {Dawkins}(2004)}]{Dawkins2004}%
  \BibitemOpen
  \bibfield  {author} {\bibinfo {author} {\bibfnamefont {C.~J.}\ \bibnamefont
  {Dawkins}},\ }\href@noop {} {\bibfield  {journal} {\bibinfo  {journal} {Urban
  Studies}\ }\textbf {\bibinfo {volume} {41}},\ \bibinfo {pages} {833}
  (\bibinfo {year} {2004})}\BibitemShut {NoStop}%
\bibitem [{\citenamefont {Dawkins}(2006)}]{Dawkins2006}%
  \BibitemOpen
  \bibfield  {author} {\bibinfo {author} {\bibfnamefont {C.}~\bibnamefont
  {Dawkins}},\ }\href@noop {} {\bibfield  {journal} {\bibinfo  {journal} {Urban
  Studies}\ }\textbf {\bibinfo {volume} {43}},\ \bibinfo {pages} {1943}
  (\bibinfo {year} {2006})}\BibitemShut {NoStop}%
\bibitem [{\citenamefont {Brown}\ and\ \citenamefont
  {Chung}(2006)}]{Brown2006}%
  \BibitemOpen
  \bibfield  {author} {\bibinfo {author} {\bibfnamefont {L.~A.}\ \bibnamefont
  {Brown}}\ and\ \bibinfo {author} {\bibfnamefont {S.-Y.}\ \bibnamefont
  {Chung}},\ }\href@noop {} {\bibfield  {journal} {\bibinfo  {journal}
  {Population, space and place}\ }\textbf {\bibinfo {volume} {12}},\ \bibinfo
  {pages} {125} (\bibinfo {year} {2006})}\BibitemShut {NoStop}%
\bibitem [{\citenamefont {Cliff}\ and\ \citenamefont {Ord}(1981)}]{Cliff1981}%
  \BibitemOpen
  \bibfield  {author} {\bibinfo {author} {\bibfnamefont {A.~D.}\ \bibnamefont
  {Cliff}}\ and\ \bibinfo {author} {\bibfnamefont {J.~K.}\ \bibnamefont
  {Ord}},\ }\href@noop {} {\emph {\bibinfo {title} {Spatial processes: models
  \& applications}}}\ (\bibinfo  {publisher} {Taylor \& Francis},\ \bibinfo
  {year} {1981})\BibitemShut {NoStop}%
\bibitem [{\citenamefont {Rey}\ and\ \citenamefont {Smith}(2013)}]{Rey2013}%
  \BibitemOpen
  \bibfield  {author} {\bibinfo {author} {\bibfnamefont {S.~J.}\ \bibnamefont
  {Rey}}\ and\ \bibinfo {author} {\bibfnamefont {R.~J.}\ \bibnamefont
  {Smith}},\ }\href@noop {} {\bibfield  {journal} {\bibinfo  {journal} {Letters
  in Spatial and Resource Sciences}\ }\textbf {\bibinfo {volume} {6}},\
  \bibinfo {pages} {55} (\bibinfo {year} {2013})}\BibitemShut {NoStop}%
\bibitem [{\citenamefont {Sethi}\ and\ \citenamefont
  {Somanathan}(2004)}]{Rajiv2004}%
  \BibitemOpen
  \bibfield  {author} {\bibinfo {author} {\bibfnamefont {R.}~\bibnamefont
  {Sethi}}\ and\ \bibinfo {author} {\bibfnamefont {R.}~\bibnamefont
  {Somanathan}},\ }\href {\doibase 10.1086/424742} {\bibfield  {journal}
  {\bibinfo  {journal} {Journal of Political Economy}\ }\textbf {\bibinfo
  {volume} {112}},\ \bibinfo {pages} {1296} (\bibinfo {year} {2004})},\ \BibitemShut {NoStop}%
\bibitem [{\citenamefont {Hendryx}\ and\ \citenamefont
  {Luo}(2020)}]{Hendryx2020}%
  \BibitemOpen
  \bibfield  {author} {\bibinfo {author} {\bibfnamefont {M.}~\bibnamefont
  {Hendryx}}\ and\ \bibinfo {author} {\bibfnamefont {J.}~\bibnamefont {Luo}},\
  }\href@noop {} {\bibfield  {journal} {\bibinfo  {journal} {Available at SSRN
  3582857}\ } (\bibinfo {year} {2020})}\BibitemShut {NoStop}%
\bibitem [{\citenamefont {Wong}\ and\ \citenamefont {Shaw}(2011)}]{Wong2011}%
  \BibitemOpen
  \bibfield  {author} {\bibinfo {author} {\bibfnamefont {D.~W.}\ \bibnamefont
  {Wong}}\ and\ \bibinfo {author} {\bibfnamefont {S.-L.}\ \bibnamefont
  {Shaw}},\ }\href@noop {} {\bibfield  {journal} {\bibinfo  {journal} {Journal
  of geographical systems}\ }\textbf {\bibinfo {volume} {13}},\ \bibinfo
  {pages} {127} (\bibinfo {year} {2011})}\BibitemShut {NoStop}%
\bibitem [{\citenamefont {Nicosia}\ \emph {et~al.}(2014)\citenamefont
  {Nicosia}, \citenamefont {Domenico},\ and\ \citenamefont
  {Latora}}]{Nicosia2014}%
  \BibitemOpen
  \bibfield  {author} {\bibinfo {author} {\bibfnamefont {V.}~\bibnamefont
  {Nicosia}}, \bibinfo {author} {\bibfnamefont {M.~D.}\ \bibnamefont
  {Domenico}}, \ and\ \bibinfo {author} {\bibfnamefont {V.}~\bibnamefont
  {Latora}},\ }\href {\doibase 10.1209/0295-5075/106/58005} {\bibfield
  {journal} {\bibinfo  {journal} {{EPL} (Europhysics Letters)}\ }\textbf
  {\bibinfo {volume} {106}},\ \bibinfo {pages} {58005} (\bibinfo {year}
  {2014})}\BibitemShut {NoStop}%
\bibitem [{Dat(2016{\natexlab{a}})}]{Data1}%
  \BibitemOpen
  \href@noop {} {\enquote {\bibinfo {title} {{Black Population in US |
  BlackDemographics.com}},}\ }\bibinfo {howpublished}
  {\url{https://blackdemographics.com/black-covid-19-tracker/}} (\bibinfo
  {year} {2016}{\natexlab{a}}),\ \bibinfo {note} {[Online; accessed
  2020-05-30]}\BibitemShut {NoStop}%
\bibitem [{Dat(2016{\natexlab{b}})}]{Data2}%
  \BibitemOpen
  \href@noop {} {\enquote {\bibinfo {title} {{COVID {Racial} {Data}
  {Tracker}}},}\ }\bibinfo {howpublished}
  {\url{https://covidtracking.com/race}} (\bibinfo {year}
  {2016}{\natexlab{b}}),\ \bibinfo {note} {[Online; accessed
  2020-05-30]}\BibitemShut {NoStop}%
\bibitem [{\citenamefont {Manson}\ \emph {et~al.}()\citenamefont {Manson},
  \citenamefont {Schroeder}, \citenamefont {Riper},\ and\ \citenamefont
  {Ruggles}}]{Ethnicity}%
  \BibitemOpen
  \bibfield  {author} {\bibinfo {author} {\bibfnamefont {S.}~\bibnamefont
  {Manson}}, \bibinfo {author} {\bibfnamefont {J.}~\bibnamefont {Schroeder}},
  \bibinfo {author} {\bibfnamefont {D.~V.}\ \bibnamefont {Riper}}, \ and\
  \bibinfo {author} {\bibfnamefont {S.}~\bibnamefont {Ruggles}},\ }\href@noop
  {} {\enquote {\bibinfo {title} {I{PUMS National Historical Geographic
  Information System: Version 14.0 [Database]. Minneapolis, MN: IPUMS.
  2019.}}}\ }\bibinfo {howpublished}
  {http://doi.org/10.18128/D050.V14.0}\BibitemShut {NoStop}%
\bibitem [{Com(2016)}]{Commuting}%
  \BibitemOpen
  \href@noop {} {\enquote {\bibinfo {title} {{Longitudinal Employer-Household
  Dynamics}},}\ }\bibinfo {howpublished} {\url{https://lehd.ces.census.gov/}}
  (\bibinfo {year} {2016}),\ \bibinfo {note} {online; accessed
  2020-05-30}\BibitemShut {NoStop}%
\bibitem [{\citenamefont {Linton}\ \emph {et~al.}(2020)\citenamefont {Linton},
  \citenamefont {Kobayashi}, \citenamefont {Yang}, \citenamefont {Hayashi},
  \citenamefont {Akhmetzhanov}, \citenamefont {Jung}, \citenamefont {Yuan},
  \citenamefont {Kinoshita},\ and\ \citenamefont {Nishiura}}]{Linton2020}%
  \BibitemOpen
  \bibfield  {author} {\bibinfo {author} {\bibfnamefont {N.~M.}\ \bibnamefont
  {Linton}}, \bibinfo {author} {\bibfnamefont {T.}~\bibnamefont {Kobayashi}},
  \bibinfo {author} {\bibfnamefont {Y.}~\bibnamefont {Yang}}, \bibinfo {author}
  {\bibfnamefont {K.}~\bibnamefont {Hayashi}}, \bibinfo {author} {\bibfnamefont
  {A.~R.}\ \bibnamefont {Akhmetzhanov}}, \bibinfo {author} {\bibfnamefont
  {S.-m.}\ \bibnamefont {Jung}}, \bibinfo {author} {\bibfnamefont
  {B.}~\bibnamefont {Yuan}}, \bibinfo {author} {\bibfnamefont {R.}~\bibnamefont
  {Kinoshita}}, \ and\ \bibinfo {author} {\bibfnamefont {H.}~\bibnamefont
  {Nishiura}},\ }\href {\doibase 10.3390/jcm9020538} {\bibfield  {journal}
  {\bibinfo  {journal} {Journal of Clinical Medicine}\ }\textbf {\bibinfo
  {volume} {9}} (\bibinfo {year} {2020}),\ 10.3390/jcm9020538}\BibitemShut
  {NoStop}%
\bibitem [{\citenamefont {Dey}\ \emph {et~al.}(2020)\citenamefont {Dey},
  \citenamefont {Frazis}, \citenamefont {Loewenstein},\ and\ \citenamefont
  {Sun}}]{Remote2020}%
  \BibitemOpen
  \bibfield  {author} {\bibinfo {author} {\bibfnamefont {M.}~\bibnamefont
  {Dey}}, \bibinfo {author} {\bibfnamefont {H.}~\bibnamefont {Frazis}},
  \bibinfo {author} {\bibfnamefont {M.~A.}\ \bibnamefont {Loewenstein}}, \ and\
  \bibinfo {author} {\bibfnamefont {H.}~\bibnamefont {Sun}},\ }\href@noop {}
  {\enquote {\bibinfo {title} {{Ability to work from home: evidence from two
  surveys and implications for the labor market in the COVID-19 pandemic}},}\
  }\bibinfo {howpublished} {https://doi.org/10.21916/mlr.2020.14} (\bibinfo
  {year} {2020})\BibitemShut {NoStop}%
\bibitem [{\citenamefont {of~Labor~Statistics}(2020)}]{LaborStats}%
  \BibitemOpen
  \bibfield  {author} {\bibinfo {author} {\bibfnamefont {B.}~\bibnamefont
  {of~Labor~Statistics}},\ }\href {https://www.bls.gov/cps/cpsaat11.htm}
  {}\bibinfo {type} {Tech. Rep.}\ (\bibinfo  {institution} {US Government},\
  \bibinfo {year} {2020})\BibitemShut {NoStop}%
\bibitem [{Mea()}]{MeatPlant}%
  \BibitemOpen
  \href@noop {} {\ }\BibitemShut {NoStop}%
\bibitem [{\citenamefont {Sy}\ \emph {et~al.}(2020)\citenamefont {Sy},
  \citenamefont {Martinez}, \citenamefont {Rader},\ and\ \citenamefont
  {White}}]{Sy2020}%
  \BibitemOpen
  \bibfield  {author} {\bibinfo {author} {\bibfnamefont {K.~T.~L.}\
  \bibnamefont {Sy}}, \bibinfo {author} {\bibfnamefont {M.~E.}\ \bibnamefont
  {Martinez}}, \bibinfo {author} {\bibfnamefont {B.}~\bibnamefont {Rader}}, \
  and\ \bibinfo {author} {\bibfnamefont {L.~F.}\ \bibnamefont {White}},\ }\href
  {\doibase 10.1101/2020.05.28.20115949} {\bibfield  {journal} {\bibinfo
  {journal} {medRxiv}\ } (\bibinfo {year} {2020}),\
  10.1101/2020.05.28.20115949},\ 
  \BibitemShut {NoStop}%
\bibitem [{\citenamefont {Farber}\ \emph {et~al.}(2012)\citenamefont {Farber},
  \citenamefont {P{\'a}ez},\ and\ \citenamefont {Morency}}]{Farber2012}%
  \BibitemOpen
  \bibfield  {author} {\bibinfo {author} {\bibfnamefont {S.}~\bibnamefont
  {Farber}}, \bibinfo {author} {\bibfnamefont {A.}~\bibnamefont {P{\'a}ez}}, \
  and\ \bibinfo {author} {\bibfnamefont {C.}~\bibnamefont {Morency}},\
  }\href@noop {} {\bibfield  {journal} {\bibinfo  {journal} {Environment and
  Planning A}\ }\textbf {\bibinfo {volume} {44}},\ \bibinfo {pages} {315}
  (\bibinfo {year} {2012})}\BibitemShut {NoStop}%
\bibitem [{\citenamefont {Ballester}\ and\ \citenamefont
  {Vorsatz}(2014)}]{Ballester2014}%
  \BibitemOpen
  \bibfield  {author} {\bibinfo {author} {\bibfnamefont {C.}~\bibnamefont
  {Ballester}}\ and\ \bibinfo {author} {\bibfnamefont {M.}~\bibnamefont
  {Vorsatz}},\ }\href {\doibase 10.1162/REST\_a\_00399} {\bibfield  {journal}
  {\bibinfo  {journal} {The Review of Economics and Statistics}\ }\textbf
  {\bibinfo {volume} {96}},\ \bibinfo {pages} {383} (\bibinfo {year}
  {2014})}\BibitemShut {NoStop}%
\bibitem [{\citenamefont {Olteanu}\ \emph {et~al.}(2019)\citenamefont
  {Olteanu}, \citenamefont {Randon-Furling},\ and\ \citenamefont
  {Clark}}]{Olteanu2019}%
  \BibitemOpen
  \bibfield  {author} {\bibinfo {author} {\bibfnamefont {M.}~\bibnamefont
  {Olteanu}}, \bibinfo {author} {\bibfnamefont {J.}~\bibnamefont
  {Randon-Furling}}, \ and\ \bibinfo {author} {\bibfnamefont {W.~A.}\
  \bibnamefont {Clark}},\ }\href@noop {} {\bibfield  {journal} {\bibinfo
  {journal} {Proceedings of the National Academy of Sciences}\ }\textbf
  {\bibinfo {volume} {116}},\ \bibinfo {pages} {12250} (\bibinfo {year}
  {2019})}\BibitemShut {NoStop}%
\bibitem [{\citenamefont {Kawabata}\ and\ \citenamefont
  {Shen}(2007)}]{Kawabata2007}%
  \BibitemOpen
  \bibfield  {author} {\bibinfo {author} {\bibfnamefont {M.}~\bibnamefont
  {Kawabata}}\ and\ \bibinfo {author} {\bibfnamefont {Q.}~\bibnamefont
  {Shen}},\ }\href@noop {} {\bibfield  {journal} {\bibinfo  {journal} {Urban
  Studies}\ }\textbf {\bibinfo {volume} {44}},\ \bibinfo {pages} {1759}
  (\bibinfo {year} {2007})}\BibitemShut {NoStop}%
\bibitem [{\citenamefont {Graif}\ \emph {et~al.}(2017)\citenamefont {Graif},
  \citenamefont {Lungeanu},\ and\ \citenamefont {Yetter}}]{Graif2017}%
  \BibitemOpen
  \bibfield  {author} {\bibinfo {author} {\bibfnamefont {C.}~\bibnamefont
  {Graif}}, \bibinfo {author} {\bibfnamefont {A.}~\bibnamefont {Lungeanu}}, \
  and\ \bibinfo {author} {\bibfnamefont {A.~M.}\ \bibnamefont {Yetter}},\
  }\href@noop {} {\bibfield  {journal} {\bibinfo  {journal} {Social Networks}\
  }\textbf {\bibinfo {volume} {51}},\ \bibinfo {pages} {40} (\bibinfo {year}
  {2017})}\BibitemShut {NoStop}%
\bibitem [{\citenamefont {Wang}\ \emph {et~al.}(2018)\citenamefont {Wang},
  \citenamefont {Phillips}, \citenamefont {Small},\ and\ \citenamefont
  {Sampson}}]{Wang2018}%
  \BibitemOpen
  \bibfield  {author} {\bibinfo {author} {\bibfnamefont {Q.}~\bibnamefont
  {Wang}}, \bibinfo {author} {\bibfnamefont {N.~E.}\ \bibnamefont {Phillips}},
  \bibinfo {author} {\bibfnamefont {M.~L.}\ \bibnamefont {Small}}, \ and\
  \bibinfo {author} {\bibfnamefont {R.~J.}\ \bibnamefont {Sampson}},\
  }\href@noop {} {\bibfield  {journal} {\bibinfo  {journal} {Proceedings of the
  National Academy of Sciences USA}\ }\textbf {\bibinfo {volume} {115}},\
  \bibinfo {pages} {7735} (\bibinfo {year} {2018})}\BibitemShut {NoStop}%
\bibitem [{\citenamefont {Bureau}(2019)}]{Population}%
  \BibitemOpen
  \bibfield  {author} {\bibinfo {author} {\bibfnamefont {U.~C.}\ \bibnamefont
  {Bureau}},\ }\href
  {https://www.census.gov/newsroom/press-kits/2019/acs-5-year.html} {\emph
  {\bibinfo {title} {American Community Survey 2014-2018 5-Year Data
  Release}}},\ \bibinfo {type} {Tech. Rep.}\ (\bibinfo  {institution} {US
  Government},\ \bibinfo {year} {2019})\BibitemShut {NoStop}%
\bibitem [{\citenamefont {Masuda}\ \emph {et~al.}(2017)\citenamefont {Masuda},
  \citenamefont {Porter},\ and\ \citenamefont {Lambiotte}}]{Masuda2017}%
  \BibitemOpen
  \bibfield  {author} {\bibinfo {author} {\bibfnamefont {N.}~\bibnamefont
  {Masuda}}, \bibinfo {author} {\bibfnamefont {M.~A.}\ \bibnamefont {Porter}},
  \ and\ \bibinfo {author} {\bibfnamefont {R.}~\bibnamefont {Lambiotte}},\
  }\href {\doibase https://doi.org/10.1016/j.physrep.2017.07.007} {\bibfield
  {journal} {\bibinfo  {journal} {Physics Reports}\ }\textbf {\bibinfo {volume}
  {716-717}},\ \bibinfo {pages} {1 } (\bibinfo {year} {2017})}\BibitemShut
  {NoStop}%
\bibitem [{\citenamefont {Getis}\ and\ \citenamefont {Ord}(2010)}]{Getis2010}%
  \BibitemOpen
  \bibfield  {author} {\bibinfo {author} {\bibfnamefont {A.}~\bibnamefont
  {Getis}}\ and\ \bibinfo {author} {\bibfnamefont {J.~K.}\ \bibnamefont
  {Ord}},\ }in\ \href@noop {} {\emph {\bibinfo {booktitle} {Perspectives on
  spatial data analysis}}}\ (\bibinfo  {publisher} {Springer},\ \bibinfo {year}
  {2010})\ pp.\ \bibinfo {pages} {127--145}\BibitemShut {NoStop}%
\end{thebibliography}
\end{document}